\documentclass[12pt,a4paper]{article}
\usepackage{indentfirst} 
\usepackage{amsmath,amssymb}   
\usepackage{latexsym} 
\usepackage{graphicx} 
\usepackage{subfigure}   
\usepackage{color}
\usepackage{CJK}
\usepackage{leftidx}
\usepackage{bm}

\usepackage[left=25mm,right=25mm,bottom=30mm,top=25mm]{geometry}

\begin{document}

\title{On the selectivity of KcsA potassium channel: asymptotic analysis and computation}
\author{Zilong Song$^{a,b}$, Xiulei Cao$^{a,b}$, Tzyy-Leng Horng$^{c,d}$, and Huaxiong Huang$^{a,b}$}
\maketitle

\noindent $^a$Department of Mathematics and Statistics, York University, Toronto, Ontario, Canada
 
\noindent  $^b$ Fields Institute for Research in Mathematical Sciences, Toronto, Ontario, Canada

\noindent  $^c$ Department of Applied Mathematics, Feng Chia University, Taichung 40724, Taiwan

\noindent $^d$ National Center for Theoretical Sciences, Taipei Office, Taipei, Taiwan 10617

\begin{abstract}

\noindent Potassium (K$^+$) channels regulate the flux of K$^+$ ions through cell membranes and plays significant roles in many physiological functions. This work studies the KcsA potassium channel, including the selectivity and current-voltage (IV) relations. A modified Poisson-Nernst-Planck system is employed, which include the size effect by Bikerman model and solvation energy by Born model. The selectivity of KcsA for various ions (K$^+$, Na$^+$, Cl$^-$, Ca$^{2+}$ and Ba$^{2+}$) is studied analytically, and the profiles of concentrations and electric potential are provided. The selectivity is mainly influenced by permanent negative charges in filter of channel and the ion sizes. K$^+$ is always selected compared with Na$^+$ (or Cl$^-$), as smaller ion size of Na$^+$ causes larger solvation energy. There is a transition for selectivity among K$^+$ and divalent ions (Ca$^{2+}$ and Ba$^{2+}$), when negative charge in filter exceeds a critical value determined by ion size. This explains why divalent ions can block the KcsA channel. The profiles and IV relations are studied by analytical, numerical and hybrid methods, and are cross-validated. The results show the selectivity of the channel and also the saturation of IV curve. A simple strategy is given to compute IV relations analytically, as first approximation.  The numerical method deals with general structure or parameters, but the limitations and difficulties of pure numerical simulation are also pointed out. The hybrid method provides IV relations most effectively for comparison. The reason for saturation of IV relation is illustrated,  and the IV curve shows agreement with the profile and scale of experimental results.

\end{abstract}

\section{Introduction}

Rapid communication in many organisms relies on fast propagation of electric signals, which in turn depend on a specialized class of protein molecules called ion channels. When the ion channels are opened on the cell membrane by either chemical ligands or membrane depolarization, they allow ionic flux across the cell membrane and lead to rapid changes of membrane potentials. Potassium (K$^+$) channels regulate the flux of K$^+$ ions through cell membranes and participate several physiological functions such as the maintenance of the resting membrane potential, the excitation of nerve and muscle cells, the secretion of hormones, and sensory transduction \cite{Hille2001,handbook2015}. When dysfunctional, ion channels would cause a number of diseases. Therefore, understanding  how the molecular structure determines channel function is profoundly interesting, both for the biological and for the medical sciences \cite{MacKinnon1998a, Ackerman1997, Catterall2010}.

The X-ray crystallographic structures of distinct potassium channels reveal a common architecture of the pore \cite{MacKinnon2004,Gouaux2005}. Four subunits are symmetrically arranged around the channel axis, with each subunit having at least two transmembrane helixes separated by a re-entrant P-loop and selectivity filter (SF). K$^+$ channels are  the most extensively studied family of ion channels, both experimentally and computationally, and the KcsA structure \cite{Thompson2008,Wu2017} has been the most popular one among K$^+$ channels since it is the first K$^+$ channel to be crystalized. Many computational and experimental data of KcsA is available for comparison.

SF of K$^+$ channels is the essential element to their permeation and selectivity mechanisms \cite{Doyle1998}. Thousands of millions of K$^+$ ions per second can diffuse in single file down their electrochemical gradient across the membrane at physiological conditions \cite{Aqvist2000,Miller2001}. Each subunit contributes to SF with a conserved signature peptide, namely TVGYG in most of the channels \cite{MacKinnon1998a}. The carbonyl oxygens of the backbone of SF point toward the lumen and orchestrate the movements of ions in and out of the channel. These carbonyl oxygens together with the side-chain hydroxyl oxygen of a threonine residue define four ion-binding sites in SF, designated S1-S4 starting at the extracellular side \cite{Zhou2001}. In addition, K$^+$ ion can bind in the central water-filled cavity of pore and two alternate positions at the extracellular side of pore \cite{Zhou2001}.

SF is generally too narrow to accommodate a K$^+$ ion with its hydration shell, and thus K$^+$ ions must be dehydrated to enter SF, when attracted by the strong negative charges of carbonyl oxygens in SF. K$^+$ ion must replace its solvation shell by the carbonyl oxygens in the backbone of SF. Each of these protein sites binds K$^+$ ions with a tight-fitting cage of 8 carbonyl oxygen atoms that resembles the solvation shell of a hydrated K$^+$ ion.

Classical Poisson-Nernst-Planck (PNP) system has been widely applied to model ionic transport in biological setting as well as other areas \cite{Gillespie2001, Markowich2013, Jasielec2013}. Various analysis and computation \cite{Singer2009,song2018b,Flavell2014} regarding this system have been attempted in the literature. Current-voltage (IV) relation is an important functional characteristic of ion channels and can be determined experimentally. PNP theory has been successfully applied to model wide ion channels, and has reproduced the experimental IV data quite successfully \cite{Zheng2011, Esienberg2003}. However, when used in narrow ion channels, such as KcsA, the classical PNP system is not suitable anymore, due to the extremely narrow SF. 
This is because classical PNP neglects the size of ions and therefore overestimates the K ion occupancy of SF. Also, classical PNP does not consider solvation energy barrier that is significantly encountered by K$^+$ ions when dehydrated to enter SF. 

Various modified PNP system have been proposed to include the steric or size effect of ions \cite{Horng2012,Gavish2018,Lin2014,Bazant2007,Lu2011}. In this study, we employed Bikerman model with specific ion sizes \cite{Horng2017}, which is one of the widely accepted models in literature. In addition, solvation energy based on Born model is included in the present formulation based on dehydration of ions and its importance emphasized above \cite{Born1920}. The adoption of Bikerman model is because of its simplicity and availability of some analytical results, which we believe can provide more physical insights into the mechanism of ion channels. 

It is well known that potassium channels have high selectivity of potassium ion over sodium ion (K$^+$ is $10^4$ times more permeant than Na$^+$) \cite{Hille2001}. Though K$^+$ and Na$^+$ have the same valence and therefore they have the same electrostatic affinity to carbonyl oxygens in SF, K$^+$ encounters less Born solvation energy barrier than Na$^+$ when passing through SF due to its size slightly larger than Na$^+$. However, few studies were found about selectivity between K$^+$ and alkaline earth ions like Ca$^{2+}$ and Ba$^{2+}$. Alkaline earth ions generally have stronger electrostatic affinity to SF than K$^+$ due to their divalence, but at the same time also bear larger Born solvation energy barrier again due to their divalence. The blockage of KcsA by Ba$^{2+}$ has demonstrated this strong competition of SF occupancy between electrostatic affinity and solvation energy \cite{Piasta2011}.

Here we employed one-dimensional asymptotic analysis and numerical simulation of present model to study (i) the mechanism of channel selectivity among K$^+$ and other ions;  (ii) the mechanism causing IV curve to be saturated when voltage gets large as recorded in experiments \cite{Miller2002}.  More precisely, we give simple explanation and analytical formulas regrading the selectivity among K$^+$, Na$^+$, Cl$^-$, Ca$^{2+}$ and Ba$^{2+}$. The selectivity is mainly influenced by permanent negative charges in SF and the ion sizes. The smaller ion size of Na$^{+}$ compared with K$^+$ gives a larger solvation energy barrier to enter filter, and hence its concentration is exponentially smaller (not selected). When negative charge in SF exceeds a critical value (given by ion size), SF starts to recruit divalent cations to coexist with K$^+$ by squeezing some K$^+$ out of SF, since divalent cations can do better in balancing the strong negative charge in narrow SF. Although Born solvation energy is increased by this recruitment of divalent cations into SF, electrostatic energy is decreased more for compensation and total energy is actually decreased then.
We have studied the IV curves by analytical, numerical, and hybrid methods, and cross-validated the results. The results have revealed the reason for saturation of IV curve and pointed out the difficulties in numerical simulations for some cases. The IV curve also shows agreement with the profile and scale of experimental results. 

The manuscript is arranged as follows. Section 2 formulates the mathematical model, i.e., the modified PNP system with Bikerman model and Born model. Section 3 deals with the the equilibrium case with zero flux, implying the selectivity of channel. Section 4 provides analytical results for IV curve for non-equilibrium case. Direct numerical simulations are conducted in Section 5 and a hybrid computational-asymptotic analysis is done in Section 6. Finally, some concluding remarks are drawn.


\section{Mathematical model}

We consider the Bikerman model with specific ion sizes \cite{Horng2017}, and include permanent charge and solvation energy into the Poisson-Nernst-Planck (PNP) formulation. The original one-dimensional (1D) system for $-L<x<L$ is 
\begin{equation}
\label{eq1}
\begin{aligned}
&- \frac{1}{A(x)} \partial_x (\epsilon_0 \epsilon_r(x) A(x) \partial_x \phi) = e_0\left( \sum_{k=1}^n z_k c_k - q(x)\right),\\
& \partial_t c_i + \frac{1}{A(x)} \partial_x J_i =0, \quad J_i = -A(x) \frac{D_i }{k_B T} c_i \partial_x \mu_i,
\end{aligned}
\end{equation}
where $c_i$ ($i=1,..,n$) denote the concentrations of ions, $\phi$ is electric potential, $A(x)$ is the cross section area, $q(x)$ is the permanent charge (positive $q$ means negative fixed charge), $\epsilon_r(x)$ is the relative permittivity, and $k_B,T,\epsilon_0,e_0$ are some constants (see Appendix \ref{appendixA}). The electro-chemical potentials are given by
\begin{equation}
\label{eq2}
\begin{aligned}
& \mu_i = k_B T \left( \log (c_i a_i^3) - \log \left(1- \sum_{k=1}^n c_k a_k^3\right) \right) + z_i e \phi + W_{i},\quad i=1,..,n,
\end{aligned}
\end{equation}
where $a_i$ are the effective diameters of ions, and $W_{i}$ is solvation energy
\begin{equation}
\label{eq3}
\begin{aligned}
& W_{i}(x)= \frac{z_i^2 e_0^2}{8 \pi \epsilon_0 a_i} \left(\frac{1}{\epsilon_r(x)}-1 \right).
\end{aligned}
\end{equation}

\begin{figure}[h]
\begin{center}
\includegraphics[width=4 in]{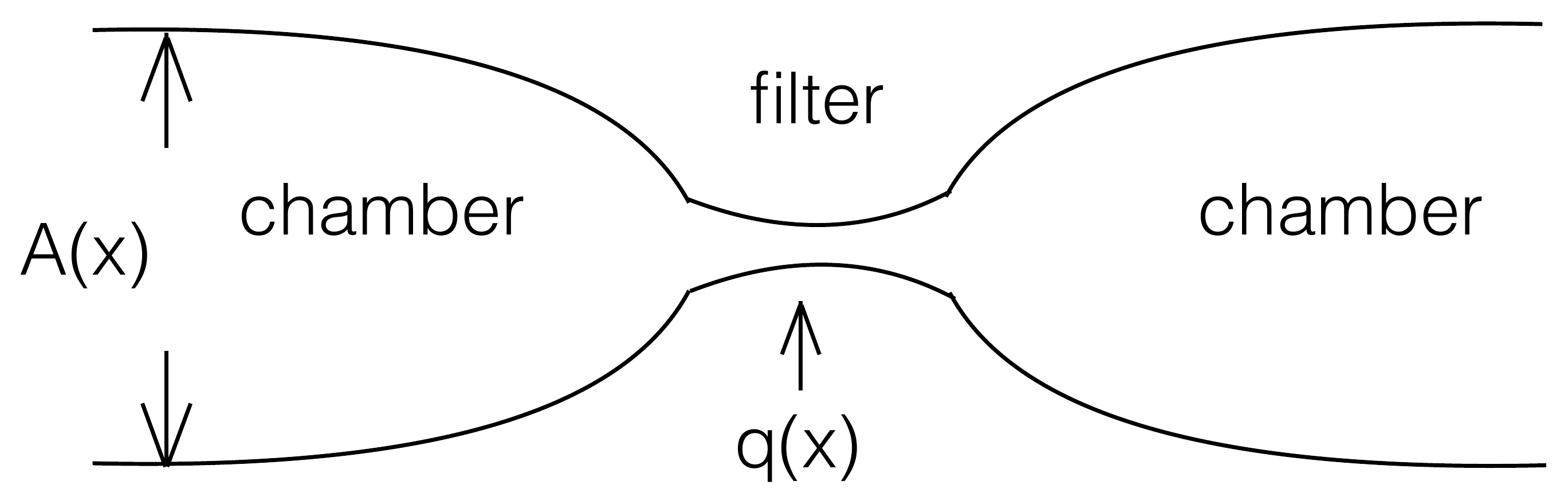}
\caption{\label{fig1} Sketch of the potassium channel.}
\end{center}
\end{figure}

Figure \ref{fig1} schematically shows the setup, where the filter of channel lies between two chambers, linking extracellular and intracellular spaces, respectively. The length of filter is  $L_f$ and the length of each chamber is set as $L_b$ (then $L=L_b + \frac{1}{2}L_f$), where some part of reservoir is included we consider a relative large $L_b$. The cross section area $A(x)$ in filter is much smaller than that of the chamber region. The permanent negative charge $q(x)$ due to carbonyl oxygens and threonine residues is confined in the small volume of the filter, so the effective $q(x)$ in the model is extremely large compared with chamber concentrations \cite{Hollerbach1999,Hollerbach2000}. This further implies that the filter attracts counter-ions and thus the saturation of ions in filter means few water molecules in filter or ions are dehydrated. Therefore, the dielectric constant $\epsilon_r(x)$ would be much smaller in filter, and this justifies the introduction of above solvation energy $W_{i}(x)$, which gives the energy barrier from chamber to filter. In later analysis, the solvation energy also causes jumps in concentrations from chamber to filter while maintaining continuous electro-chemical potentials $\mu_i$.

In the chamber, the model is approximately the classic PNP system, where the size effect is negligible and $q(x)=0$.  With this in mind, we do a traditional nondimensionalization with reference scales in the chamber. We set
\begin{equation}
\label{eq4}
\begin{aligned}
& \tilde{x} = \frac{x}{L}, \quad \tilde{c}_i = \frac{c_i}{c_0},\quad \tilde{\phi} = \frac{\phi}{\phi_0},\quad \tilde{D}_i = \frac{D_i}{D_0},  \quad \tilde{t} = \frac{t}{L^2/D_0}, \quad \tilde{\epsilon}_r = \frac{\epsilon_r}{\epsilon_{rb}},\\
& \tilde{a}_i = \frac{a_i}{a_0},\quad \tilde{A} =\frac{A}{A_b},\quad \tilde{L}_f =\frac{L_f}{L},\quad \tilde{q} = \frac{q}{c_0}, \quad \tilde{W}_{i} = \frac{W_{i}}{k_B T},
\end{aligned}
\end{equation}
where $a_0$ is a reference diameter, $D_0$ is a reference diffusion constant, $A_b$ is reference (maximum) cross section area in chamber and $\epsilon_{rb}$ is (maximum) relative permittivity at farther end of chamber (see (\ref{eqA2}) in  Appendix \ref{appendixA} for their values).

By removing the tilde, the dimensionless system in $-1<x<1$ is
\begin{equation}
\label{eq5}
\begin{aligned}
&- \epsilon^2 \frac{1}{A(x)} \partial_x ( \epsilon_r(x) A(x) \partial_x \phi) = \sum_{k=1}^n z_k c_k - q(x),\\
& \partial_t c_i + \frac{1}{A(x)} \partial_x J_i =0, \quad J_i = -A(x) D_i  c_i \partial_x \mu_i,
\end{aligned}
\end{equation}
where $i=1,..,n$ and
\begin{equation}
\label{eq6}
\begin{aligned}
& \mu_i (x)= \log [c_i(x)] - \log \left(1- \sum_{k=1}^n c_k(x) a_k^3 \delta \right) + z_i \phi(x) + W_{i}(x),\\
& W_{i} (x)= \frac{z_i^2}{a_i} \left(\frac{1}{\epsilon_{rb} \epsilon_r(x)}-1\right) W_0,
\end{aligned}
\end{equation}
where the first term in $\mu_i$ is originally  $\log (c_i a_i^3 \delta)$ by dimensionalization but we removed the constant $\log (a_i^3 \delta)$ from $\mu_i$ since this would not affect the system. The dimensionless parameters are 
\begin{equation}
\label{eq7}
\begin{aligned}
&\epsilon = \sqrt{\frac{\epsilon_0 \epsilon_{rb} k_B T}{e_0^2 c_0 L^2}},\quad \delta = a_0^3 c_0,\quad W_0 =  \frac{e^2}{8 \pi \epsilon_0 a_0 k_B T}.
\end{aligned}
\end{equation}
Please refer to Appendix \ref{appendixA} for the estimates of parameters in this system.
One easily see that with $W_i=$constant and as $c_k \delta$ tends to 0, the above $\mu_i$ goes back to that in classical PNP system. This is the case in chamber region, whereas in filter region $c_k$ is quite large and $c_k \delta$ terms can not be neglected.

\section{Equilibrium case with zero flux}

In this section, we study the selectivity of the channel in equilibrium case, for simplicity. We will see the conclusions also hold for non-equilibrium case with finite fluxes. This is seen in analysis of this section that the boundary conditions (inducing finite flux when different) have negligible or exponential small impact on the results. This is also verified by analytical and numerical results in non-equilibrium case, as the selected ions are in equilibrium in filter (non-equlibrium outside), see Figures of $\mu_i$ in Sections 4 \& 5 and analysis before (\ref{Eq63}).

The same  boundary conditions at two ends of chamber are used
\begin{equation}
\label{eq10}
\begin{aligned}
& c_{i} (x) = c_{ib},\quad \phi(x) = \phi_b, \quad \textrm{at} \quad x=\pm1,
\end{aligned}
\end{equation}
where $i=1,..,n$ and the electro-neutrality (EN) condition $\sum z_i c_{ib} = 0$ is satisfied. Therefore, there's no flux across the filter. The aim is to study the relative concentrations of ions in filter under different situations, which would imply the selectivity.

In general case, we notice that by definition of $\mu_i$ in (\ref{eq6}) we can solve $c_i$ ($i=1,..,n$) in terms of $\phi$ and $\mu_i$ (see Appendix \ref{appendixB})
\begin{equation}
\label{eq12}
\begin{aligned}
 c_i = \frac{e^{ \mu_i-W_i- z_i \phi }}{(1+ F \delta)}, \quad F= \sum_{k=1}^n a_k^3 e^{ \mu_k - W_k- z_k \phi}.
\end{aligned}
\end{equation}
For equilibrium case, by $J_i=0$, we conclude that $\mu_i$ is constant throughout filter and chamber
\begin{equation}
\label{eq11}
\begin{aligned}
& \mu_i(x)= B_i = \log (c_{ib}) - \log \left(1- \sum_{k=1}^n  c_{ib} a_k^3 \delta\right) + z_i \phi_b + W_i(1)
\end{aligned}
\end{equation}
where the constant $B_i$ is determined by boundary conditions in (\ref{eq10}). In this case, by substituting (\ref{eq11}) into (\ref{eq12}), $c_i$ is expressed explicitly in terms of $\phi$.

Since the filter region is quite small, it is natural to adopt some effective charge  \cite{Hollerbach2000,Nadler2003}.  We assume $q(x)=q$ is a large constant in filter, and treat $q$ as a crucial parameter. Depending on the relative magnitude of $q$, we have either electro-neutral (EN) case or non-EN case in filter. We also assume $\epsilon_r(x) = \epsilon_{r0}$ in filter, where $\epsilon_{r0}$ is constant (say $1/40$, corresponding to original $\epsilon_r=2$). Note by choice of scaling in (\ref{eq4}), we have $\epsilon_r(\pm1)=1$. 

\subsection{K$^+$/Na$^+$ selectivity}

In this subsection, we consider the case with three ions K$^+$, Na$^+$ and Cl$^-$ (respectively $c_1,c_2$ and $c_3$), and study the selectivity between Na$^+$ and K$^+$. 

From the expression of $c_1$ and $c_2$ in (\ref{eq12},\ref{eq11}), we get in the filter
\begin{equation}
\label{eq13}
\begin{aligned}
 \frac{c_1}{c_2} = \frac{e^{ B_1-W_1(0)}}{e^{ B_2-W_2(0)}}.
\end{aligned}
\end{equation}
Thus, the ratio $c_1/c_2$ is a constant independent of $\phi$ and $x$ in filter. More precisely, we have
\begin{equation}
\label{eq14}
\begin{aligned}
& B_i- W_i(0) \approx -\Delta W_i + \log c_{ib} + z_i \phi_b, \quad i=1,..,n,
\end{aligned}
\end{equation}
where the $O(\delta)$ term in chamber has been dropped and $ \Delta W_i $ is the barrier from chamber to filter due to solvation energy
\begin{equation}
\label{eq15}
\begin{aligned}
& \Delta W_i= W_i(0)-W_i(1) = \frac{z_i^2}{a_i} (\frac{1}{\epsilon_{rb} \epsilon_{r0}}-\frac{1}{\epsilon_{rb}}) W_0,\quad i=1,..,n.
\end{aligned}
\end{equation}

Since the diameter of K$^+$ is larger than that of Na$^+$, the barrier of K$^+$ from chamber to filter is smaller, i.e.,  
\begin{equation}
\label{eq16}
\begin{aligned}
& a_1>a_2 \quad \Rightarrow \Delta W_1<\Delta W_2.
\end{aligned}
\end{equation}
From the data in (\ref{eqA3}) of Appendix \ref{appendixA}, we get $\Delta W_i \sim W_0 \sim O(10^2)$, thus the term $\Delta W_i$ dominates the ratio (\ref{eq13}), and hence $c_2$ is always exponentially smaller than $c_1$. As both $c_1$ and $c_2$ are at most at the order of $O(q)$, the concentration $c_2$ is exponentially small and negligible in the filter. This means that K$^+$ is favored or selected in filter compared with Na$^+$, and this fact is independent of $q$. Based on data in (\ref{eqA2},\ref{eqA3}) of Appendix \ref{appendixA}, we get
\begin{equation}
\label{eq17}
\begin{aligned}
& \Delta W_1=49.2,\quad \Delta W_2= 66.6.
\end{aligned}
\end{equation}
This implies that, the term $\Delta W_i$ dominates in (\ref{eq13},\ref{eq14}) unless the chamber contraction $c_{2b}$ is $10^7$ times larger than $c_{1b}$. Since boundary values $c_{ib},\phi_b$ have negligible effect, one can imagine this conclusion holds for non-equilibrium case. One can rigorously prove this by noting $\mu_i$ is monotone in non-equilibrium case with finite flux.

The high selectivity of SF for larger K$^+$ over smaller Na$^+$ has been also intensively studied by molecular dynamics (MD) \cite{egwolf2010,luzhkov2001,shrivastava2002,Miller2002} and experiments \cite{heginbotham1999,lemasurier2001}, just to name a few. Experiments show Na$^+$ can block KcsA K$^+$ current from intracellular side but not from extracellular side \cite{heginbotham1999}. This observation is explained by MD studies that Na+ would encounter a much larger energy barrier than K$^+$ when approaching S2 binding site in multi-cation knock-on entering SF from extracellular side \cite{egwolf2010}, and all binding sites are more selective to K+ than Na+ except the internal water cavity site lying at the entrance of SF from intracellular side \cite{luzhkov2001}.

\subsubsection{EN case}
From the data (\ref{eqA3},\ref{eqA5}) of Appendix \ref{appendixA}, we get $q\sim O(1/\delta)$. When $q$ does not exceed the critical value
\begin{equation}
\label{eq18}
\begin{aligned}
&  q<\frac{1}{a_1^3 \delta},
\end{aligned}
\end{equation} 
we have the EN condition in filter
\begin{equation}
\label{eq19}
\begin{aligned}
& c_1 + c_2 -c_3=q,
\end{aligned}
\end{equation}
which provides a nonlinear equation for $\phi$, with the help of (\ref{eq12}, \ref{eq11}). In fact, this is a quadratic equation for $e^\phi$. The analytic solution involves many exponential large and exponential small terms, and can easily lead to wrong or complex solutions by direct compuation with softwares (like Mathematica). It's easy to prove that $c_3$ is also exponentially small. As a leading order approximation,  we get
\begin{equation}
\label{eq20}
\begin{aligned}
& c_1^\ast =q,\quad c_k^\ast=0,\quad k=2,3,\\
& \phi^\ast =  -\Delta W_1 + \log c_{1b} + \phi_b+ \log (1- a_1^3 q\delta) - \log q.
\end{aligned}
\end{equation} 
which can also be obtained directly from analytical solution by keeping essential exponential small terms and dropping high-order exponentially small terms.

\noindent \textbf{Remark:} In above analysis, by EN condition we mean that it is valid  in most middle part of filter region. Actually, near the two edges of filter (or interface of filter and chamber), say $x=\pm s$, there is a tiny boundary layer due to large $q$ and small $\epsilon_r$ in $(\ref{eq5})_1$, where the variation of $\phi,c_i$ is quite large. In the approximation (\ref{eq20}), only some exponentially small terms are dropped, so the expressions are accurate enough.

\begin{figure}[h]
\begin{center}
\includegraphics[width=2.8 in]{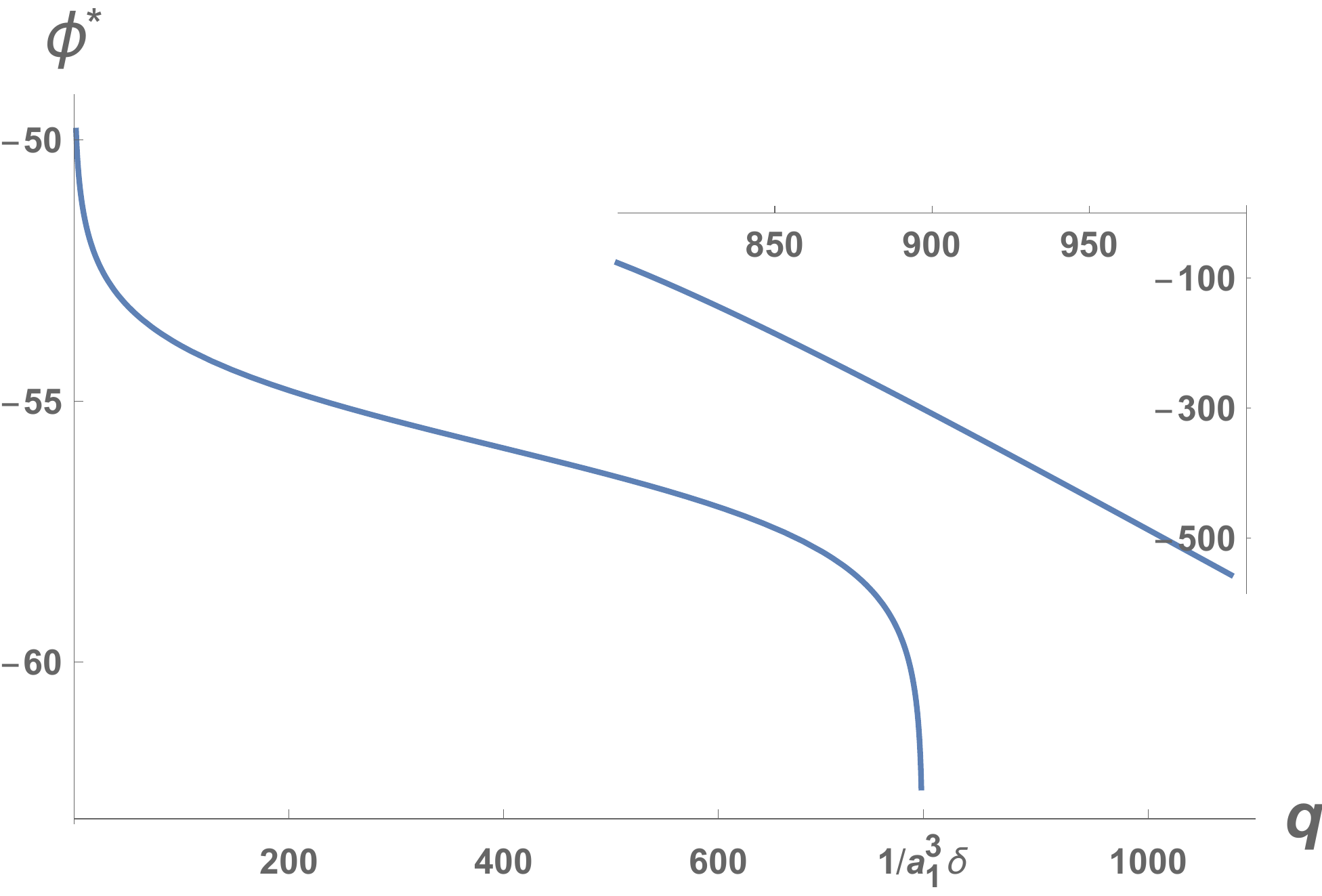} \includegraphics[width=2.8 in]{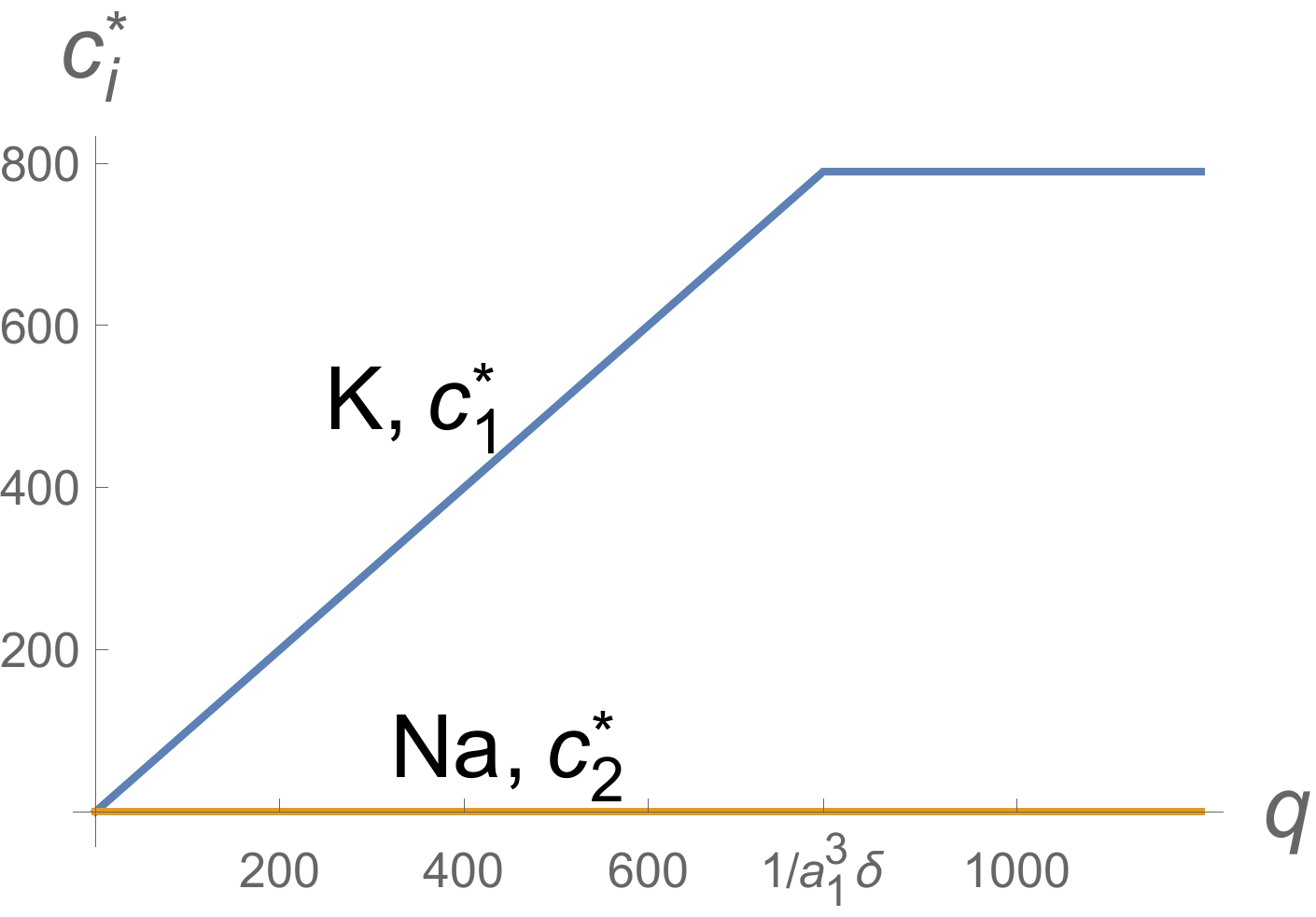}
\caption{\label{fig2}Dependence of $\phi^\ast$, $c_1^\ast,c_2^\ast$ in filter on permanent charge $q$.}
\end{center}
\end{figure}

Figure \ref{fig2} shows the dependence of above solution on $q$, with $c_{1b} =1,\phi_b= 0$ and some data in (\ref{eqA3}) of Appendix \ref{appendixA}. The value of $c_{2b}$ (assumed as $O(1)$), the profiles of $A(x)$ and $\epsilon_r(x)$, as long as $\epsilon_r(0)=1/40$,  will not affect the above approximation in filter. This will be verified in direct numerical simulations. Based on selected values of parameters, the critical value is $q=1/a_1^3\delta \approx 790$ in Figure \ref{fig2}. When $q$ is near 0 or near this critical value, the above solution is not valid, as indicated by some singularity in Figure \ref{fig2}a. When $q$ exceeds the critical value, one should solve the full equation $(\ref{eq5})_1$ in filter instead of EN condition, as we see in next subsection. For $q>1/a_1^3\delta$, the subfigure in Figure \ref{fig2}a (the minimum of $\phi$) and the curve  in Figure \ref{fig2}b are based on next subsection. Figure \ref{fig2}b shows the selectivity of K$^+$ and Na$^+$ in filter.

\subsubsection{Non-EN case}

For this case, we can not use EN condition and instead we should solve the full equation (\ref{eq5}).  Since length of filter is at the same scale of boundary layer in classical PNP of chamber region, we will introduce a new scale $X =x/ \epsilon$ to study the system. For the equilibrium case, the equation for $\phi$ is
\begin{equation}
\label{eq9}
\begin{aligned}
&- \frac{1}{A(X)} [ \epsilon_r(X) A(X) \phi'(X)]' = \sum_{k=1}^n z_k c_k - q(X),\quad -\infty <X<\infty,\\
\end{aligned}
\end{equation}
where  prime denotes the derivative with respect to $X$. In above, we consider a relatively long chamber region, so the domain is set as $\infty$ as approximation (this causes essentially no difference). The position of interface between filter and chamber is $X=S\equiv {L_f}/{2 \epsilon}$, where $S\sim O(1)$.

\begin{figure}[h]
\begin{center}
\includegraphics[width=4 in]{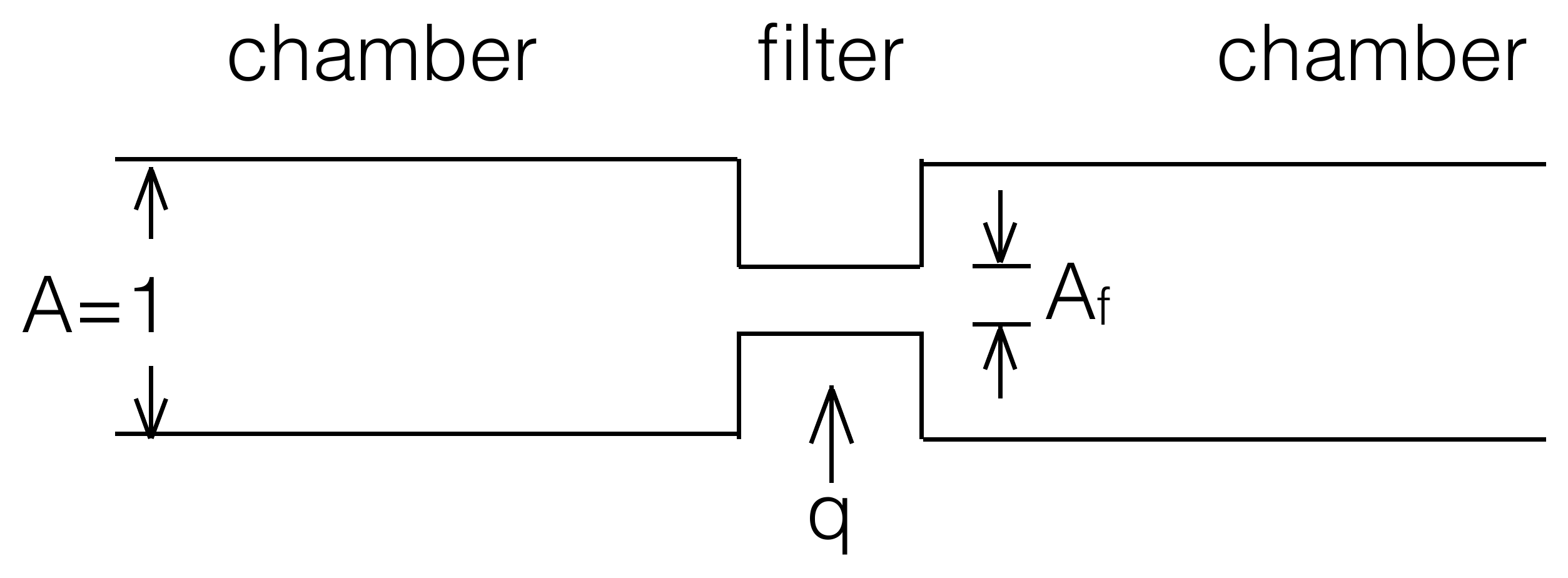}
\caption{\label{fig3} Sketch of the potassium channel with simplified geometry.}
\end{center}
\end{figure}

For simplicity, we consider a simple geometry (see Figure \ref{fig3}) that the cross section area $A(X)$, the fixed charge $q(X)$ and the relative permittivity $\epsilon_r(X)$ are constants in either chamber or filter region. We denote
\begin{equation}
\label{eq21}
\begin{aligned}
&A(X)=A_f,\quad \epsilon_r(X)= \epsilon_{r0},\quad q(X) = q,\quad -S<X<S,\\
&A(X) = 1,\quad  \epsilon_r(X)= 1,\quad q(X) = 0,\quad |X|>S.
\end{aligned}
\end{equation}
Note that some typical values are
\begin{equation}
\label{eq22}
\begin{aligned}
& A_f = 1/30,\quad q=10^3,\quad \epsilon_{r0} = 1/40,
\end{aligned}
\end{equation}
which will be used to show the results.

Because of symmetry,  we only consider the interval $X\in [0,\infty)$.  For the chamber region, equation (\ref{eq9}) reduces to classical Poisson-Boltzmann equation by neglecting the  $O(\delta)$ term,
\begin{equation}
\label{eq24}
\begin{aligned}
&- \phi''(X)= e^{-\phi} -e^{\phi},\quad S<X<\infty,\\
\end{aligned}
\end{equation}
where we have assumed the boundary conditions at $\infty$
\begin{equation}
\label{eq25}
\begin{aligned}
&\phi(\infty)=0,\quad c_1(\infty)+ c_2(\infty)=c_3(\infty)=1.
\end{aligned}
\end{equation}
It is easy to get
\begin{equation}
\label{eq27}
\begin{aligned}
& \phi' = \sqrt{2} (e^{-\phi/2} - e^{\phi/2}), \quad S<X<\infty,
\end{aligned}
\end{equation}
and hence obtain the solution $\phi(X)$ in chamber region (see (\ref{eqB1}) in Appendix \ref{appendixB}). 

In filter, we have from equation (\ref{eq9}) and symmetry condition that
\begin{equation}
\label{eq23}
\begin{aligned}
&-  \epsilon_{r0}  \phi''(X)= c_1 +c_2-c_3 - q,\quad 0<X<S,\\
& \phi'(0)=0,\quad \textrm{at} \quad X=0,
\end{aligned}
\end{equation}
where $c_i$ ($i=1,2,3$) are given by (\ref{eq12}, \ref{eq11}). One can easily prove that the function $\phi(X)$ is monotonically increasing throughout the interval $[0,\infty)$, since $c_2+c_1<q$ in filter. For filter region, since $c_2$ is exponentially small  (see the analysis below (\ref{eq13})), it can be neglected. In addition, by expression of $c_3$ in (\ref{eq12},\ref{eq11}) and some data (\ref{eqA3}) in Appendix \ref{appendixA}, we get 
\begin{equation}
\label{eq28}
\begin{aligned}
&c_3 < e^{B_3-W_3(0)+\phi},\quad B_3 - W_3(0) \approx -\Delta W_3 \approx-37.5.
\end{aligned}
\end{equation}
From the fact that $\phi$ is increasing, we get that $\phi<0$, thus $c_3$ is always exponentially small and can be neglected. Therefore, the filter equation $(\ref{eq23})_1$ is simplified to 
\begin{equation}
\label{eq29}
\begin{aligned}
&-  \epsilon_{r0}  \phi''(X)= c_1 - q,\quad c_1 = \frac{e^{B_1-W_1(0)-\phi}}{1+ \delta a_1^3 e^{B_1-W_1(0)-\phi}}.
\end{aligned}
\end{equation}
By integration, we easily get
\begin{equation}
\label{eq30}
\begin{aligned}
& \sqrt{\epsilon_{r0}} \phi' = \sqrt{2(G(\phi) - G(\phi_0))}, \\
& X= \sqrt{\frac{\epsilon_{r0}}{2}} \int_{\phi_0}^{\phi} \frac{1}{\sqrt{G(\phi) - G(\phi_0)}} d\phi
\end{aligned}
\end{equation}
where $\phi_0\equiv\phi(0)$ is to be determined, and the function $G(\phi)$ is given by
\begin{equation}
\label{eq31}
\begin{aligned}
& G(\phi) = \int^\phi q-c_1 d\phi = q\phi +\frac{1}{a_1^3 \delta} \log(1+a_1^3 \delta e^{B_1-W_1(0) - \phi}).
\end{aligned}
\end{equation} 

At interface $X=S$, we have
\begin{equation}
\label{eq26}
\begin{aligned}
&\phi(S-)=\phi(S+),\quad \epsilon_{r0} A_f \phi' (S-) = \phi'(S+).
\end{aligned}
\end{equation} 
Denote $\phi(S \pm)=\phi_s$, then the two quantities $\phi_0,\phi_s$ are determined by
\begin{equation}
\label{eq32}
\begin{aligned}
&A_f \sqrt{\epsilon_{r0}(G(\phi_s) - G(\phi_0))} = e^{-\phi_s/2} - e^{\phi_s/2},\\
& \sqrt{\frac{\epsilon_{r0}}{2}}  \int_{\phi_0}^{\phi_s} \frac{1}{\sqrt{G(\phi) - G(\phi_0)}} d\phi =S.
\end{aligned}
\end{equation}
Once they are found, we get the explicit solutions for filter and chamber.

\begin{figure}[h]
\begin{center}
\includegraphics[width=2in]{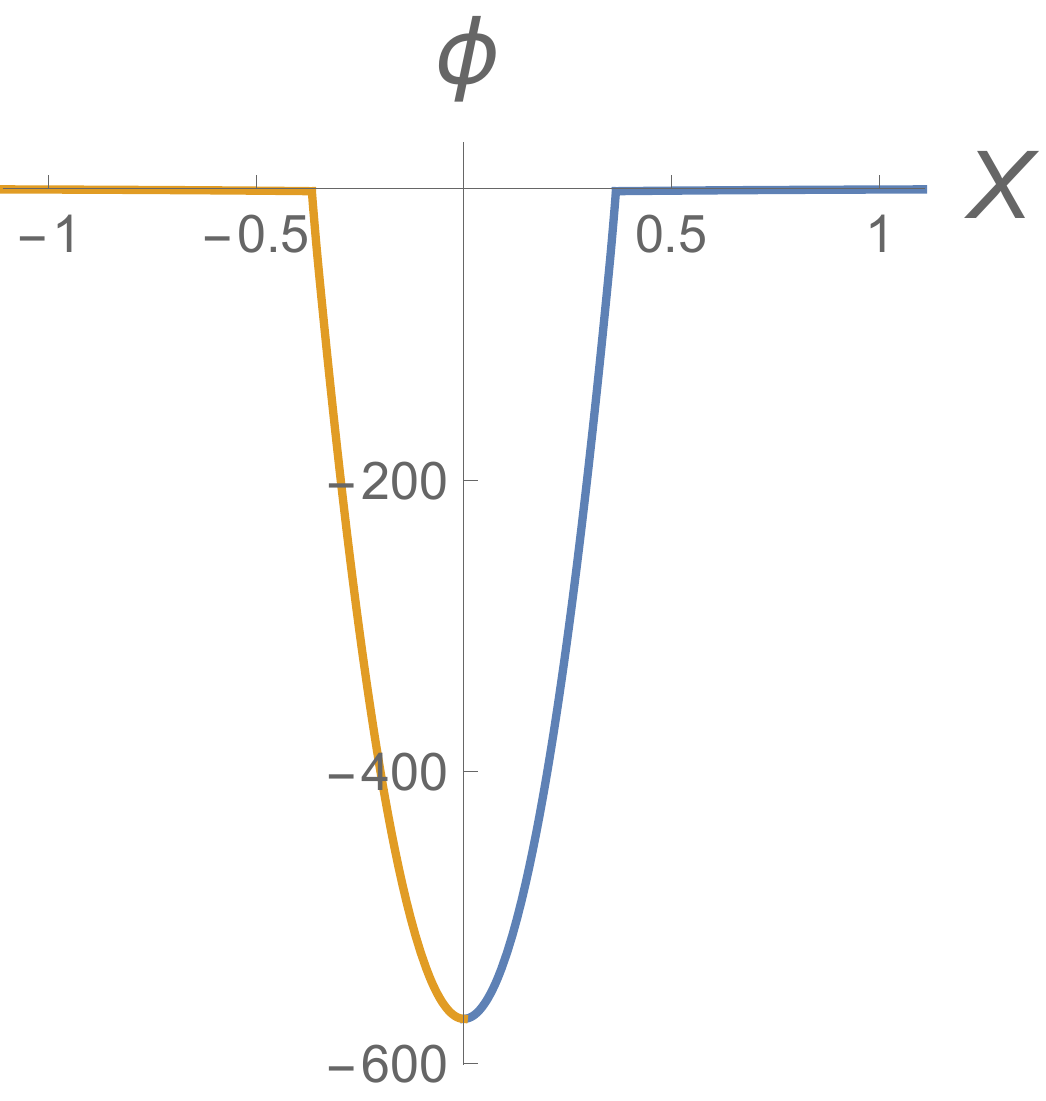} ~ \includegraphics[width=2 in]{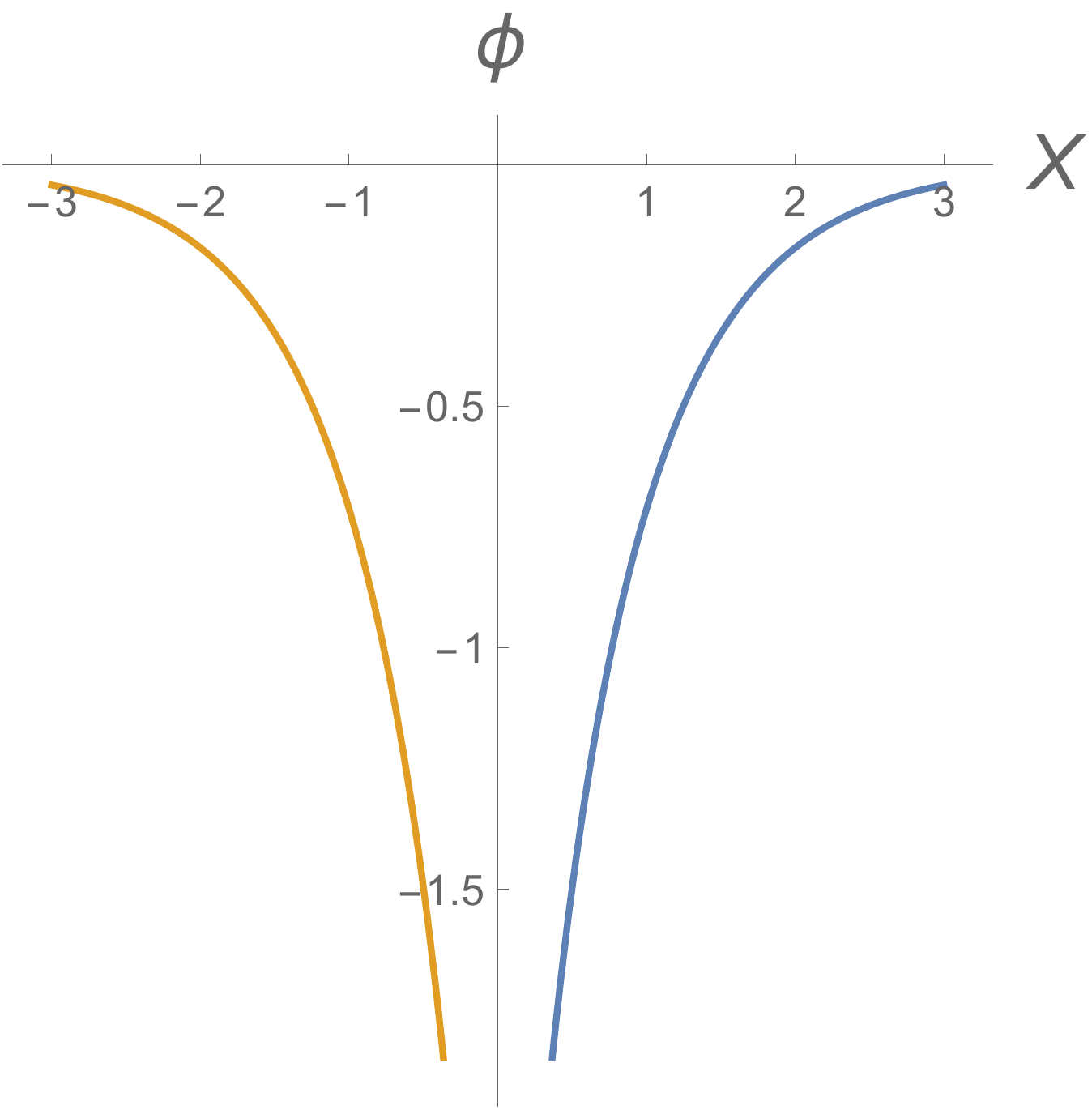} ~\includegraphics[width=2 in]{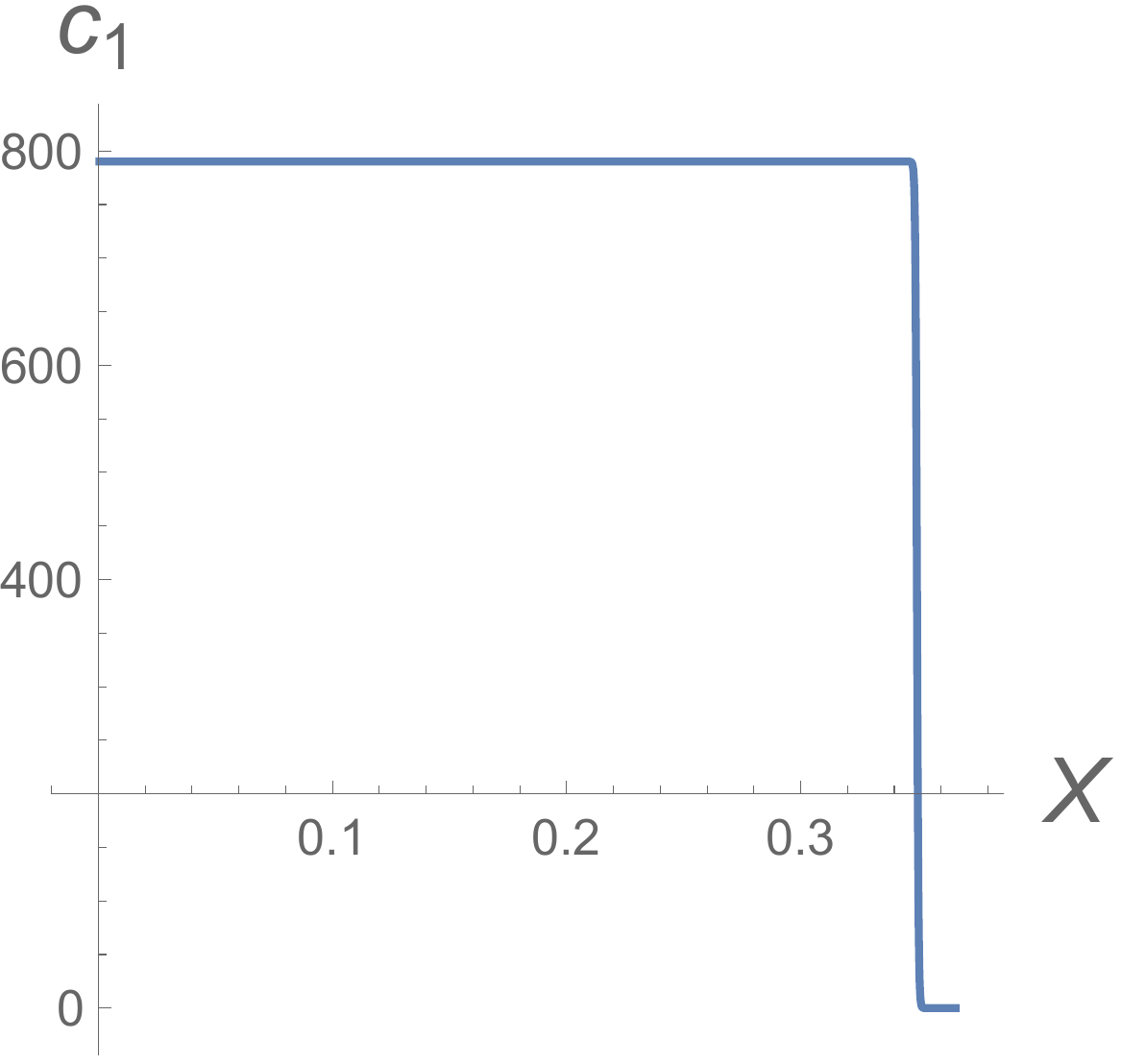} 
\caption{\label{fig4} The profiles for case $q=1000>1/a_1^3\delta$: (a) $\phi$ in both filter and chamber, (b) $\phi$ in chamber, (c) $c_1$ in right-half interval of filter.}
\end{center}
\end{figure}

Figure \ref{fig4} shows the profiles of $\phi(X)$ and $c_1(X)$, with values in (\ref{eq22}, \ref{eqA3}) and $c_1(\infty) = 1$. In filter region, Figure \ref{fig4}(a) shows that the minimum value of $\phi$ is much smaller than the EN case, and Figure \ref{fig4}(b) shows that $\phi\sim O(1)$ in chamber region. Figure \ref{fig4}(c) shows the profile of $c_1$ in right-half filter region, indicating that $c_1=1/a_1^3\delta$ in most middle part of filter and there is a inner transition point from exponential small to that value. This means that in most part of filter, it is fully packed
\begin{equation}
\label{eq32_1}
\begin{aligned}
&1- \delta \sum_{i=1}^3 a_i^3 c_i =0,
\end{aligned}
\end{equation}
but it still needs the derivatives $\phi''(X)$ to balance the large $q$. The solutions (e.g., minimum $\phi_0$ and interface value $\phi_s$) are most influenced by dimensionless quantities $L_f$ (position $S$), $A_f$ and $\epsilon_r$ in filter.

\begin{figure}[h]
\begin{center}
\includegraphics[width=2in]{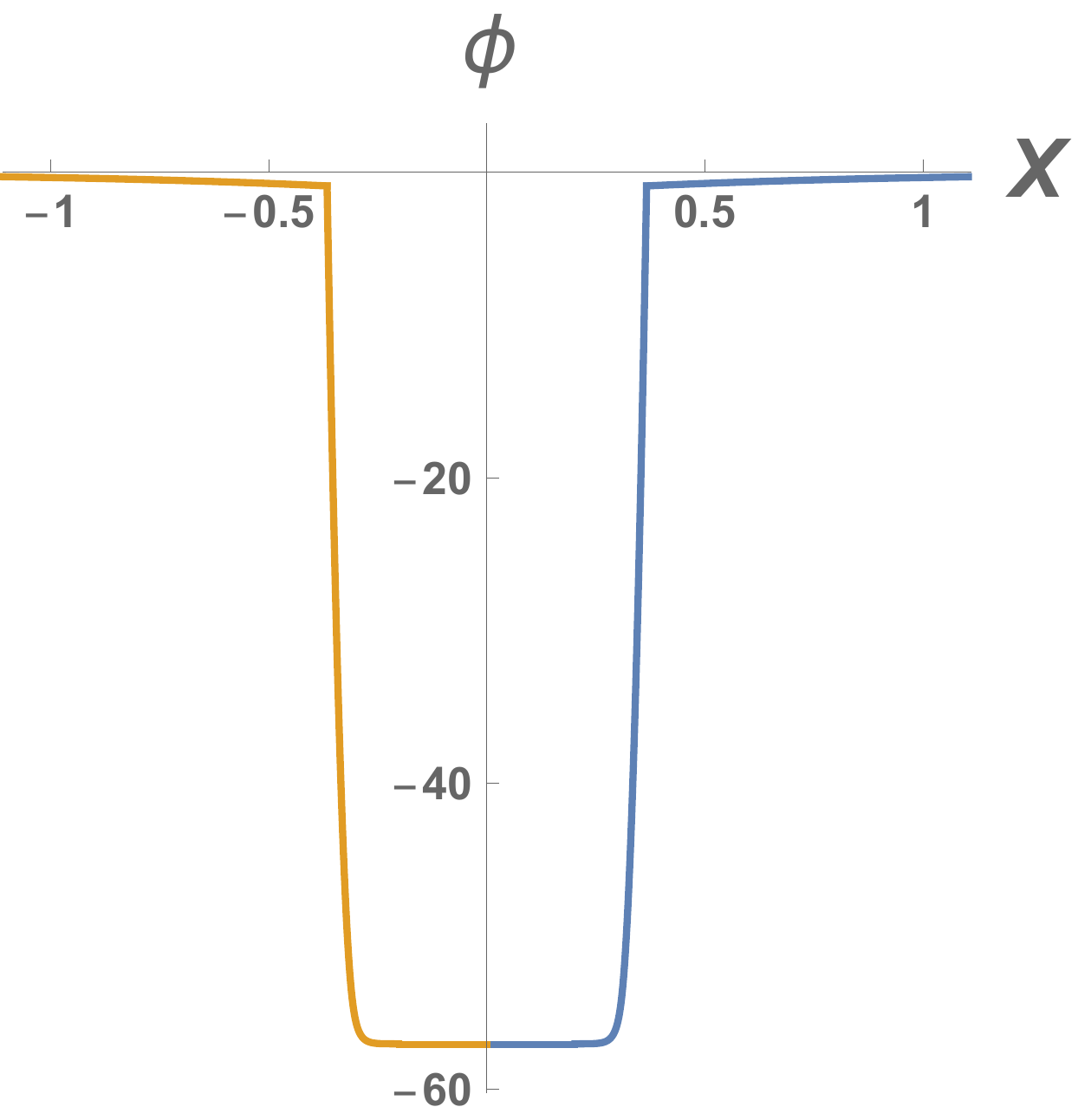} ~ \includegraphics[width=2 in]{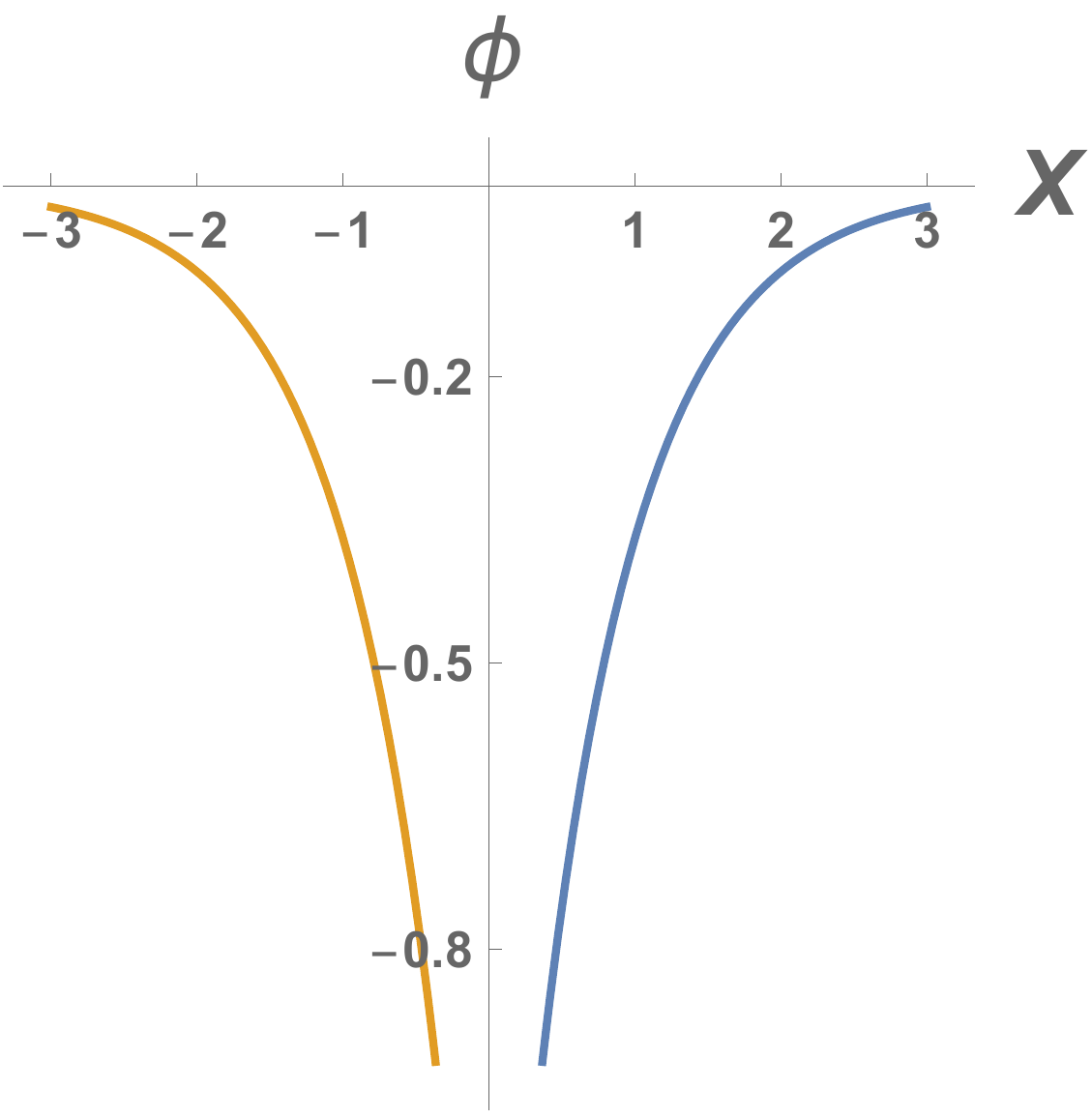} ~\includegraphics[width=2 in]{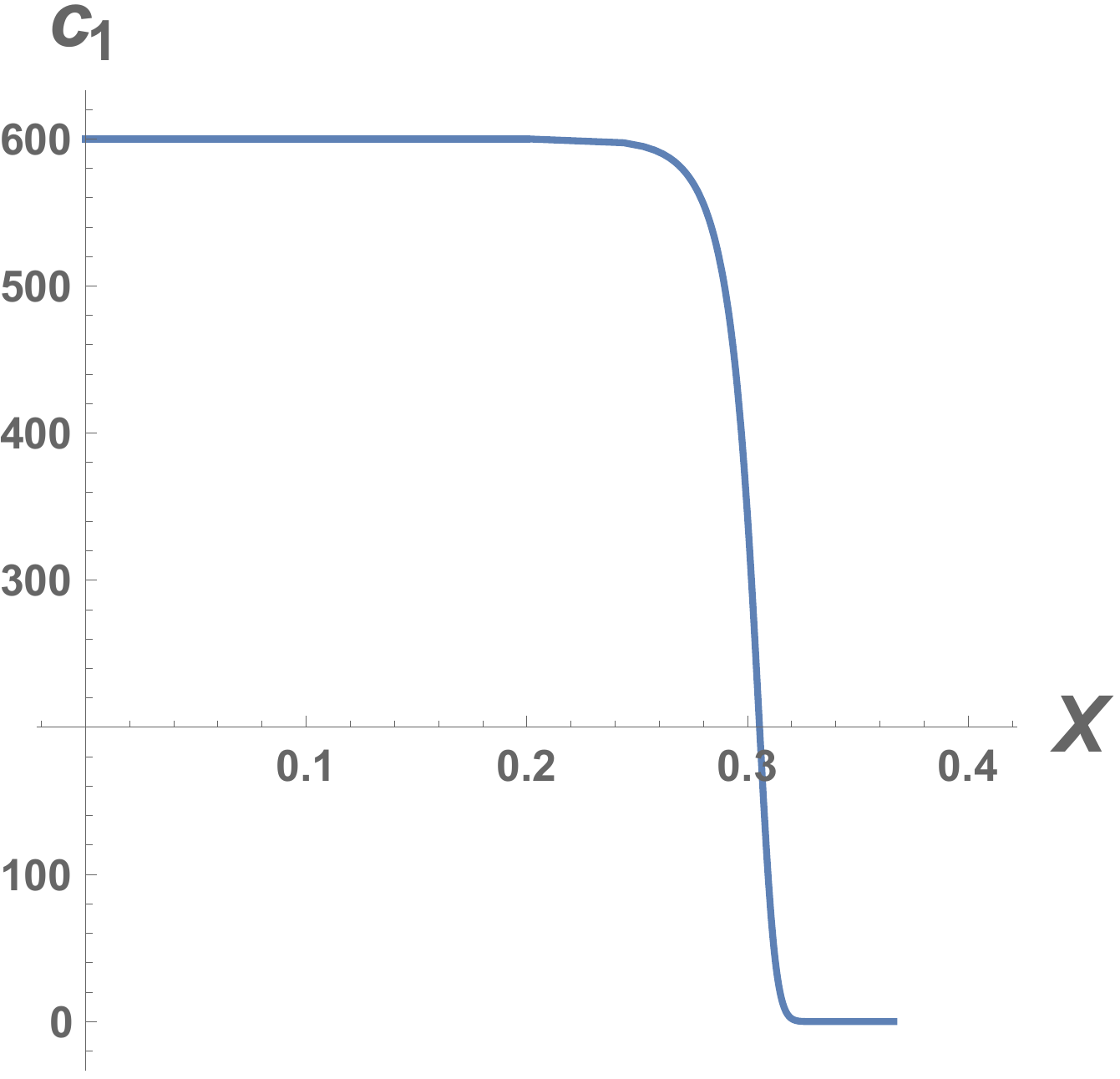} 
\caption{\label{fig4en} The profiles for case $q=600<1/a_1^3\delta$: (a) $\phi$ in both filter and chamber, (b) $\phi$ in chamber, (c) $c_1$ in right-half interval of filter.}
\end{center}
\end{figure}

\noindent \textbf{Remark}: In Section 3.1.1, we only considered the constant solution $\phi=\phi^\ast$ in middle part of filter. Actually the constant $\phi^\ast$ is connected to the chamber by a standard boundary layer (BL) in filter and near two edges. The solution in filter can be easily constructed similar to above analysis, and is given
\begin{equation}
\label{eq32_2}
\begin{aligned}
& X= S+\sqrt{\frac{\epsilon_{r0}}{2}} \int_{\phi_s}^{\phi} \frac{1}{\sqrt{G(\phi) - G(\phi^\ast)}} d\phi, \quad 0<X<S,
\end{aligned}
\end{equation}
where $\phi_s$ determined by $(\ref{eq32})_1$ with $\phi_0=\phi^\ast$ there. The results for $\phi$ and $c_1$ are shown in Figure \ref{fig4en} for the case $q=600<1/a_1^3\delta$ with other parameters as before. One clearly see the typical BLs of $\phi$ near two edges in filter.




\subsection{K$^+$/Ca$^{2+}$ selectivity}

In this subsection, we consider the case with three ions K$^+$, Ca$^{2+}$ and Cl$^-$ (respectively $c_1,c_2$ and $c_3$), and study the selectivity between K$^+$ and Ca$^{2+}$ (or Ba$^{2+}$). 

In this case, one can not directly analyze the ratio $c_1/c_2$ anymore, since they have difference valences. Due to the factor $z_i^2$ in $\Delta W_i$ in (\ref{eq15}), the barrier $\Delta W_2\approx 274$ for Ca$^{2+}$ is much larger. Now, we consider the EN case in filter 
\begin{equation}
\label{eq33}
\begin{aligned}
& c_1 + 2 c_2 -c_3=q.
\end{aligned}
\end{equation}
With the help of (\ref{eq12}, \ref{eq11}), this is a cubic equation for $e^\phi$ and once $\phi$ is solved all $c_i$ can be recovered.  The analytic solution for $\phi$ is quite complicated, and involves many exponentially large and small terms. One can not get right answer unless making proper approximations in different situations by keeping only leading exponential terms and neglecting high-order exponential terms. There are two situations. When $q$ satisfies (\ref{eq18}), we get the same approximation as in (\ref{eq20}). When $q$ is relatively large,
\begin{equation}
\label{eq34}
\begin{aligned}
& \frac{1}{a_1^3 \delta} <q< \frac{2}{a_2^3 \delta},
\end{aligned}
\end{equation} 
we get the leading-order approximation
\begin{equation}
\label{eq35}
\begin{aligned}
c_1^\ast & =\frac{2-a_2^3 q \delta}{(2 a_1^3 -a_2^3) \delta}, \quad  c_2^\ast =\frac{a_1^3 q \delta-1}{(2 a_1^3 -a_2^3) \delta} ,\quad c_3^\ast = 0,\\
\phi^\ast & = B_2 -W_2(0) - (B_1-W_1(0))  - \log c_2^\ast+ \log c_1^\ast\\
& =\Delta W_1 -\Delta W_2 + \log c_{2b} - \log c_{1b} + \phi_b  - \log (a_1^3 q \delta-1)+ \log (2-a_2^3 q \delta).
\end{aligned}
\end{equation} 
In above, $\phi^\ast$ depends on the calculated $c_1^\ast$ and $c_2^\ast$, thus the size effect on $\phi^\ast$ is through these two quantities. One can see that the boundary conditions affects $\phi^\ast$, but have negligible influence on the selectivity. The conclusion on selectivity also applies to non-equilibrium case. 

\noindent \textbf{Remark:} In above approximation (\ref{eq35}), $c_1^\ast$ and $c_2^\ast$ are determined by the constraints
\begin{equation}
\label{eq36}
\begin{aligned}
& c_1^\ast + 2 c_2^\ast = q, \quad \delta(a_1^3 c_1^\ast + a_2^3 c_2^\ast )= 1.
\end{aligned}
\end{equation} 
This implies that EN condition is satisfied, and at the same time SF is saturated with K$^+$ and Ca$^{2+}$. These two combined effects determine concentrations of K$^+$ and Ca$^{2+}$. It further implies, in the case of (\ref{eq34}), the concentration of K$^+$ itself can not balance $q$ in SF, and SF needs to recruit Ca$^{2+}$ (by squeezing out some K+ at the same time) to help out the electrostatic balancing since Ca$^{2+}$ has a larger valence in spite of its larger Born solvation energy as well.


\begin{figure}[h]
\begin{center}
\includegraphics[width=2.8 in]{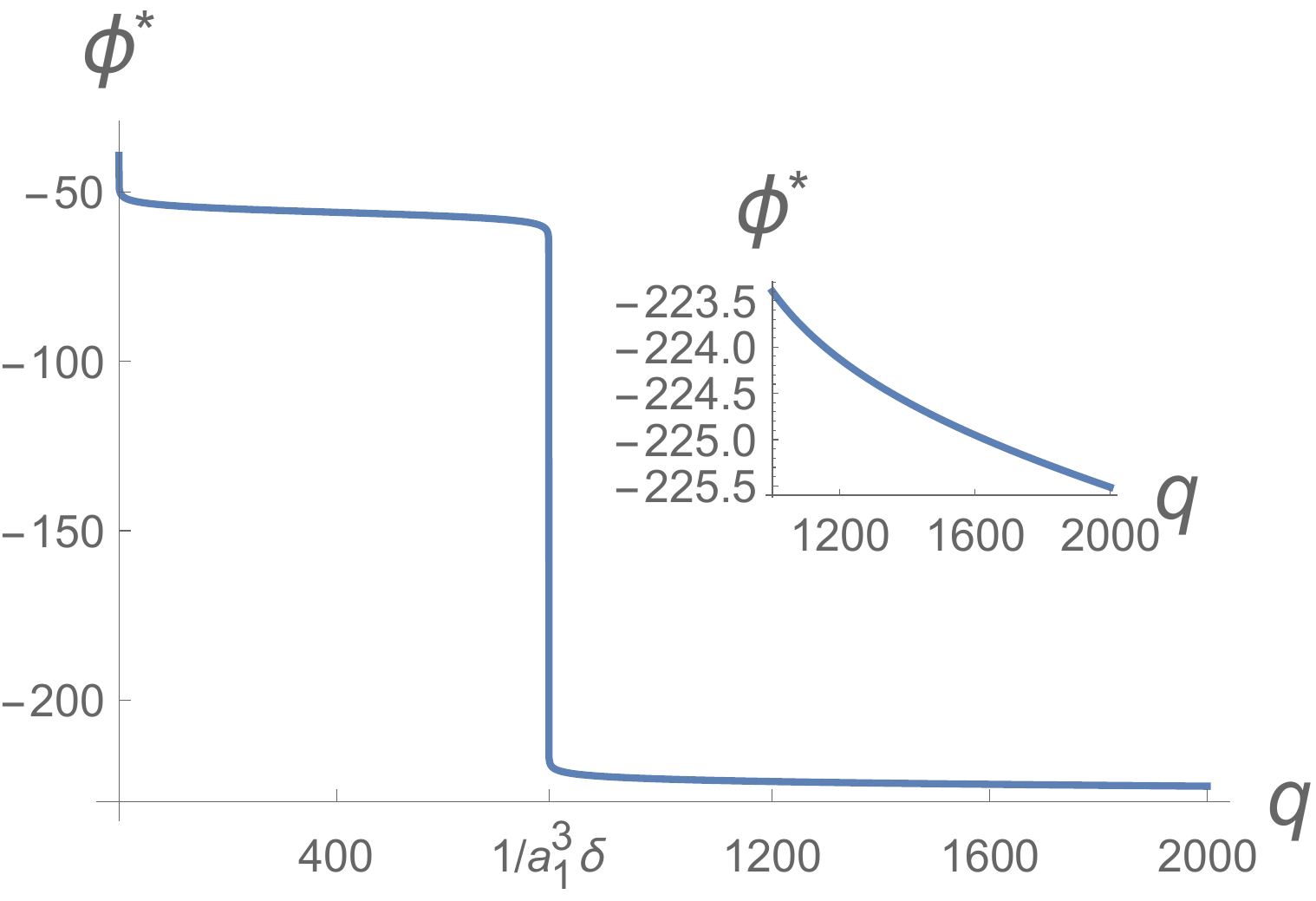}~ \includegraphics[width=2.8 in]{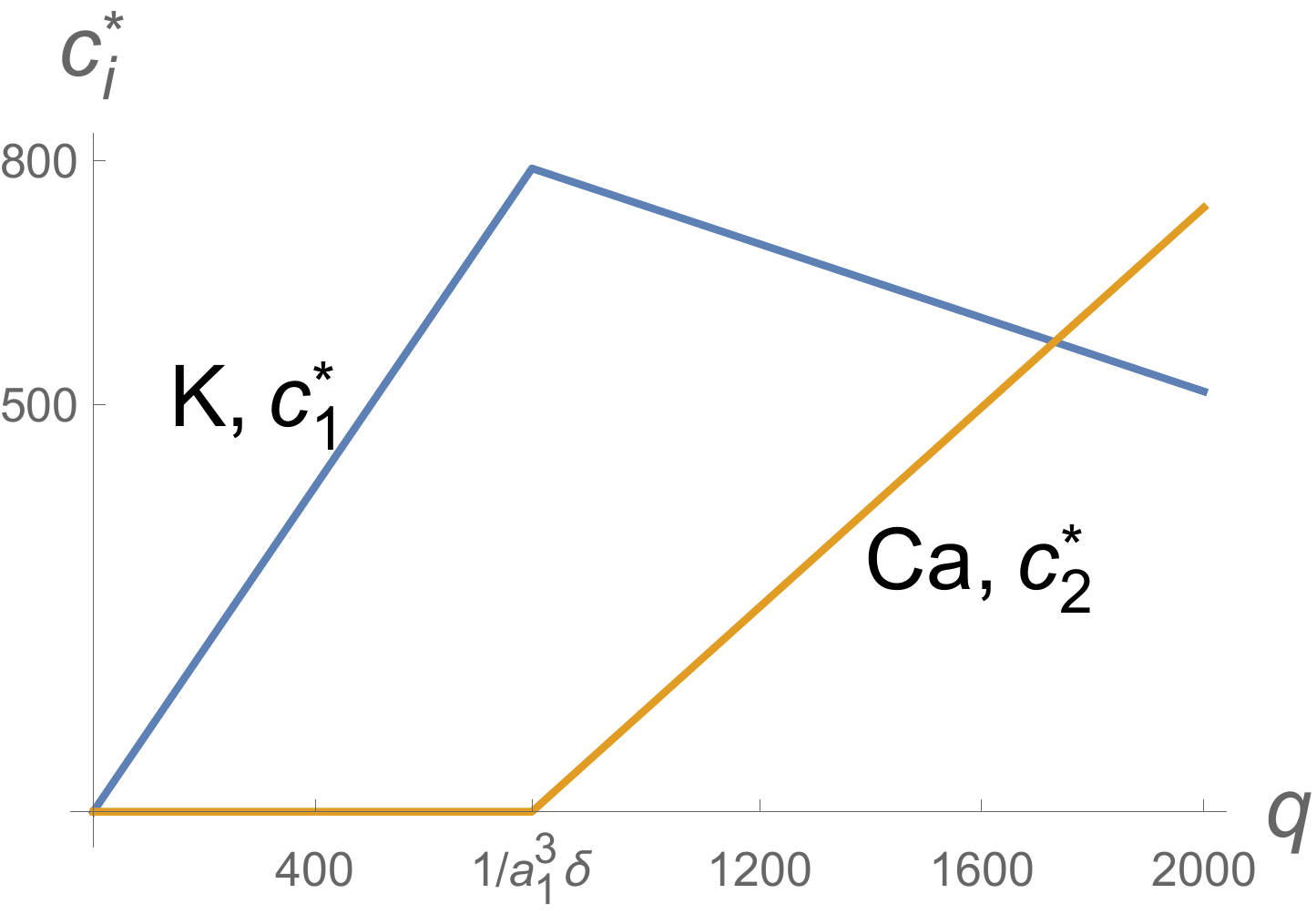}
\caption{\label{fig5} Dependence of $\phi$ and $c_i$ ($i=1,2$) in filter on charge $q$.}
\end{center}
\end{figure}

Figure \ref{fig5} shows the dependence of above solution (\ref{eq35}) on $q$, with $c_{1b}=c_{2b} =1,\phi_b= 0$ and some data (\ref{eqA3}) in Appendix \ref{appendixA}. The first part of the curves is the same as Figure \ref{fig2}, when $q<1/a_1^3\delta\approx 790$. When $q$ exceeds this critical value, the concentration of Ca$^{2+}$ increases while that of K$^{+}$ decreases. When $q$ crosses the critical value, the constant $\phi^\ast$ in filter transits from previous state at about -55 to another state at about -225, see the embedded figure in Figure \ref{fig5}a.  Based on the data (\ref{eqA3}) in Appendix \ref{appendixA}, the next critical value for saturation of $c_2$ is $q={2}/{a_2^3 \delta}\approx 4280$.  Figure \ref{fig5} does not reach this value.

\begin{figure}[h]
\begin{center}
\includegraphics[width=2.8 in]{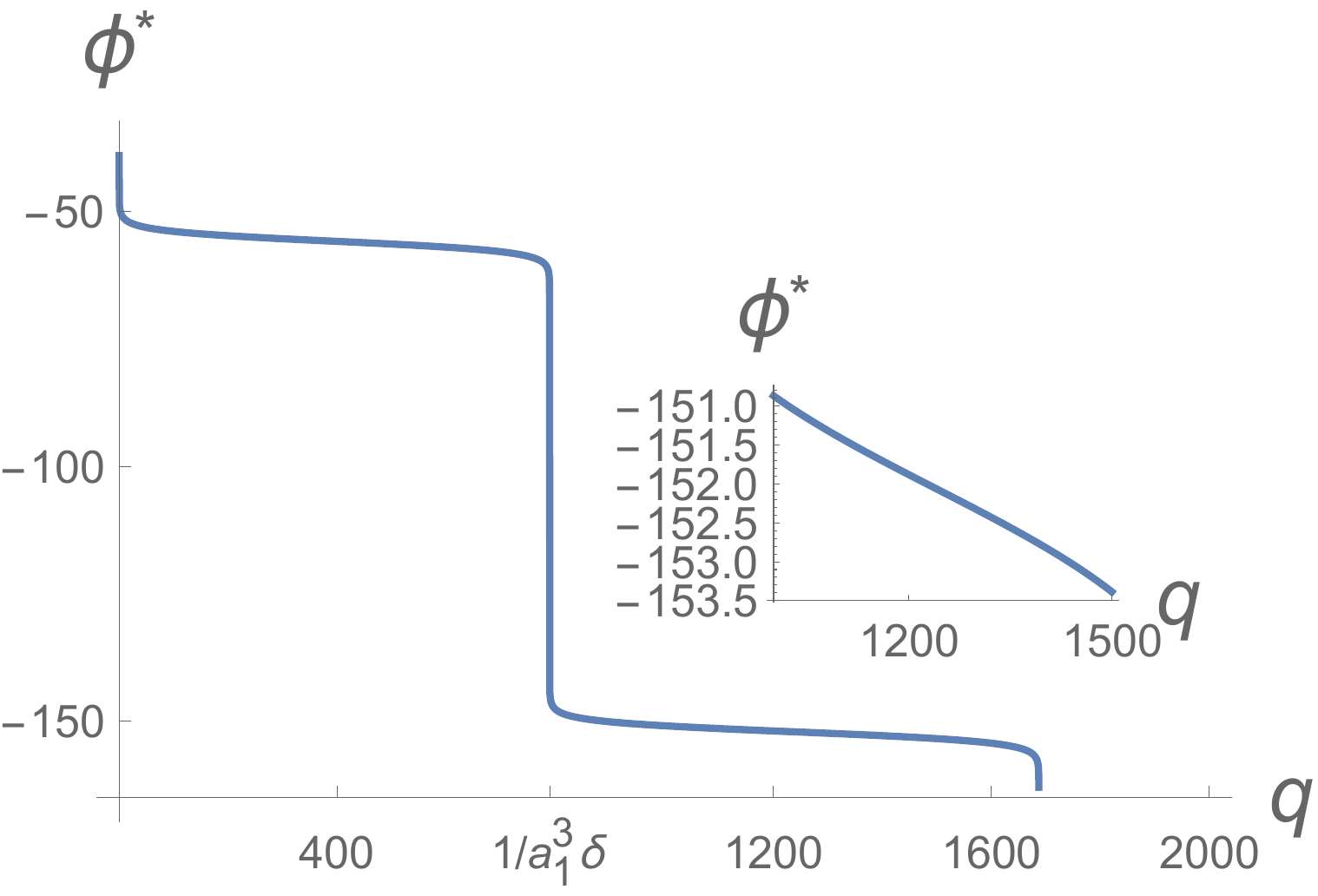}~ \includegraphics[width=2.8 in]{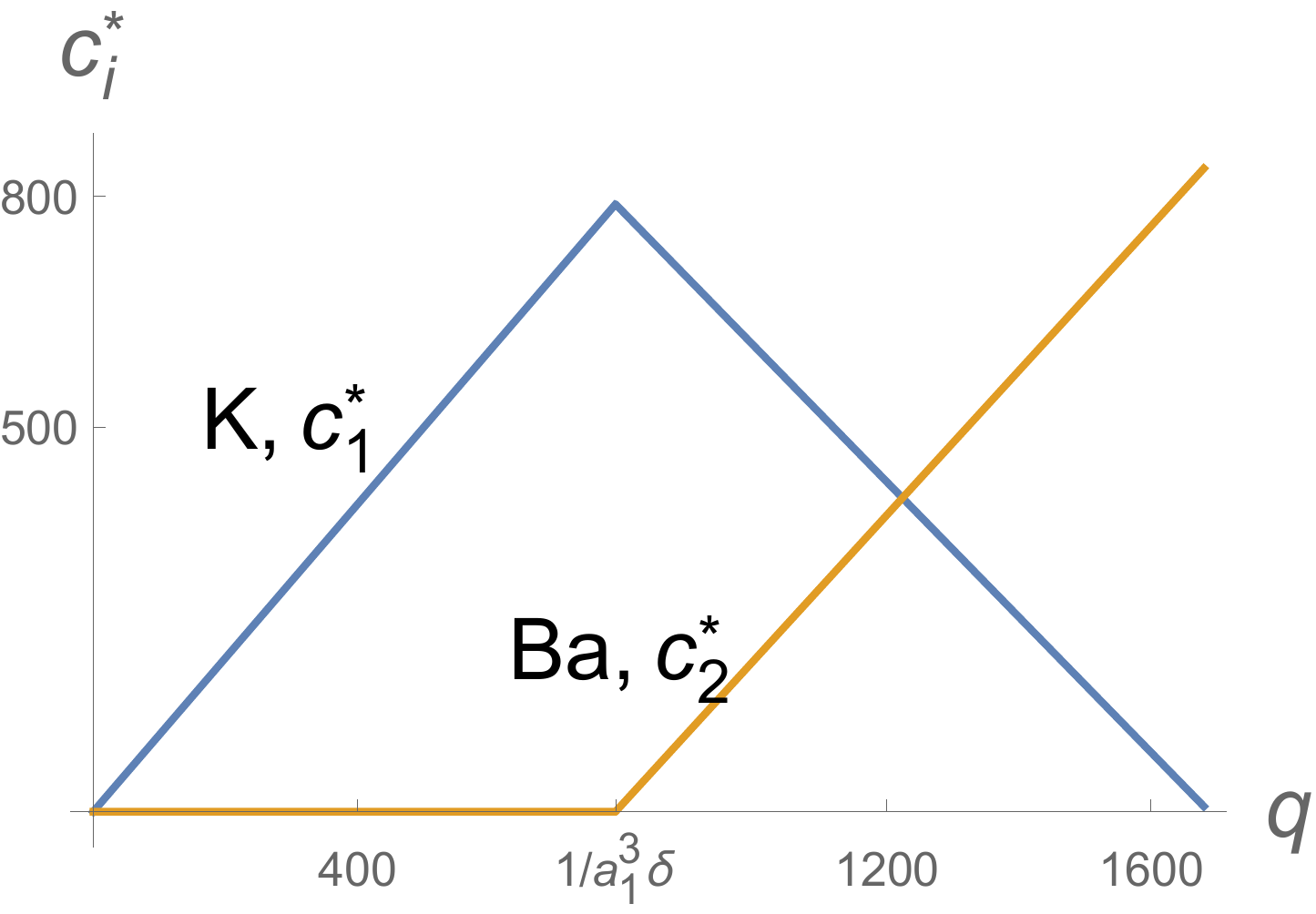}
\caption{\label{fig5_1} Dependence of $\phi$ and  $c_i$ ($i=1,2$) in filter on charge $q$.}
\end{center}
\end{figure}

The Barium Ba$^{2+}$ has been used to block K$^+$ channel for a long time \cite{Piasta2011}. The size of Ba$^{2+}$ is larger than Ca$^{2+}$, given in Appendix \ref{appendixA}. Since it also has +2 valence, the energy barrier ($\sim 201.2$) is still much larger than that of K$^{+}$. The above analysis will not change, and in this case the critical value is $q=2/a_{Ba}^3\delta\sim 1688$. Figure \ref{fig5_1} shows the results and dependence on $q$, with same data as before. Figure \ref{fig5_1}b indicates that Ba$^{2+}$  is more effective to block K$^+$ due to larger size.

Ba$^{2+}$ specifically blocks K$^+$ channels via electrostatic stabilization in the permeation pathway. At high concentrations of external K$^+$, the block-time distribution of Ba$^{2+}$ is double exponential, implies at least two Ba$^{2+}$ binding sites in SF \cite{Piasta2011}. This coexistence of Ba$^{2+}$ and K$^+$ inside SF was also observed in MD computation [6] with Ba$^{2+}$ at binding site S2 and K$^+$ at binding site S0 forming a lock-in state impeding the translocation of Ba$^{2+}$ \cite{rowley2013}. 

\noindent \textbf{Remark:}  The above analysis is also valid for the case with four ions: K$^+$, Na$^+$, Ca$^{2+}$ and Cl$^-$. Based on the analysis in Section 3.1, the concentration of Na$^+$ is always exponentially smaller than K$^+$. Thus, K$^+$ is favored compared with Na$^+$, and adding Na$^+$ will make no difference. The non-EN case will not be discussed here, since for relatively large $q$ the two ions K$^+$ and Ca$^{2+}$ can coexist to maintain EN. For even larger $q>{2}/{a_2^3 \delta}$ or near transition point $q = 1/a_1^3\delta$, we need to consider the non-EN case. The analysis is similar to Section 3.1.2, except that we have a more complicated $G(\phi)$ in (\ref{eq31}) for this case.



\section{Non-equilibrium case and flux-voltage relation}

In this section, we assume the same concentrations at two ends of chamber but with different electric potential. Then there is variation in electro-chemical potential $\mu_i$ across interval $x\in [-1,1]$, and we intend to study the flux-voltage relations at steady state for previous two cases. This section is restricted to relative long chamber region (length $L$), where some analytical flux-voltage relations are available. For general cases, numerical or semi-analytical solutions will be shown in the next section.

\subsection{Fluxes of K$^+$/Na$^+$ case}

In this subsection, we consider the three-ion case with K$^+$,Na$^+$ and Cl$^-$.  At two ends $x=\pm 1$, we impose 
\begin{equation}
\label{eq37}
\begin{aligned}
&c_1(\pm 1) = 1,\quad c_2(\pm1) = c_{2b}, \quad c_3(\pm 1) = 1+ c_{2b},\\
& \phi(-1) = V,\quad \phi(1) = 0.
\end{aligned}
\end{equation} 

In this case, the results in Section 3.1 about selectivity of K$^+$ and Na$^+$ are still valid. Although $B_i$ in (\ref{eq11}) is not an exact constant anymore, the variation is small since $\mu_i$ is monotone.  We have also pointed out in Section 3.1 that  $c_2$ is exponentially small unless $c_2$ is $10^7$ times larger than that of $c_1$ near filter. Based on results on selectivity, now we study the relative variation $\Delta \mu_i$ for each $\mu_i$ ($i=1,2,3$) in chamber and filter. Since in chamber it is almost the classical PNP system, we get $c_i \sim O(1)$, implying
\begin{equation}
\label{eq38}
\begin{aligned}
&\Delta \mu_i = O(J_i), \quad \textrm{in chamber}.
\end{aligned}
\end{equation} 
In filter, we have either $c_i\sim O(q)$ or $c_i$ is exponentially small. Since the filter interval is small, as a first approximation, we have 
\begin{equation}
\label{eq39}
\begin{aligned}
&\Delta \mu_i \approx \frac{L_f J_i}{ A_f c_i^\ast}, \quad \textrm{in filter},
\end{aligned}
\end{equation} 
where $L_f$ and $A_f$ are dimensionless quantities already. We know that the total variation (sum of above two, (\ref{eq38}) and (\ref{eq39})) from left end to right end is $O(1)$ with $V\sim O(1)$. From Section 3.1, we have $c_1 \sim O(q)$ in filter, and then we get the estimate from some data (\ref{eqA3},\ref{eqA5}) in Appendix \ref{appendixA}
\begin{equation}
\label{eq40}
\begin{aligned}
&\frac{L_f}{ A_f c_1^\ast} \sim 10^{-3} - 10^{-2}.
\end{aligned}
\end{equation} 
This implies that $J_1\sim O(1)$, and the filter region can be neglected for variation of $\mu_1$. On the other hand, $c_2^\ast$ and $c_3^\ast$ are exponentially small in filter, thus $J_2$ and $J_3$ can only be exponentially small, but this still gives finite variation $\Delta\mu_2,\Delta\mu_3$ in filter by (\ref{eq39}). In this context, we can treat $J_2=J_3=0$ when studying the chamber region, and therefore we only need to concentrate on $J_1$-$V$ relation.

In the chamber, it is eligible to use the EN condition as first approximation for relative long chamber.  We take constant cross section $A(x)=1$ for illustration. By neglecting $O(\delta)$ term, we get the classic system
\begin{equation}
\label{eq41}
\begin{aligned}
& c_1' (x)+ c_1  \phi'(x) = -J_1/D_1\equiv -J,\\
& c_2' (x)+ c_2  \phi'(x) = 0,\\
& c_3'(x) - c_3  \phi' (x) = 0,\\
& c_1 + c_2 =c_3.
\end{aligned}
\end{equation} 
This can be solved explicitly for left half chamber $-1<x<0$ and right half chamber $0<x<1$, given in Appendix \ref{appendixB}. Here $x=0$ is treated as filter. By the continuity of $\mu_1$ at filter, we get (see Appendix \ref{appendixB})
\begin{equation}
\label{eq42}
\begin{aligned}
& \log \left(\frac{(1+c_{2b} - J/2)^2}{1+c_{2b}} - c_{2b} \right) +V =  \log \left(\frac{(1+c_{2b} + J/2)^2}{1+c_{2b}} - c_{2b} \right),
\end{aligned}
\end{equation} 
which provides the $J$-$V$ relation. This can be obtained by solving a quadratic equation, and we select the reasonable root that satisfies $J=0$ at $V=0$,
\begin{equation}
\label{eq43}
\begin{aligned}
& J = \frac{2(1+c_{2b}) (1+e^V) - 2 \sqrt{1+c_{2b}} \sqrt{4e^V + c_{2b} (1+e^V)^2} }{e^V-1}.
\end{aligned}
\end{equation} 
For the special case $c_{2b}=0$, we have 
\begin{equation}
\label{eq44}
\begin{aligned}
 J = \frac{2(e^{V/2}-1)}{(e^{V/2} +1)}.
\end{aligned}
\end{equation} 
The general case of $A(x)$ causes no essential problem (see Appendix \ref{appendixB}), and finally we get
\begin{equation}
\label{eq43_1}
\begin{aligned}
& J \int_{L_f/2}^1 \frac{1}{A(s)} ds = \frac{2(1+c_{2b}) (1+e^V) - 2 \sqrt{1+c_{2b}} \sqrt{4e^V + c_{2b} (1+e^V)^2} }{e^V-1}.
\end{aligned}
\end{equation} 
Since $L_f \sim O(\epsilon)$, for special case $A(x)=1$, this factor after $J$ degenerates to $1- L_f/2 \sim 1$.

\begin{figure}[h]
\begin{center}
\includegraphics[width=3 in]{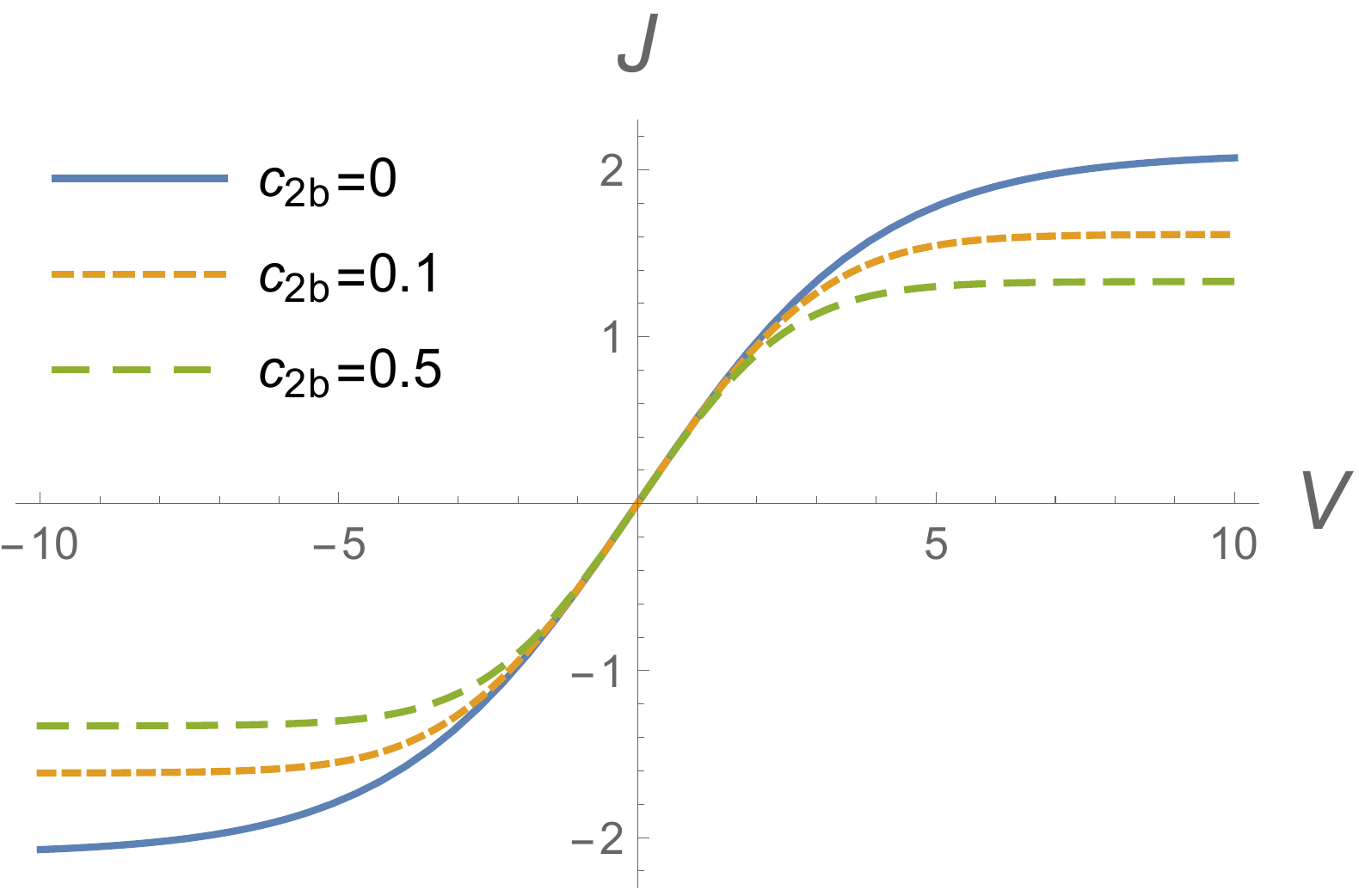}
\caption{\label{fig6} Flux-voltage $J$-$V$ relations with different boundary concentrations $c_{2b}$.}
\end{center}
\end{figure}

\noindent \textbf{Remark}: We have used EN condition in above system, which causes an $O(\epsilon J)$ error in estimate of the variation $\Delta \mu_1$,  due to classical BL near filter edge in chamber (see \cite{song2018}). Also in (\ref{eq43}),  there is $O(\epsilon)$ error by treating the filter as a point $x=0$ as filter length is $O(\epsilon)$, but in (\ref{eq43_1}) the exact point $x=L_f/2$ of filter edge is used. Later numerical calculations show that the above approximation is good for small $V$, and it slightly underestimates the flux for relatively large $V$.

Figure \ref{fig6} shows the $J$-$V$ relations (note $J_1 = D_1 J$) in (\ref{eq43_1}) with $A=1$ and different boundary concentrations $c_{2b}$. It indicates that the flux J tends to saturate for relatively large V (the reason will be illustrated in later section), which agrees well with experimental measurements \cite{Miller2002}. The presence of Na$^+$ reduces the flux of K$^+$ with the still tendency to saturate at large V. These generally agree well with experiment measurements in \cite{Miller2002} except that there is a dip in experimental IV curves at moderate V corresponding to the blockage by Na$^+$ and it becomes relieved at high V by a ``punch-through" mechanism. The failure to predict the dip of IV curve caused by Na$^+$ is due to the limitation of current analysis. Na$^+$ is expected to bind at the water cavity site near the intracellular entrance of SF, and this peak of  Na$^+$ concentration at water cavity site is totally overlooked by current asymptotic analysis which assumes EN over there.

\begin{figure}[h]
\begin{center}
\includegraphics[width=2.5in]{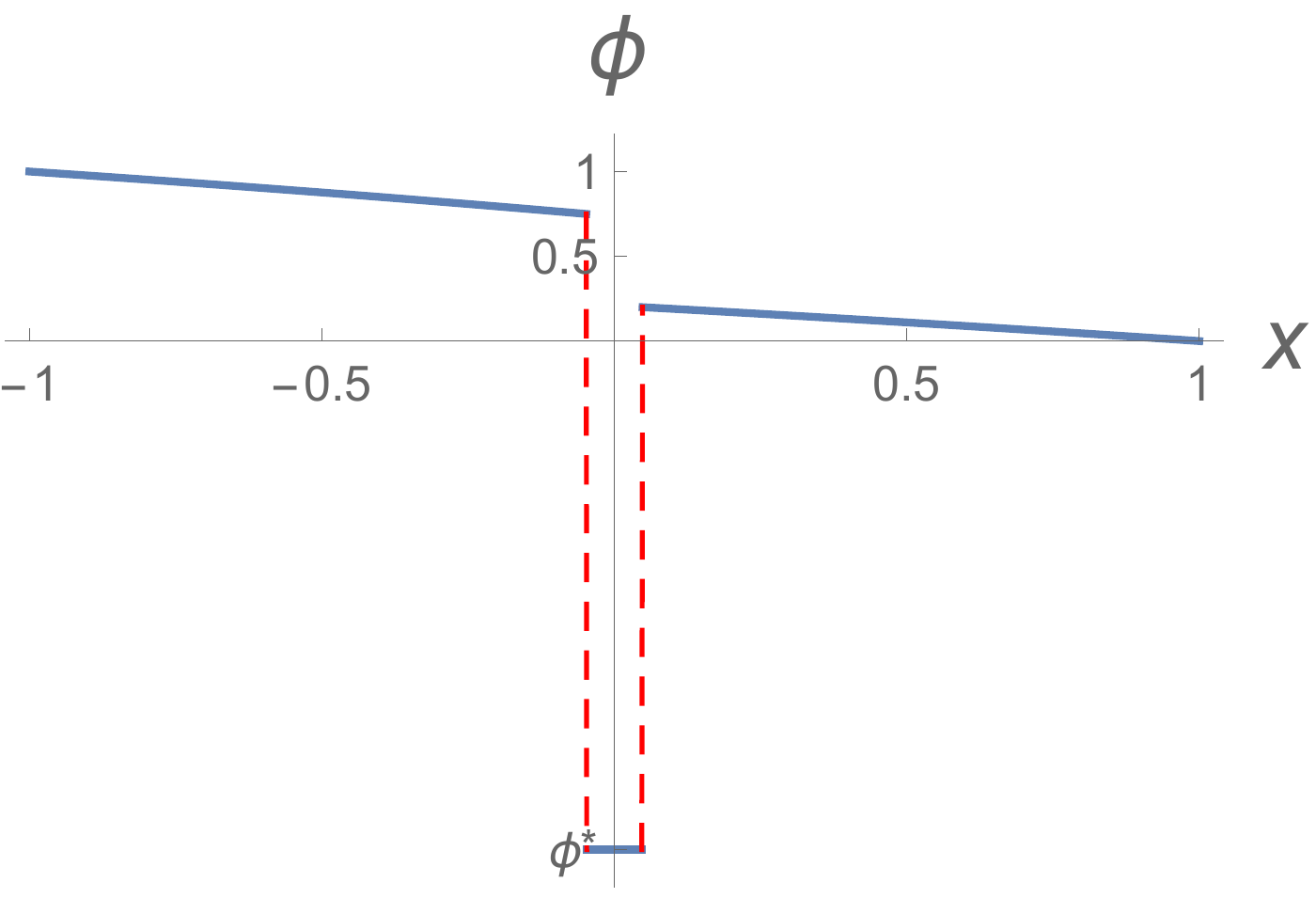}\quad \includegraphics[width=2.5in]{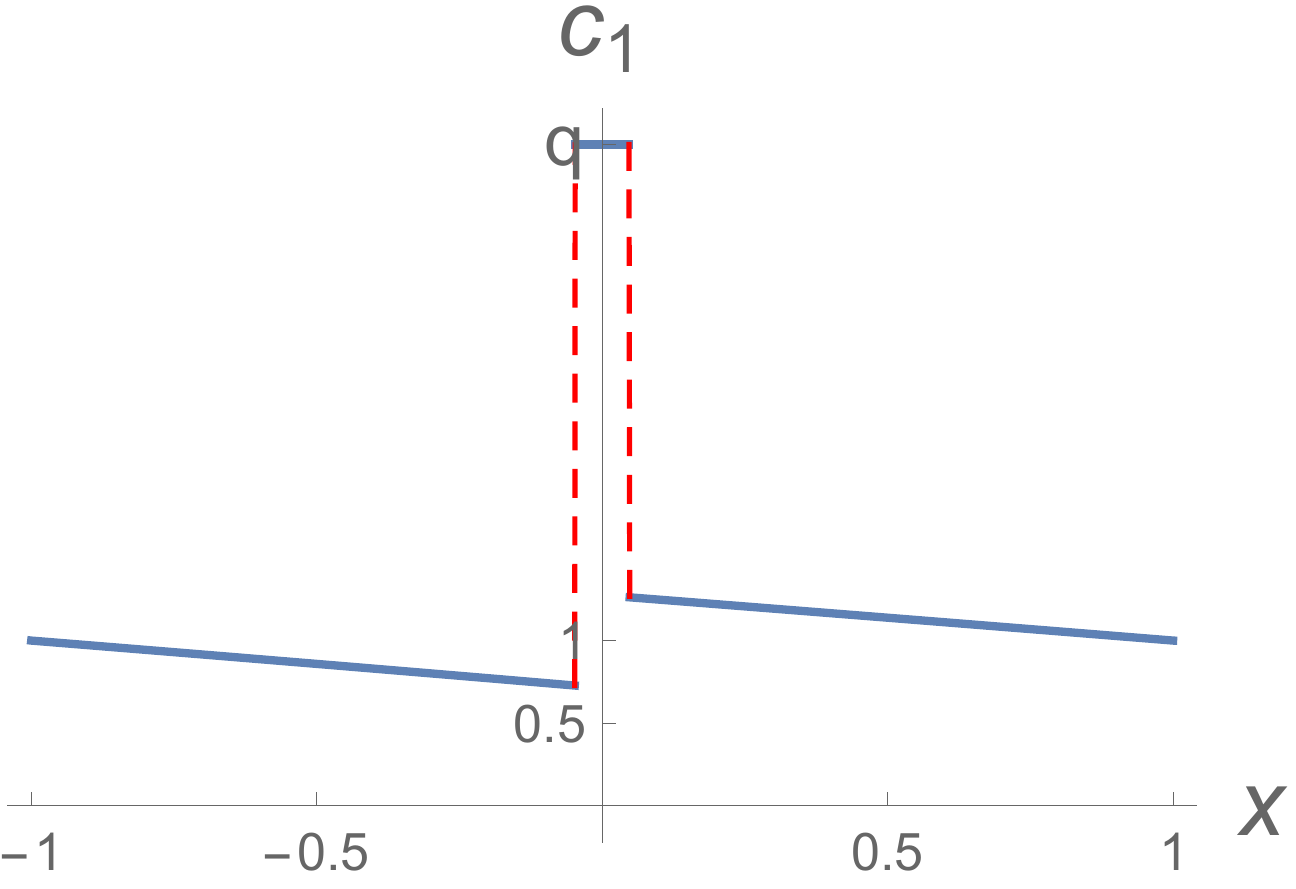} \\
\vspace{0.5cm}
\includegraphics[width=2.5in]{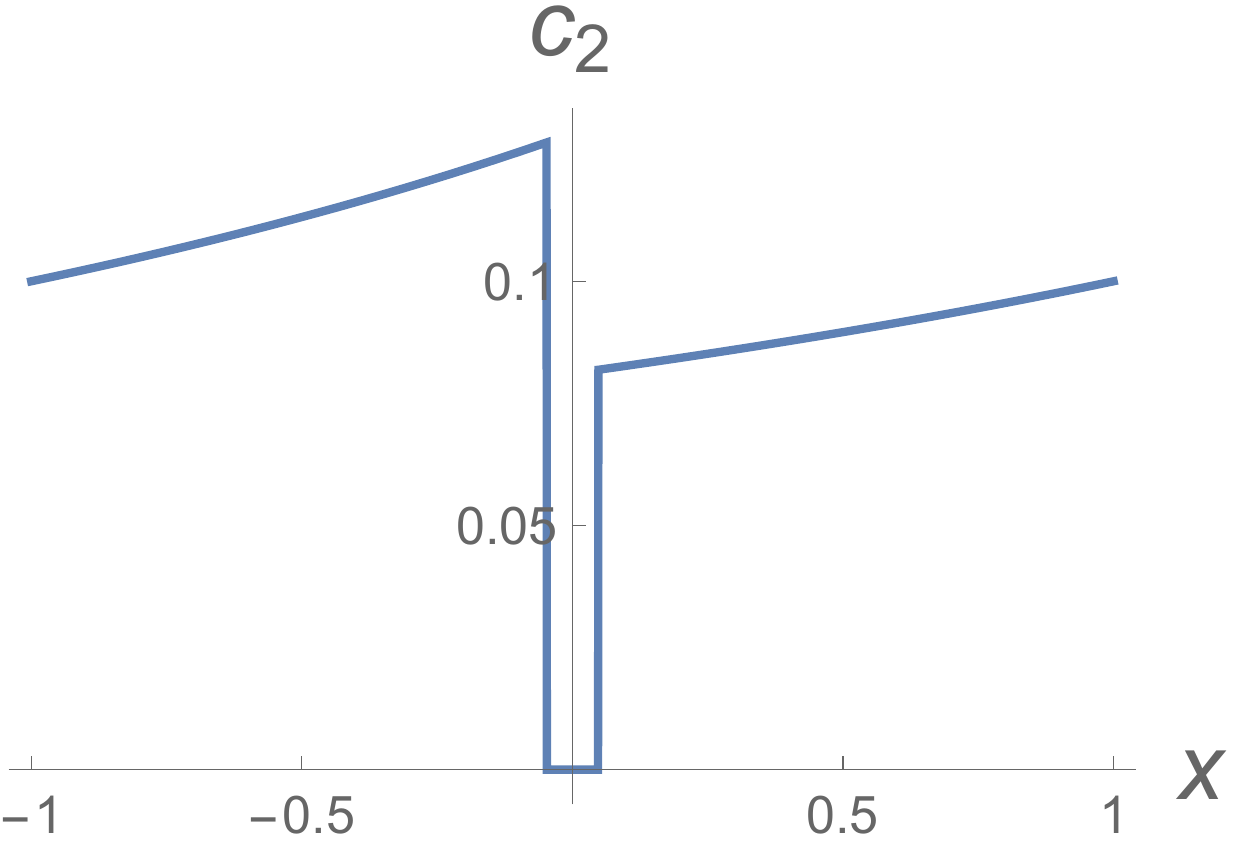}\quad \includegraphics[width=2.5in]{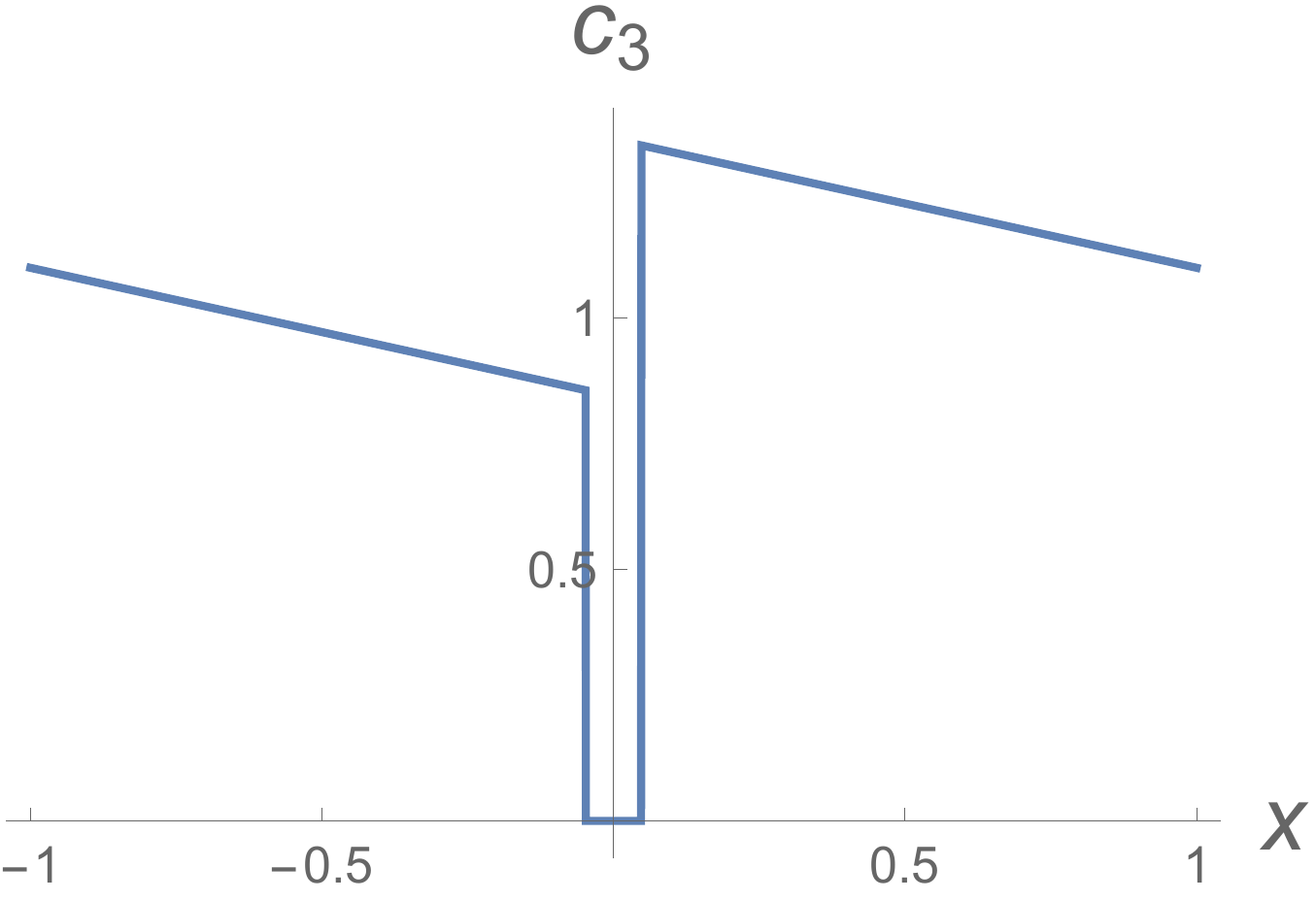}
\caption{\label{fig7} Profiles of $\phi(x)$ and $c_i(x)$ ($i=1,2,3$) with $c_{2b}=0.1,V=1$.}
\end{center}
\end{figure}

\begin{figure}[h]
\begin{center}
\includegraphics[width=2in]{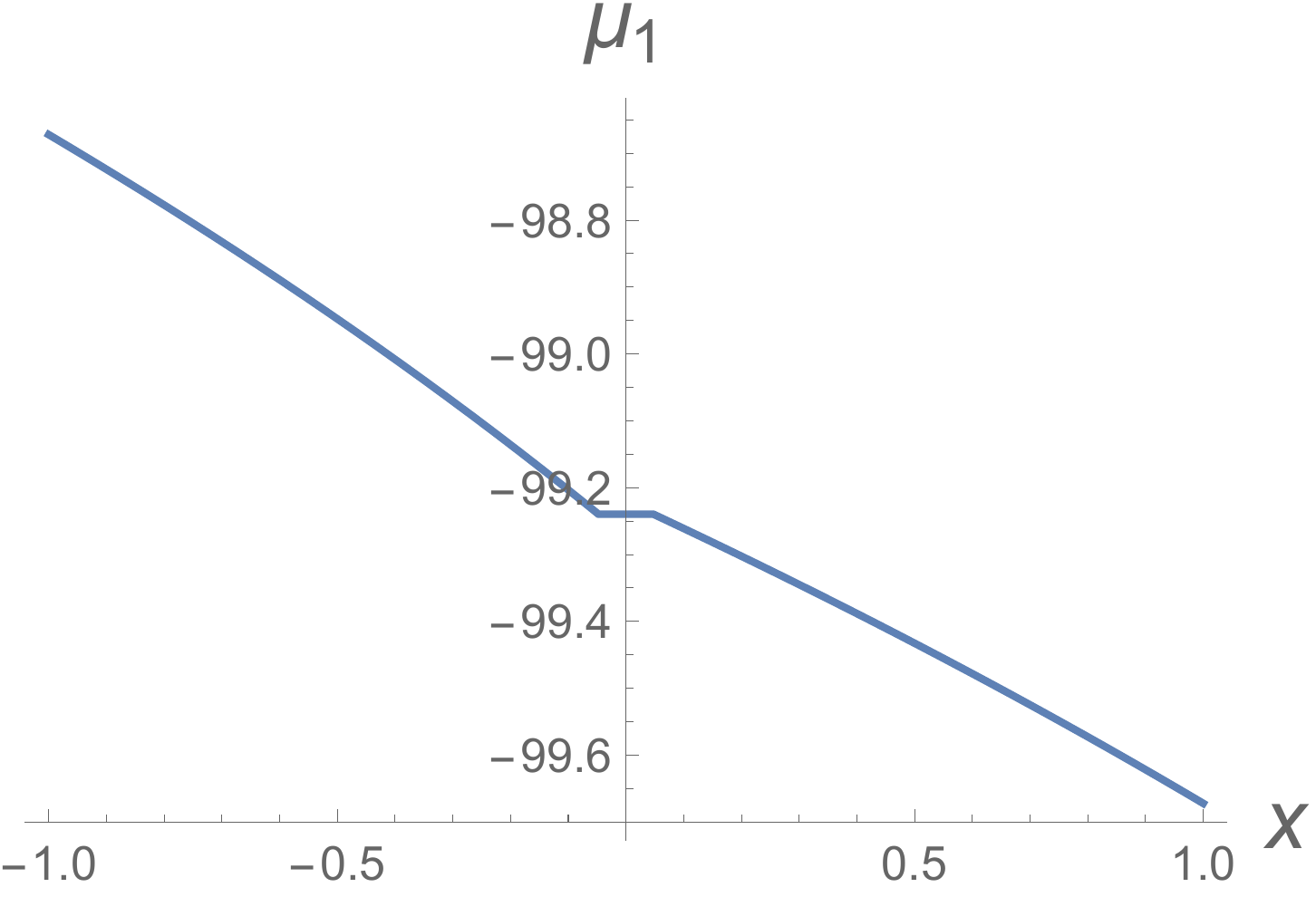} \includegraphics[width=2in]{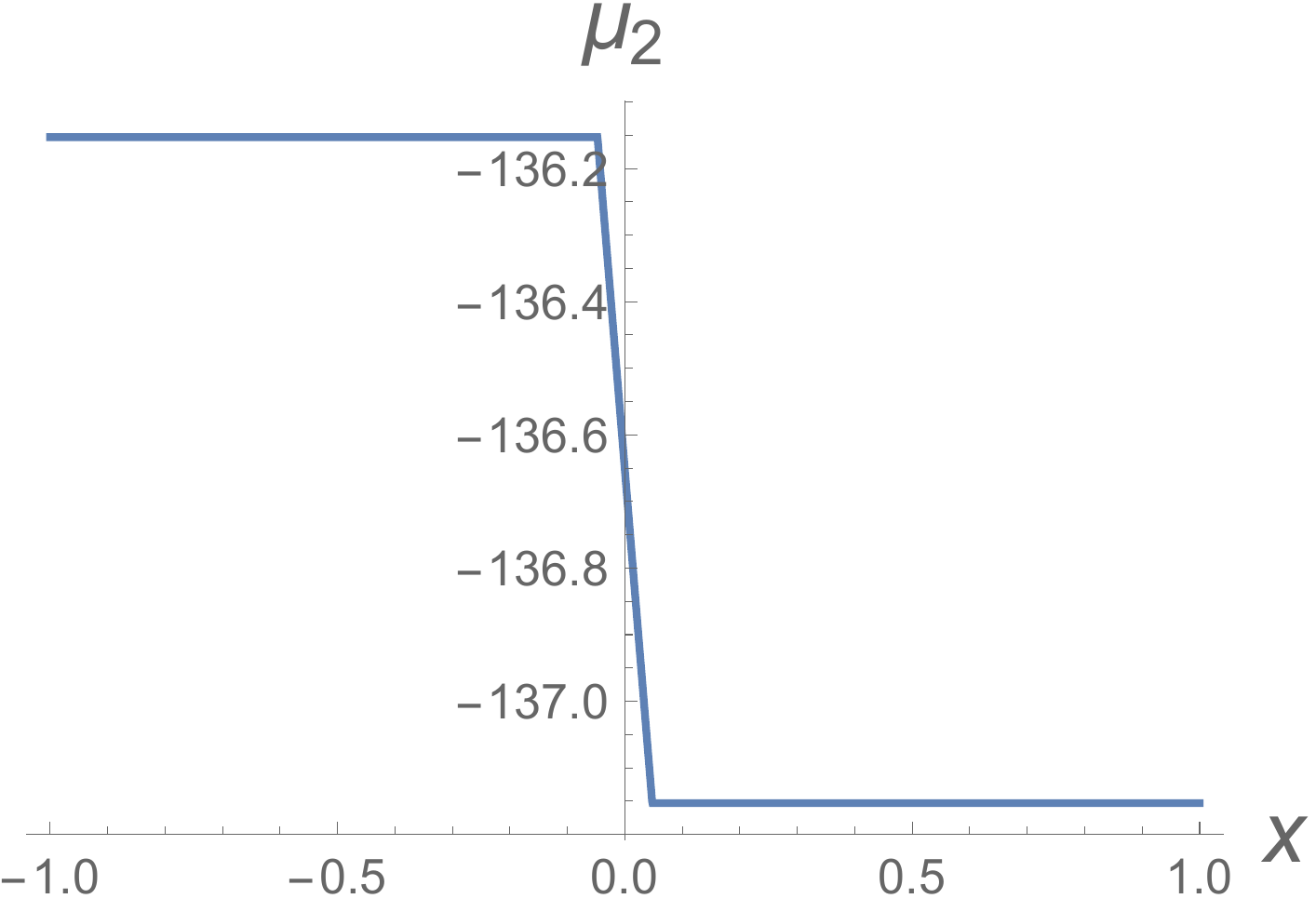}\includegraphics[width=2in]{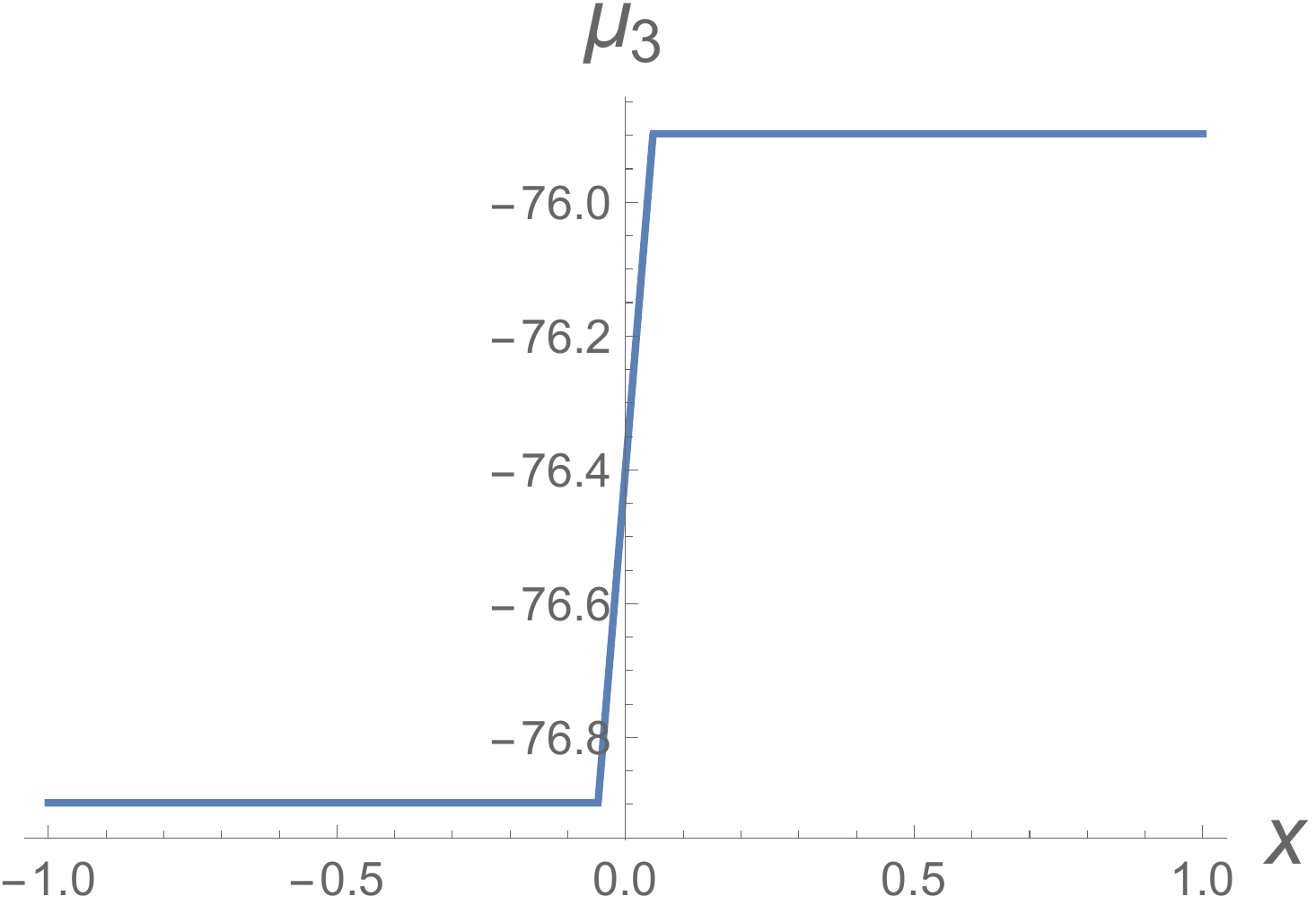}
\caption{\label{fig8} Profiles of $\mu_i(x)$ ($i=1,2,3$) with $c_{2b}=0.1,V=1$.}
\end{center}
\end{figure}

Figure \ref{fig7} shows the profiles of $\phi(x)$ and $c_i(x)$ ($i=1,2,3$) with boundary values $c_{2b}=0.1,V=1$ in (\ref{eq37}) and parameter $q<1/a_1^3\delta$. The choice of $q<1/a_1^3\delta$ is for illustration purpose as $\phi$ in most part of filter is approximated by $\phi^\ast$. The exact values of $q$ and $\phi^\ast$ are not used in Figure \ref{fig7} because they are so large, and the red dashed vertical lines mean a big jump to the two values. For larger $q>1/a_1^3\delta$ the results will not change much except that $\phi$ in filter has a  profile like Figure \ref{fig4}(a).  Figure \ref{fig8} shows the profiles of $\mu_i(x)$ ($i=1,2,3$) for each ion species. There is finite variation for $\mu_1$ in chamber, which causes the finite flux of $c_1$. The $\mu_2$ and $\mu_3$ are constant in chamber, leading to 0 fluxes. Even though there's finite variation for $\mu_2$ and $\mu_3$ in filter, there's no flux since the concentrations $c_2$ and $c_3$ are essentially 0 in filter.


Now we summarize the strategy for determining $J$-$V$ relations, which also applies to other cases like next subsection. 
\begin{itemize}
\item from the equilibrium case, determine which ions (here K$^+$, next subsection K$^+$ and Ca$^{2+}$) are prevalent in filter and which (here Na$^+$ and Cl$^-$) are 0 in filter.
\item set finite flux for only those ions prevalent in filter and set 0 flux for others, and then solve the chamber equations for left and right chamber regions
\item determine the $J$-$V$ relations by using continuity of $\mu_i$ at filter for only those ions prevalent in filter (note that other $\mu_i$ are constant in chamber and have jumps at filter).
\end{itemize}
It appears that we have only used chamber equations to approximation the $J$-$V$ relations, but actually it is totally different to directly solve chamber equations without the filter since all fluxes and variation of all $\mu_i$ would be finite and continuous in that case. It is also clear that, for the present case with filter, the chamber solutions of $\phi, c_i$ ($i=1,2,3$)  in Figure \ref{fig7} have  jumps at filter, and the $\mu_2$ and $\mu_3$ in Figure \ref{fig8} are constants in each chamber. 


\subsection{Fluxes of K$^+$/Ca$^{2+}$ case}

In this subsection, we consider the three-ion case with K$^+$,Ca$^{2+}$ and Cl$^-$ (the case for Ba$^{2+}$ is similar).  At two ends $x=\pm 1$, we impose 
\begin{equation}
\label{eq45}
\begin{aligned}
&c_1(\pm 1) = 1,\quad c_2(\pm1) = c_{2b}, \quad c_3(\pm 1) = 1+2 c_{2b},\\
& \phi(-1) = V,\quad \phi(1) = 0.
\end{aligned}
\end{equation} 

The analysis on the variation of $\Delta \mu_i$ ($i=1,2,3$) are similar to the preceding subsection, and we can follow the preceding strategy to determine the flux-voltage relations. Depending on the parameter $q$ and results in Section 3.2 about selectivity of K$^+$ and Ca$^{2+}$, there are two cases.  (1) When $q<{1}/{a_1^3 \delta}$, we get $J_2=0,J_3=0$ and finite $J_1$. Then the results of $J_1$-$V$ relation will be similar to preceding subsection, and the profiles of $c_i$ and $\mu_i$ are similar. (2) When ${1}/{a_1^3 \delta} <q< {2}/{a_2^3 \delta}$ is relatively large as in (\ref{eq34}), we have $J_3=0$ and finite $J_1$ and $J_2$, since both K$^+$ and Ca$^{2+}$ can exist in filter.

\begin{figure}[h]
\begin{center}
\includegraphics[width=2.5 in]{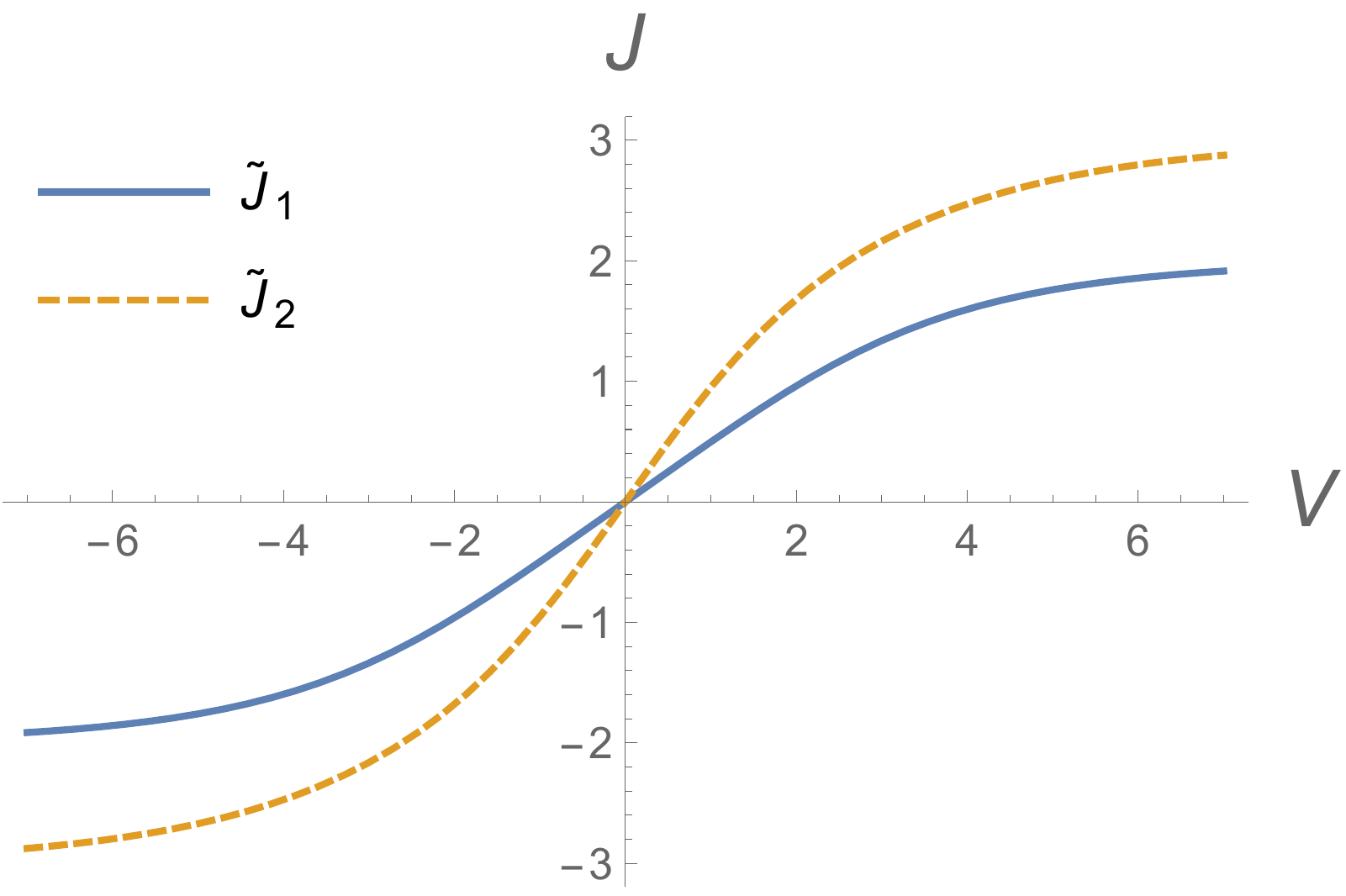}\includegraphics[width=2.5 in]{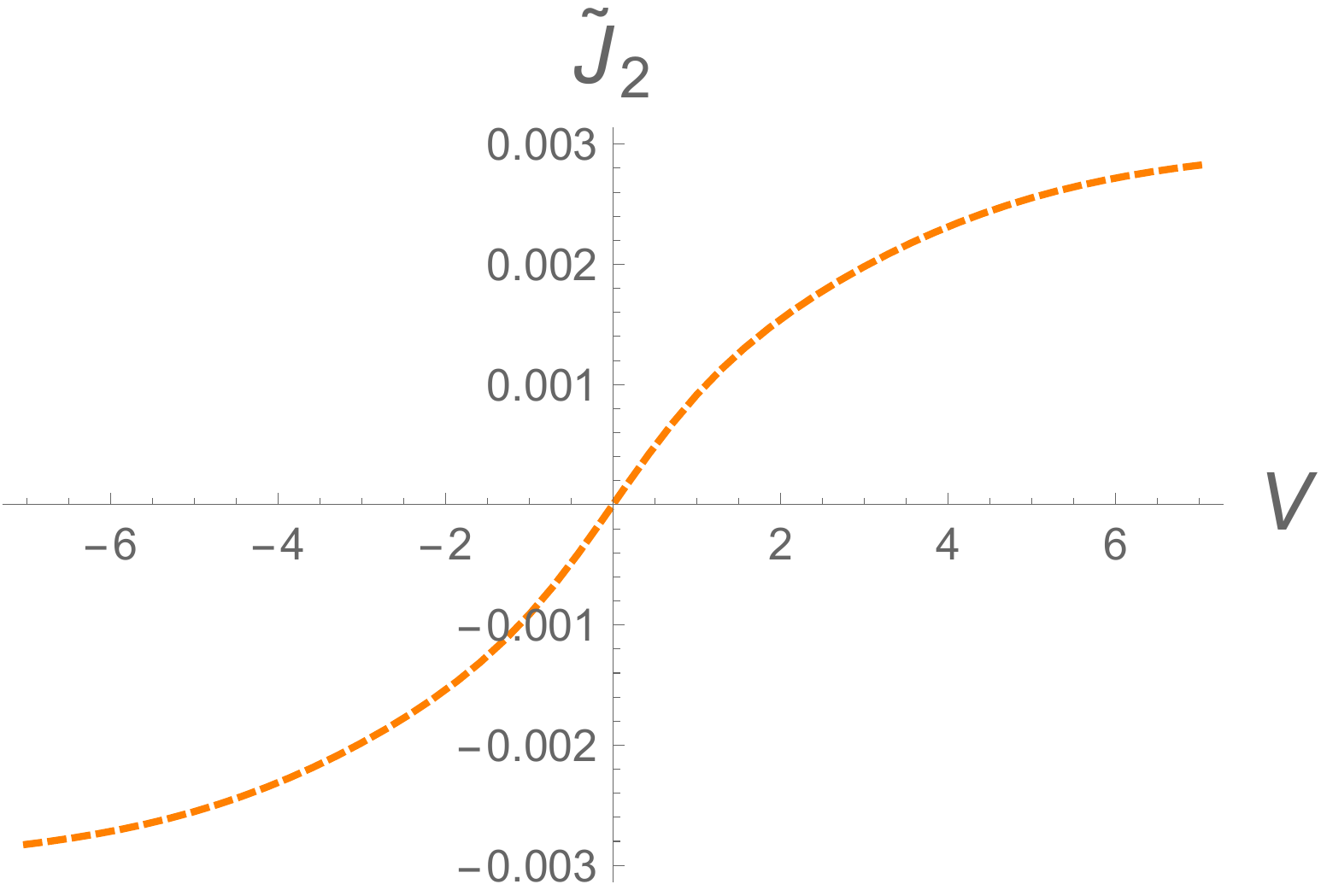}
\caption{\label{fig9} Flux-voltage $J$-$V$ relations with $c_{2b}=1$ and $c_{2b}=10^{-3}$.}
\end{center}
\end{figure}

Now we focus on the second case and take $A(x)=1$ for illustration. We solve the following system in chamber 
\begin{equation}
\label{eq46}
\begin{aligned}
& \partial_x c_1 + c_1 \partial_x \phi = -J_1/D_1 \equiv \tilde{J}_1,\\
& \partial_x c_2 +2 c_2 \partial_x \phi = -J_2/D_2 \equiv \tilde{J}_2,\\
& \partial_x c_3 - c_3 \partial_x \phi = 0,\\
& c_1 + 2c_2 - c_3 = 0.
\end{aligned}
\end{equation} 
It is not easy to solve $\phi(x)$ and $c_i(x)$ ($i=1,2,3$) directly, instead if we treat $\phi$ as the independent variable, we can solve $x(\phi)$ and $c_i(\phi)$ ($i=1,2,3$) explicitly. We denote solutions by $x_R(\phi),c_{iR}(\phi)$ for the right half interval $0<x<1$ and by  $x_L(\phi),c_{iL}(\phi)$ for $-1<x<0$, given in Appendix \ref{appendixB}. Then, by the continuity of $\mu_1$ and $\mu_2$ at $x=0$, we get
\begin{equation}
\label{eq47}
\begin{aligned}
& \phi_{0L} +  \ln c_{1L}(\phi_{0L}) = \phi_{0R} +  \ln c_{1R}(\phi_{0R}),\\
& 2 \phi_{0L} + \ln c_{2L}(\phi_{0L}) = 2 \phi_{0R} +  \ln c_{2R}(\phi_{0R}),\\
\end{aligned}
\end{equation} 
where $ \phi_{0L}$ and $\phi_{0R}$ are left and right limit values of $\phi$ at $x=0$, which are defined by 
\begin{equation}
\label{eq48}
\begin{aligned}
& x_L(\phi_{0L})=0,\quad  x_R(\phi_{0R})=0.
\end{aligned}
\end{equation} 
All these four equations involve the fluxes $\tilde{J}_1,\tilde{J}_2$ and $V$, thus they determine $\tilde{J}_1,\tilde{J}_2,\phi_{0L},\phi_{0R}$ in terms of $V$. The general case of $A(x)$ needs only slight modifications, see Appendix \ref{appendixB}.

\begin{figure}[h]
\begin{center}
\includegraphics[width=2.5in]{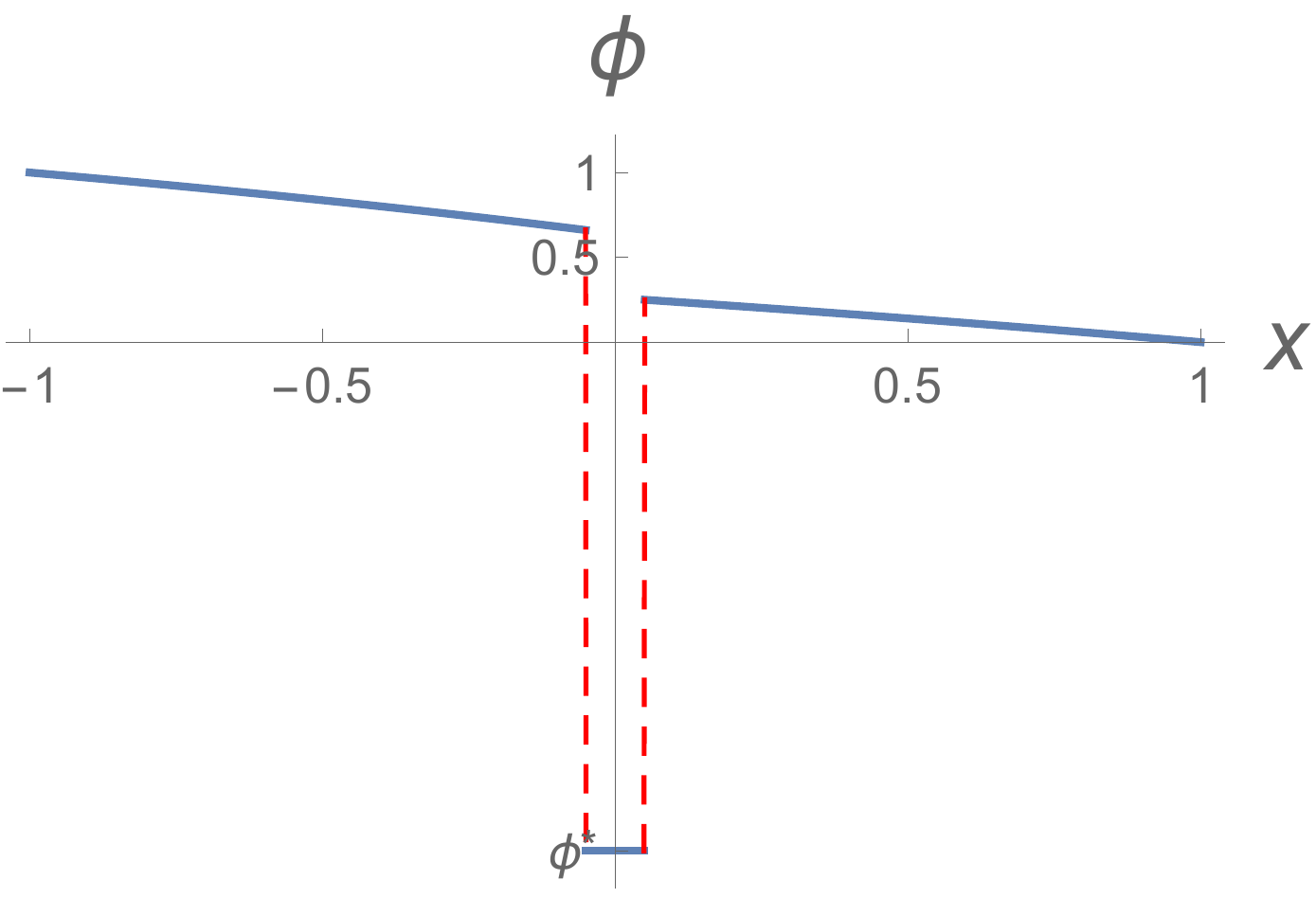}\quad \includegraphics[width=2.5in]{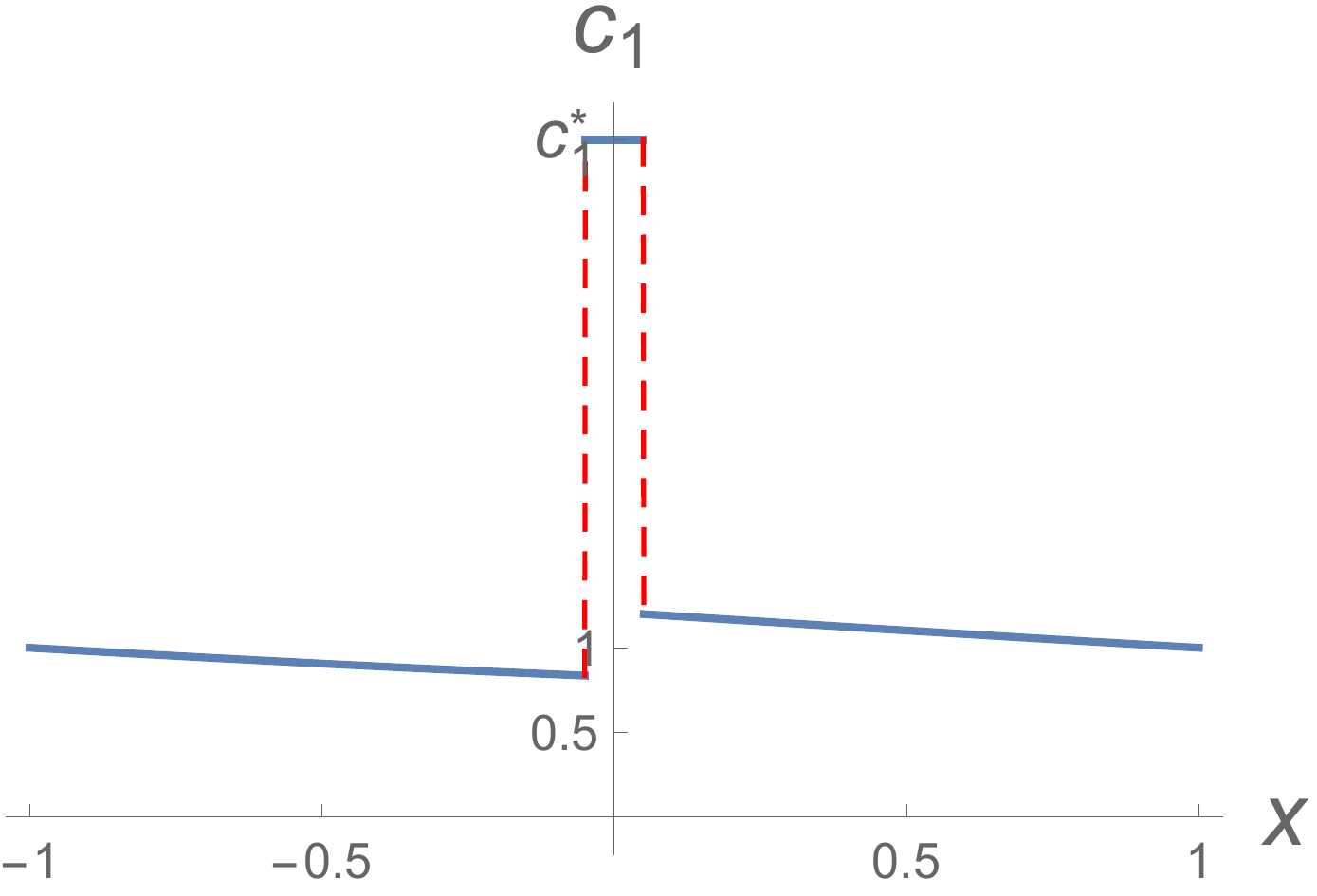} \\
\vspace{0.5cm}
\includegraphics[width=2.5in]{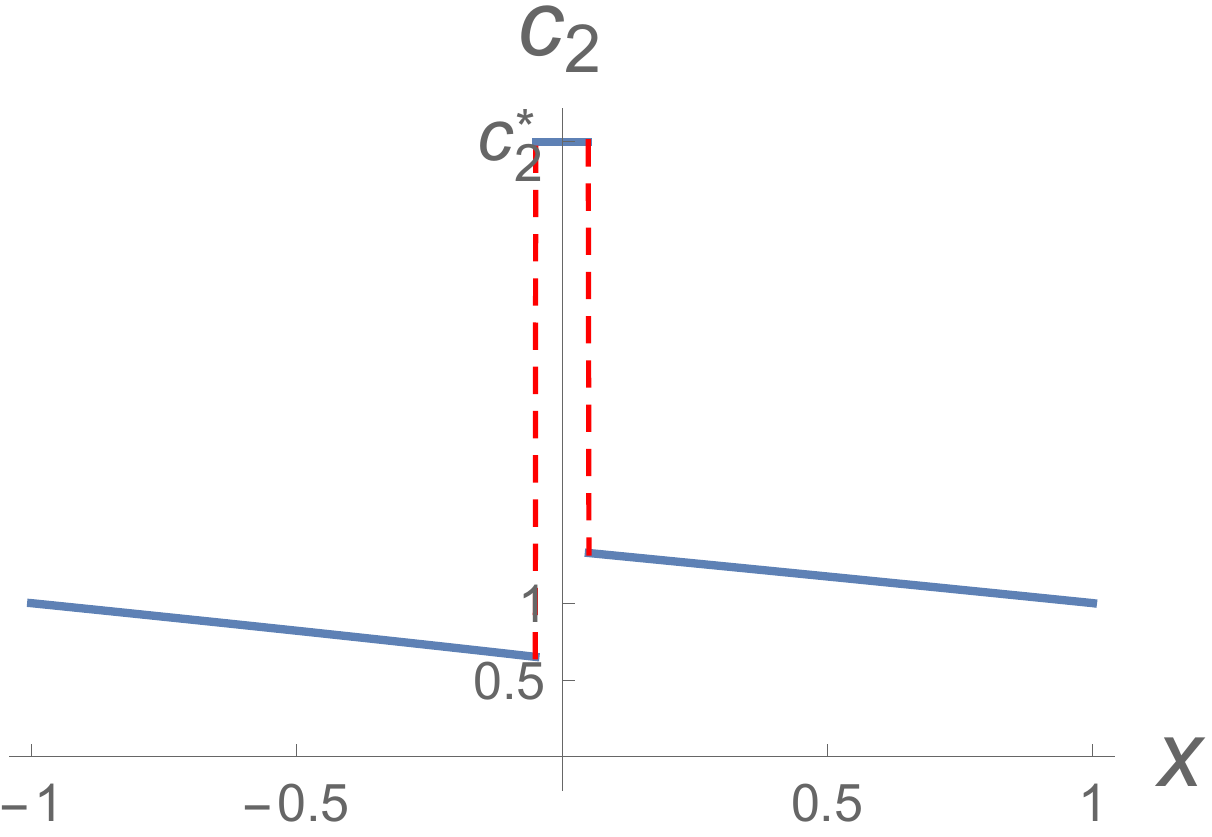}\quad \includegraphics[width=2.5in]{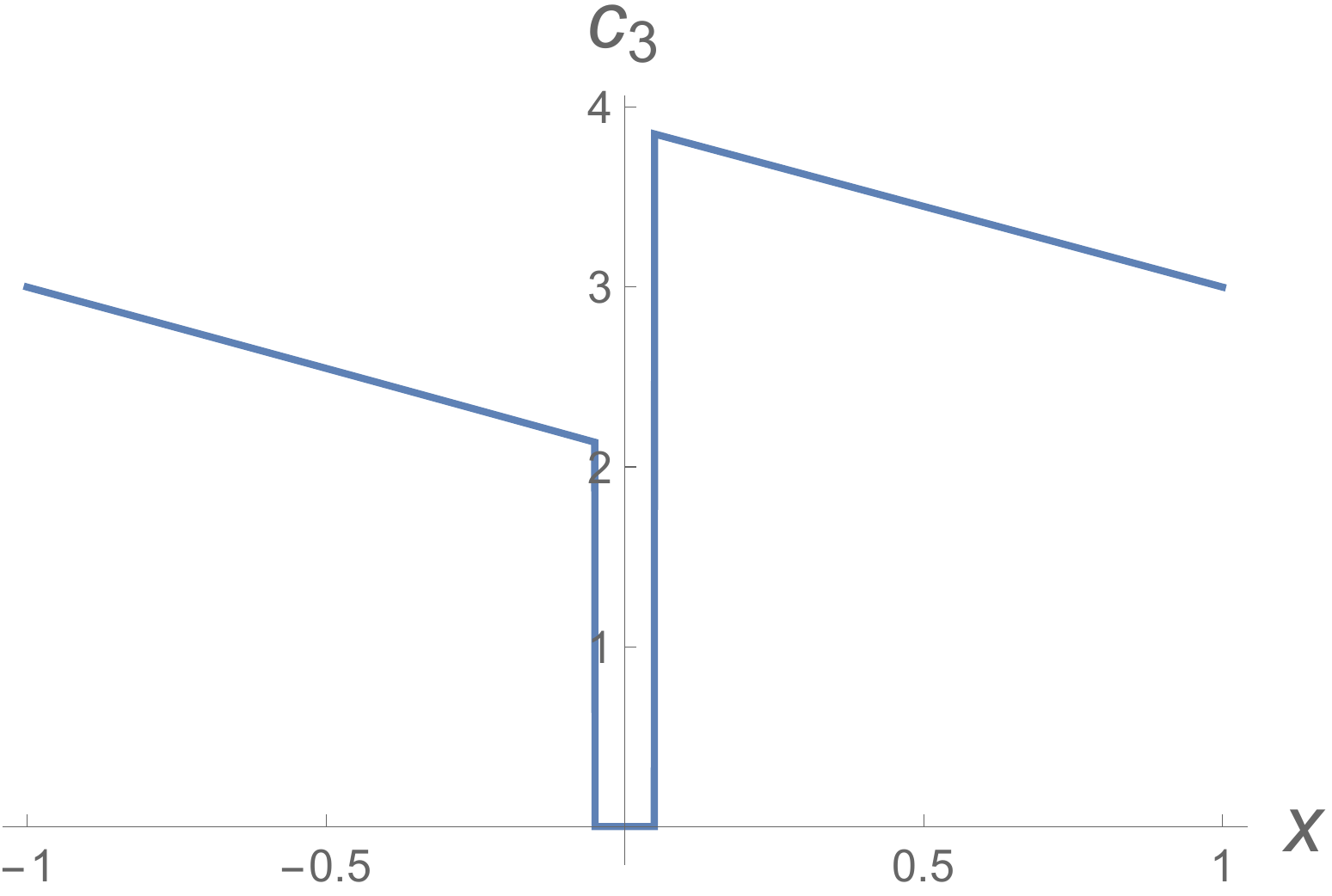}
\caption{\label{fig10} Profiles of $\phi(x)$ and $c_i(x)$ ($i=1,2,3$) with $c_{2b}=1,V=1$ and $1/a_1^3<q<2/a_2^3 \delta$.}
\end{center}
\end{figure}

\begin{figure}[h]
\begin{center}
\includegraphics[width=2in]{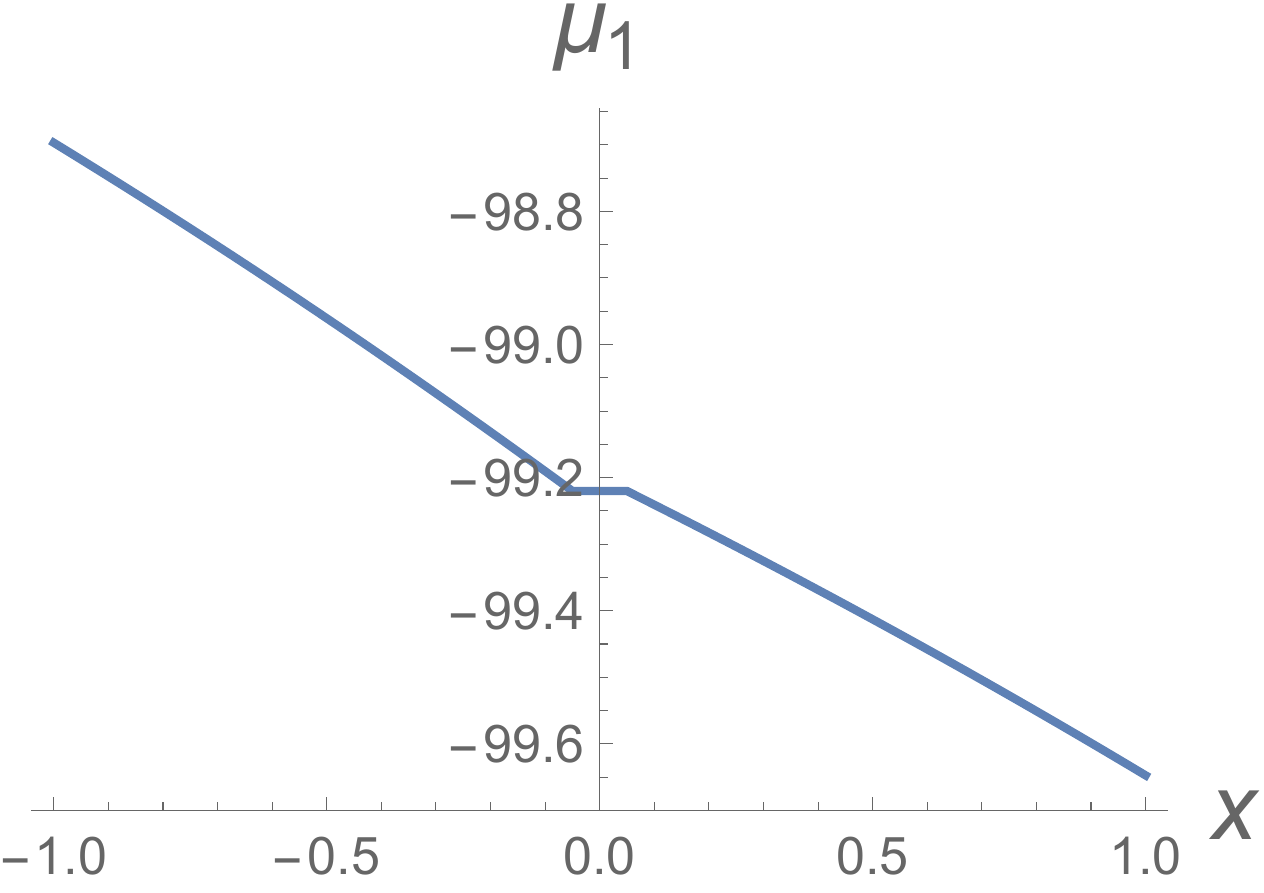} \includegraphics[width=2in]{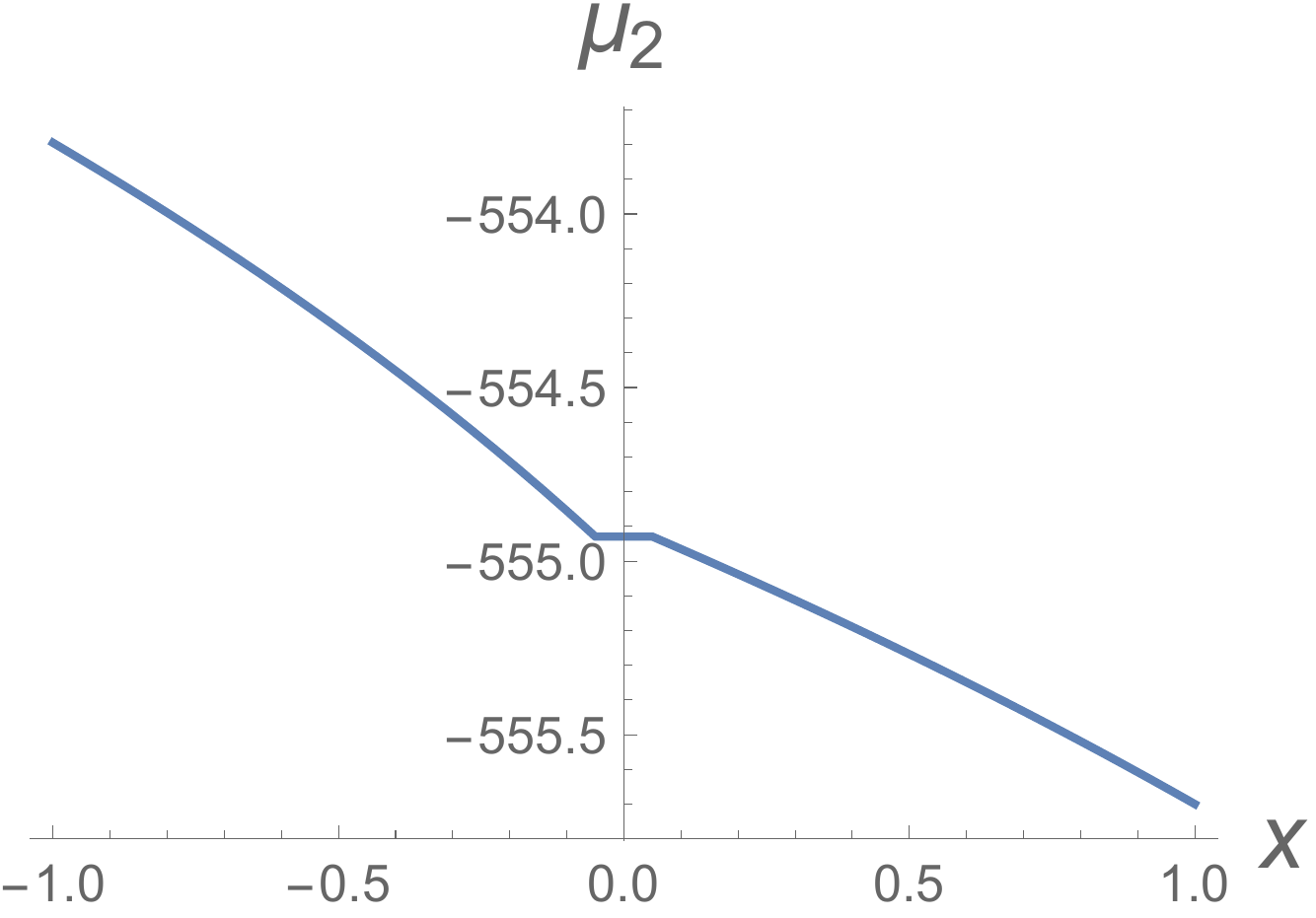}\includegraphics[width=2in]{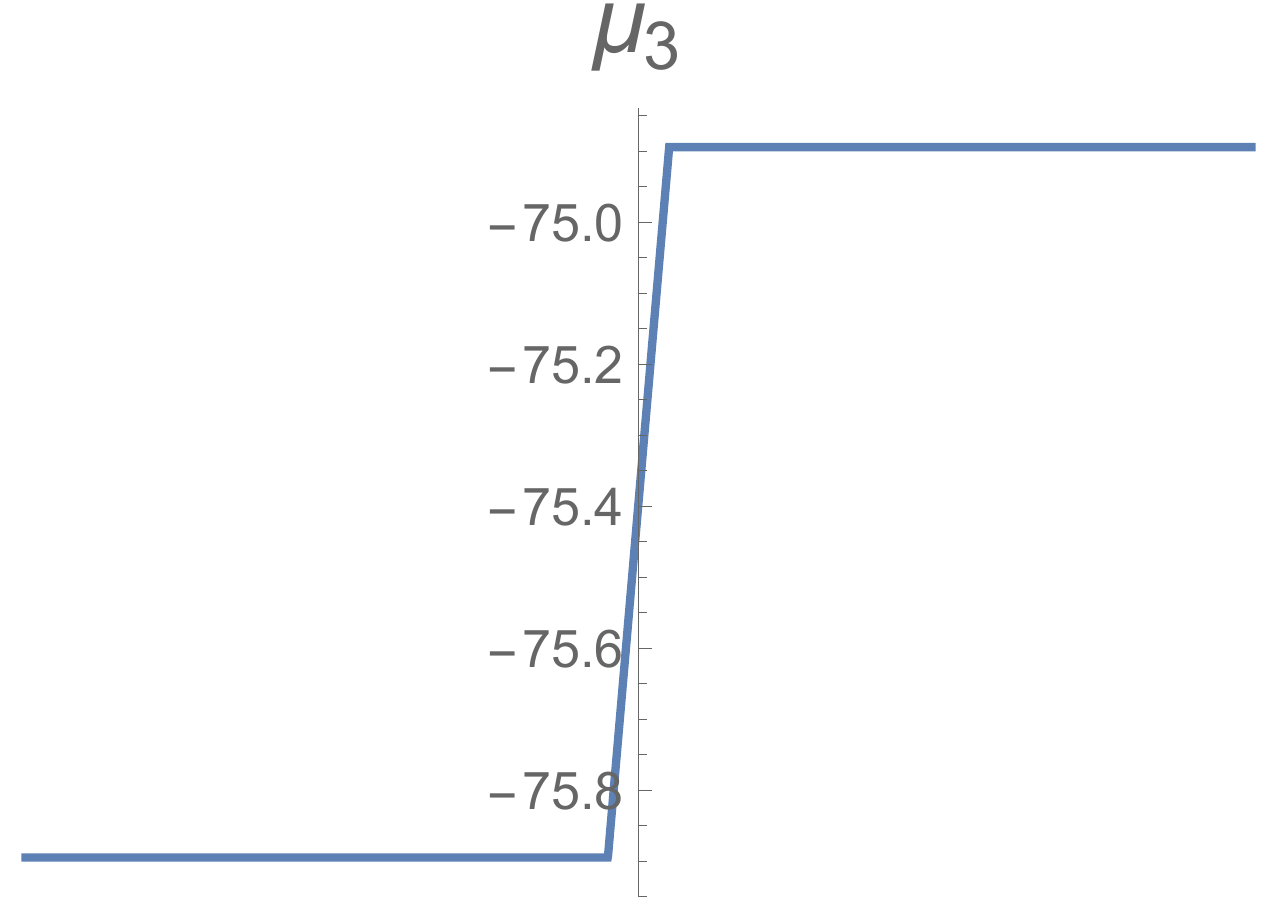}
\caption{\label{fig11} Profiles of $\mu_i(x)$ ($i=1,2,3$) with $c_{2b}=1,V=1$ and $1/a_1^3<q<2/a_2^3 \delta$.}
\end{center}
\end{figure}

Figure \ref{fig9}(a) shows flux-voltage $J$-$V$ relations with $c_{2b}=1$, indicating that both fluxes $\tilde{J}_1$ and $\tilde{J}_2$ saturate for relatively large $V$. Figure \ref{fig9}(b) shows the flux $\tilde{J}_2$ when $c_{2b}= 10^{-3}$ is set very small, indicating the flux almost proportionally gets smaller as chamber concentration gets smaller. In Figure \ref{fig9}(b),  the flux $\tilde{J}_1$ is omitted since it is almost the same as in \ref{fig9}(a) and in much larger scale. Figure \ref{fig10} shows the profiles of $\phi(x)$ and $c_i(x)$ ($i=1,2,3$) with boundary values $c_{2b}=1,V=1$ and parameter $1/a_1^3<q<2/a_2^3 \delta$. Figure \ref{fig11} shows the profiles of $\mu_i(x)$ ($i=1,2,3$) for each ion species. The finite variation of $\mu_1$ and $\mu_2$ in chamber causes the finite flux of $c_1$ and $c_2$, while $\mu_3$ is constant in chamber.

\section{Computational analysis}

In this section, we solve the modified PNP system numerically. Our main objective is to verify our asymptotic analysis under simplifying conditions.

\begin{figure}[h]
\begin{center}
\includegraphics[width=2.3in]{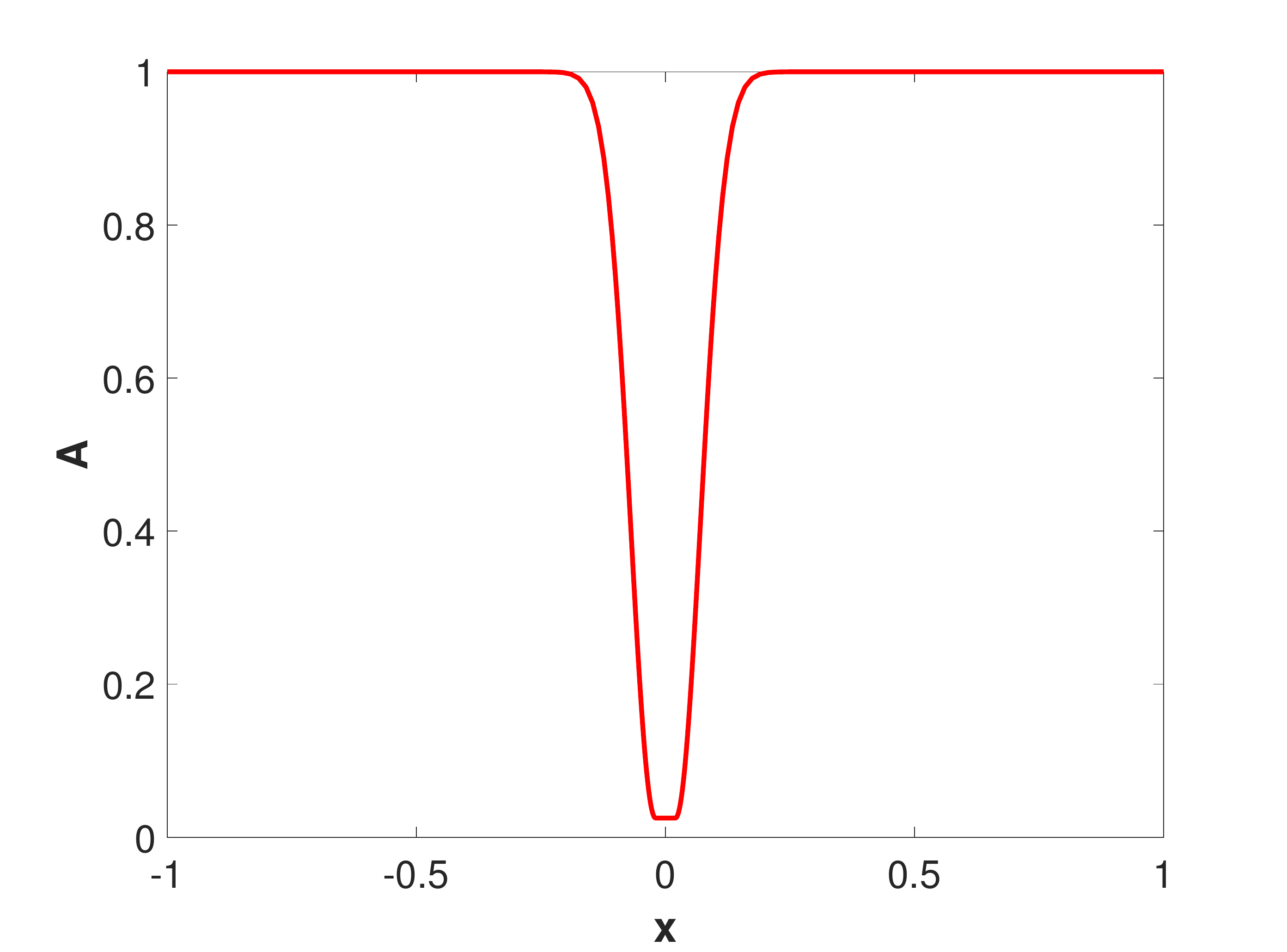} \includegraphics[width=2.3in]{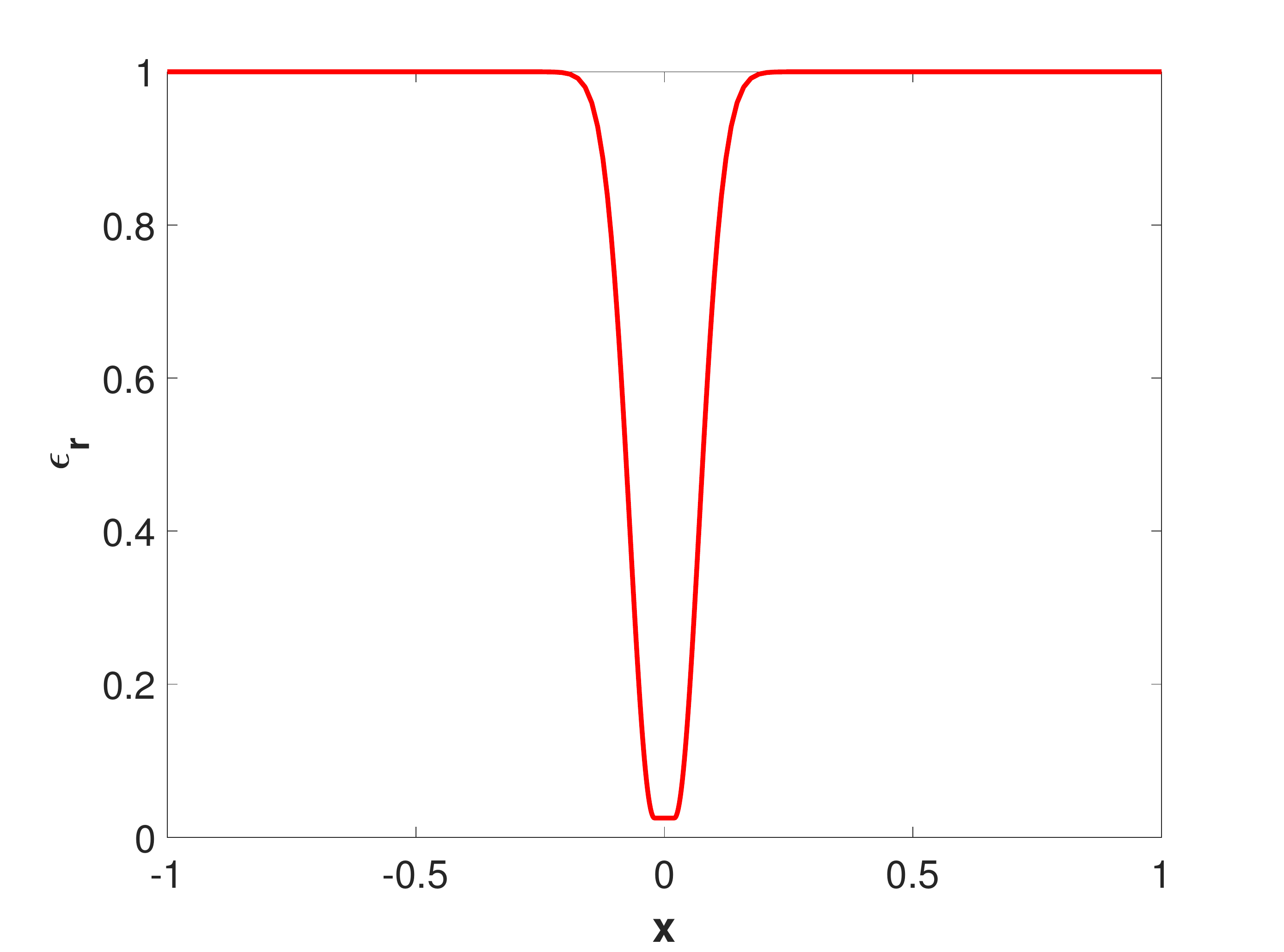} 
\caption{\label{Fig14}Smooth functions $A(x)$ and $\epsilon_r(x)$ used in simulation.}
\end{center}
\end{figure}

We use the dynamic process to simulate the steady state solutions for $\phi,c_i$ and associated fluxes. Some smooth dimensionless functions $\epsilon_r(x)$ (connecting $1/40$ and $1$) and $A(x)$ (connecting $1/30$ and $1$) will be used in the simulation, see Figure \ref{Fig14}. Now we illustrate it by considering the 3-ions case with $c_i$ ($i=1,2,3$) for K$^+$, Na$^+$, Cl$^{-}$. This is to verify previous analytical results for both equilibrium and non-equilibrium cases. We adopt the initial conditions at $t=0$,
\begin{equation}
\label{Eq51}
\begin{aligned}
&c_1(x,0) = 1,\quad c_2(x,0) = 0.1,\quad c_3(x,0) = 1.1.
\end{aligned}
\end{equation} 
The boundary conditions are 
\begin{equation}
\label{Eq52}
\begin{aligned}
&c_1(\pm1,t) = 1,\quad c_2(\pm 1,t) = 0.1, \quad c_3(\pm 1,t) = 1.1, \\
& \phi(-1) = V, \quad \phi(1)=0.
\end{aligned}
\end{equation} 

\begin{figure}[h]
\begin{center}
\includegraphics[width=2.2in]{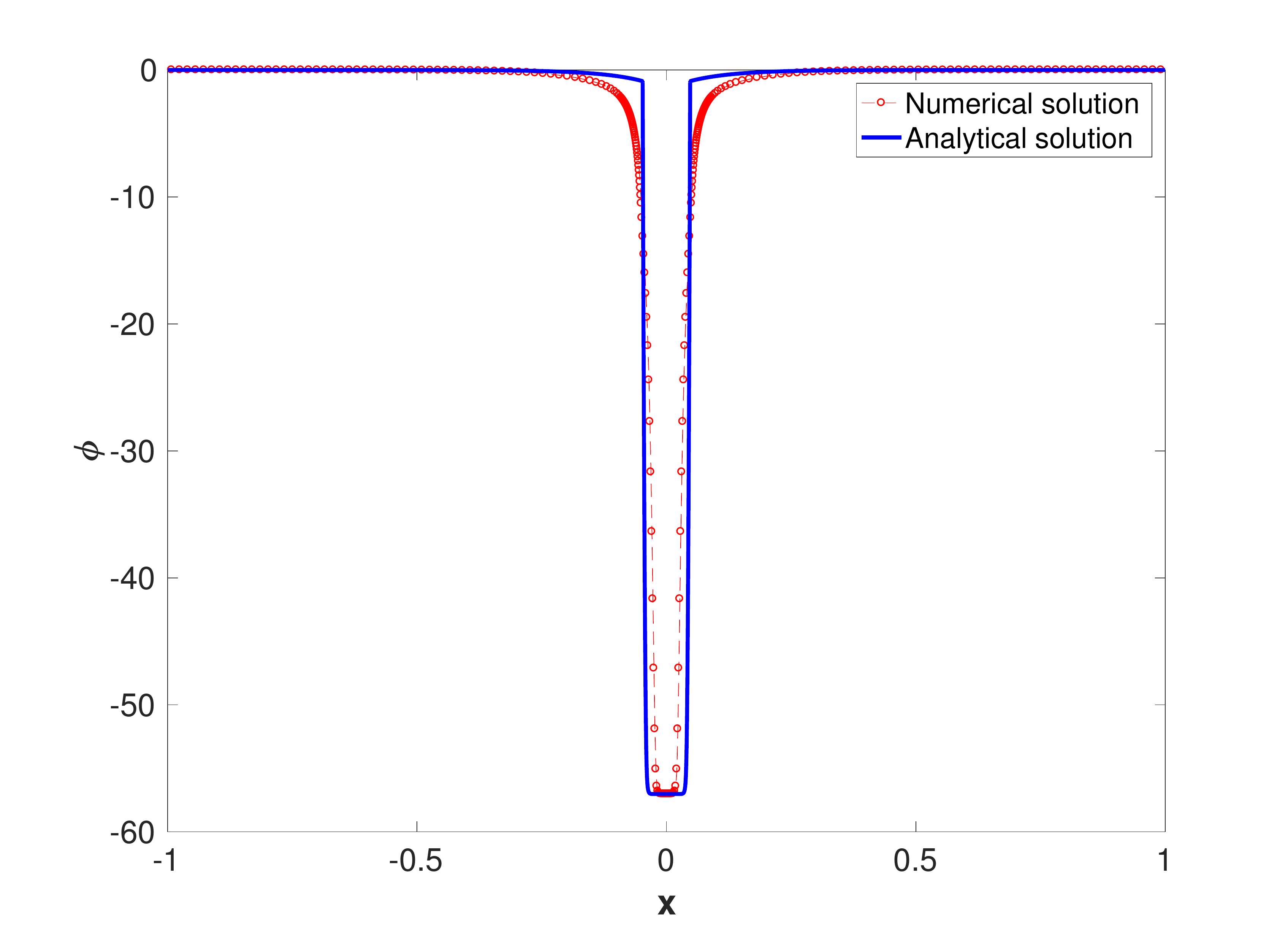}
\includegraphics[width=2.2in]{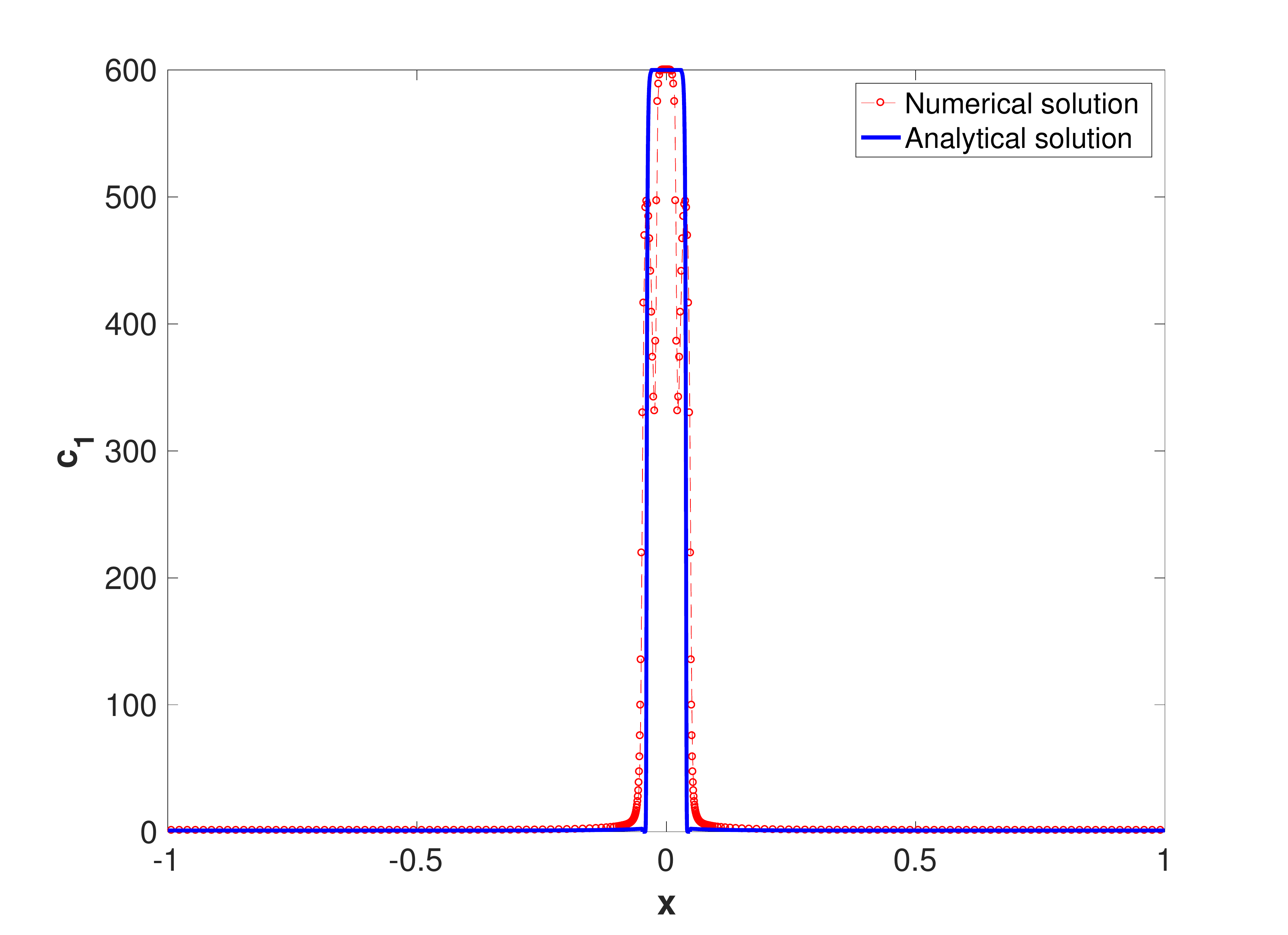} \\
\includegraphics[width=2.2in]{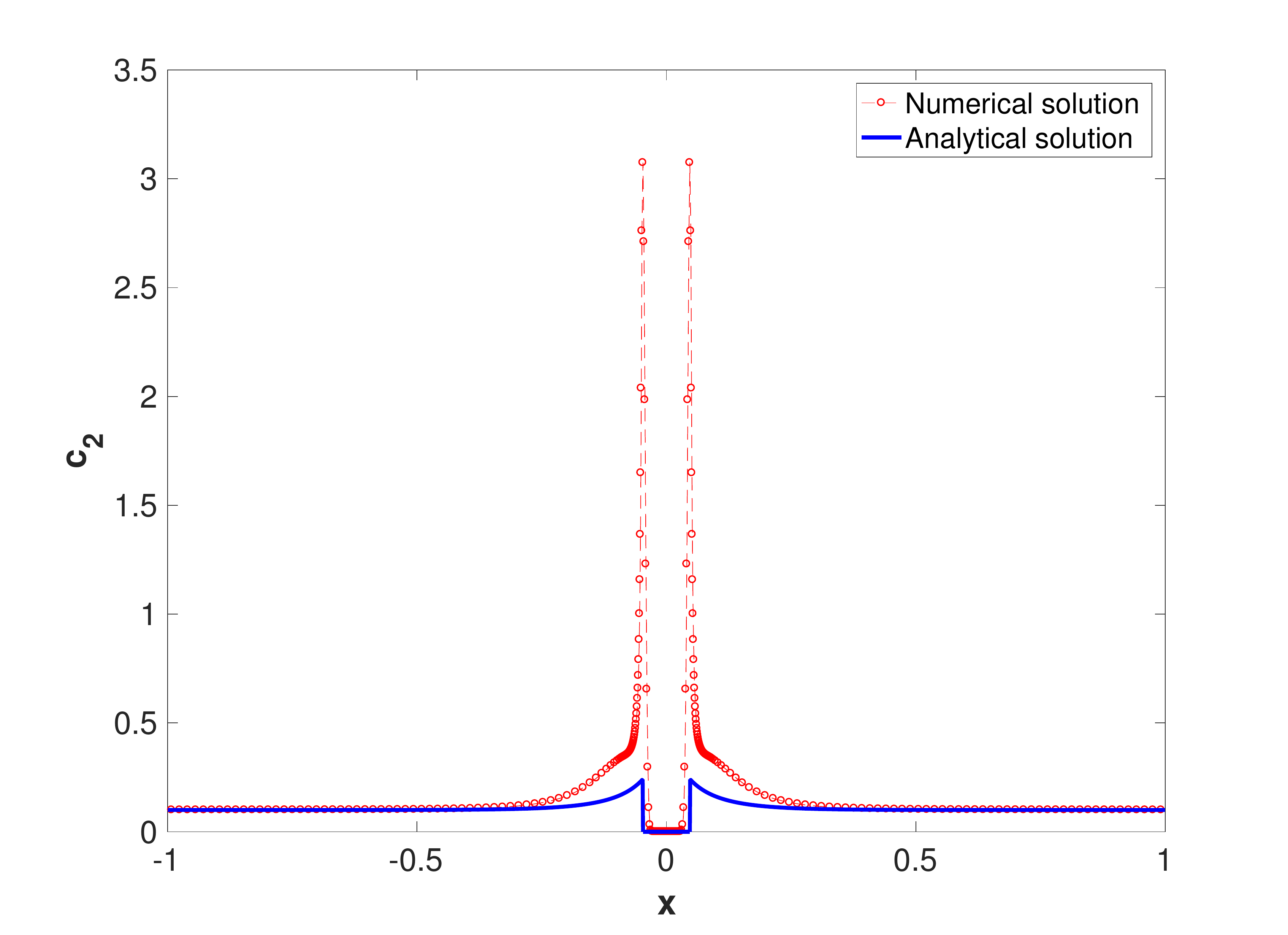} \includegraphics[width=2in]{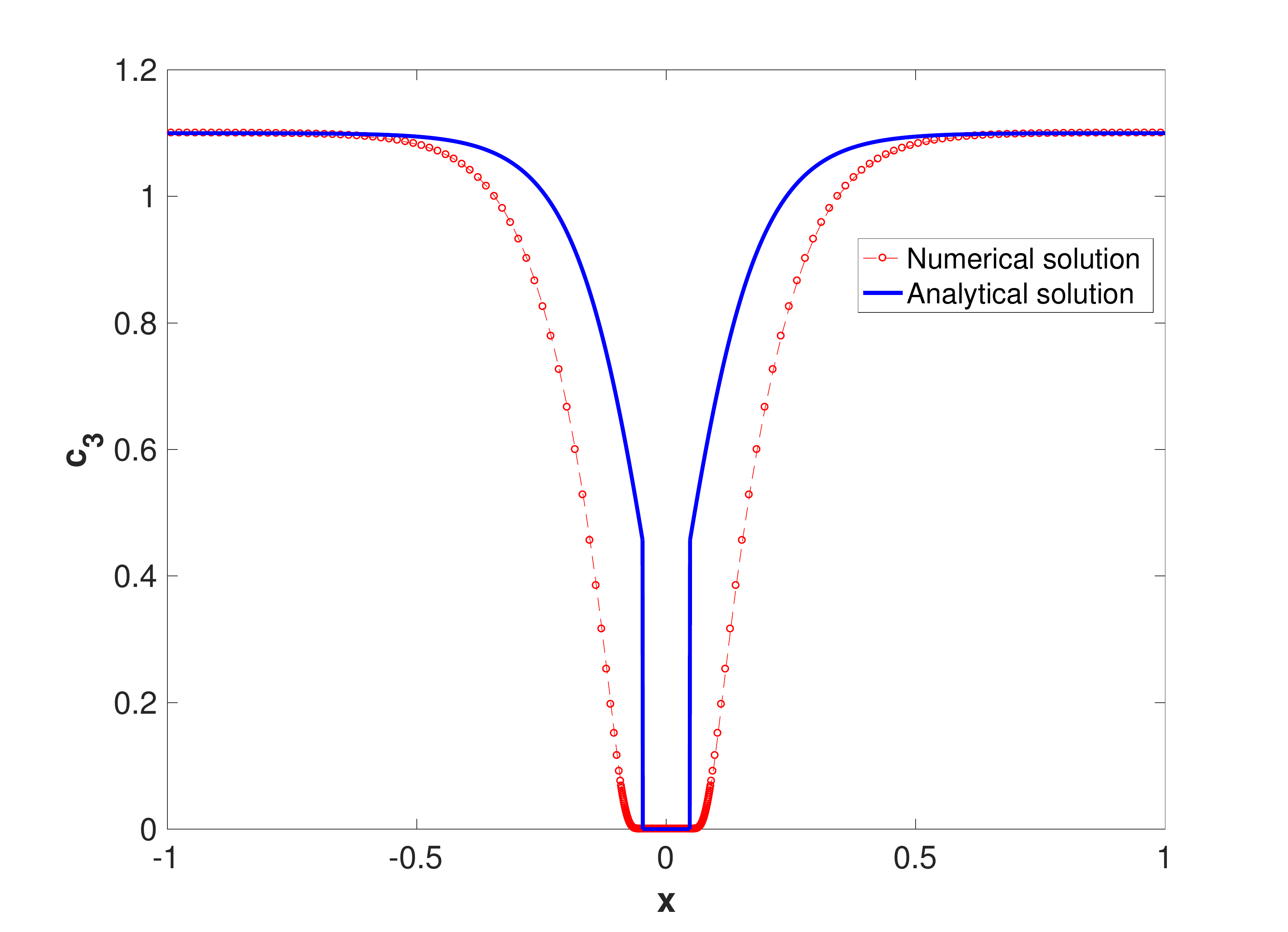}
\caption{\label{Fig15}Profiles of $\phi$ and $c_i$ ($i=1,2,3$) near steady state for $V=0$ and $q=600$.}
\end{center}
\label{fig21}
\end{figure}

\begin{figure}[h]
\begin{center}
\includegraphics[width=2in]{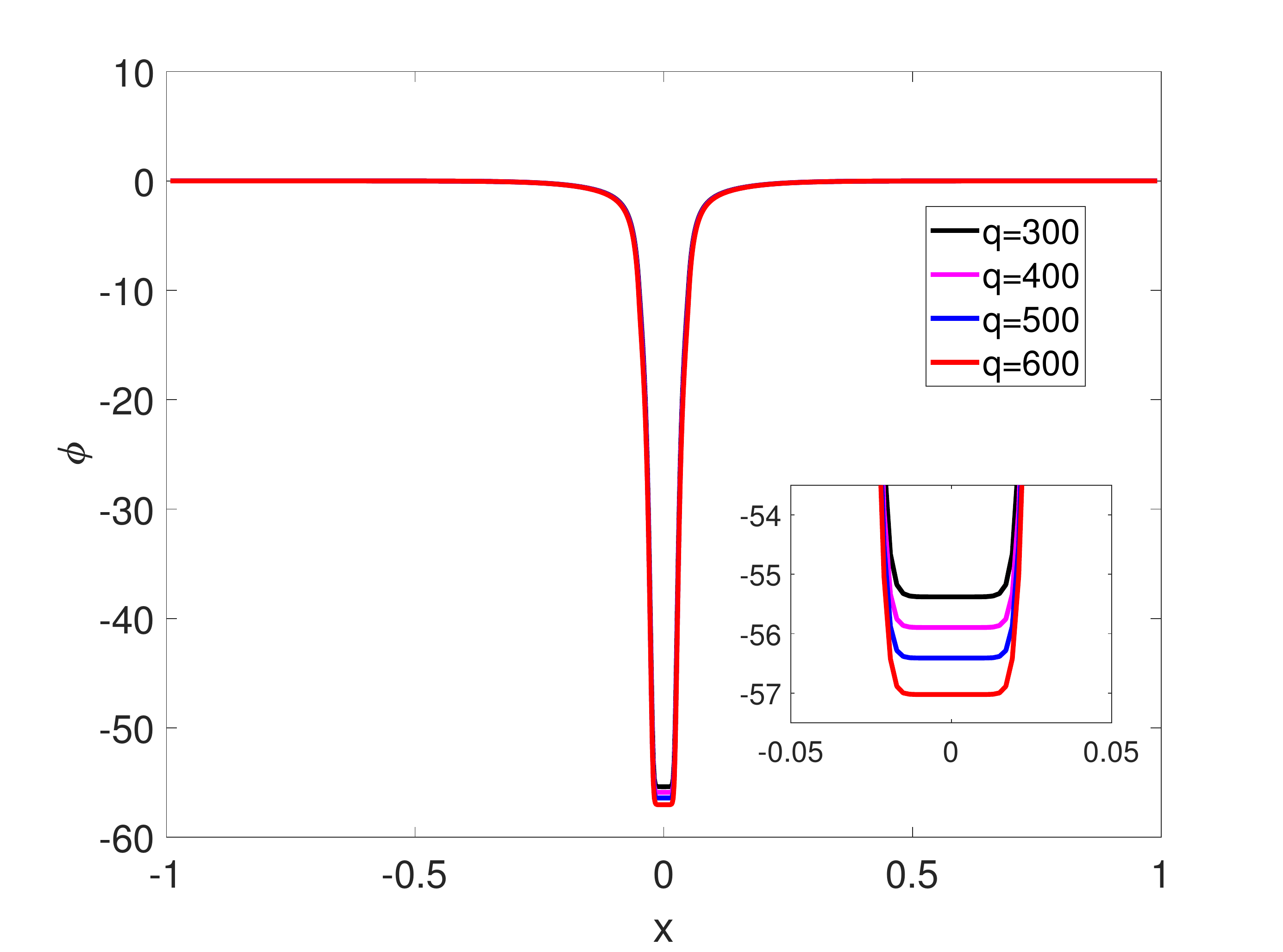} \includegraphics[width=2in]{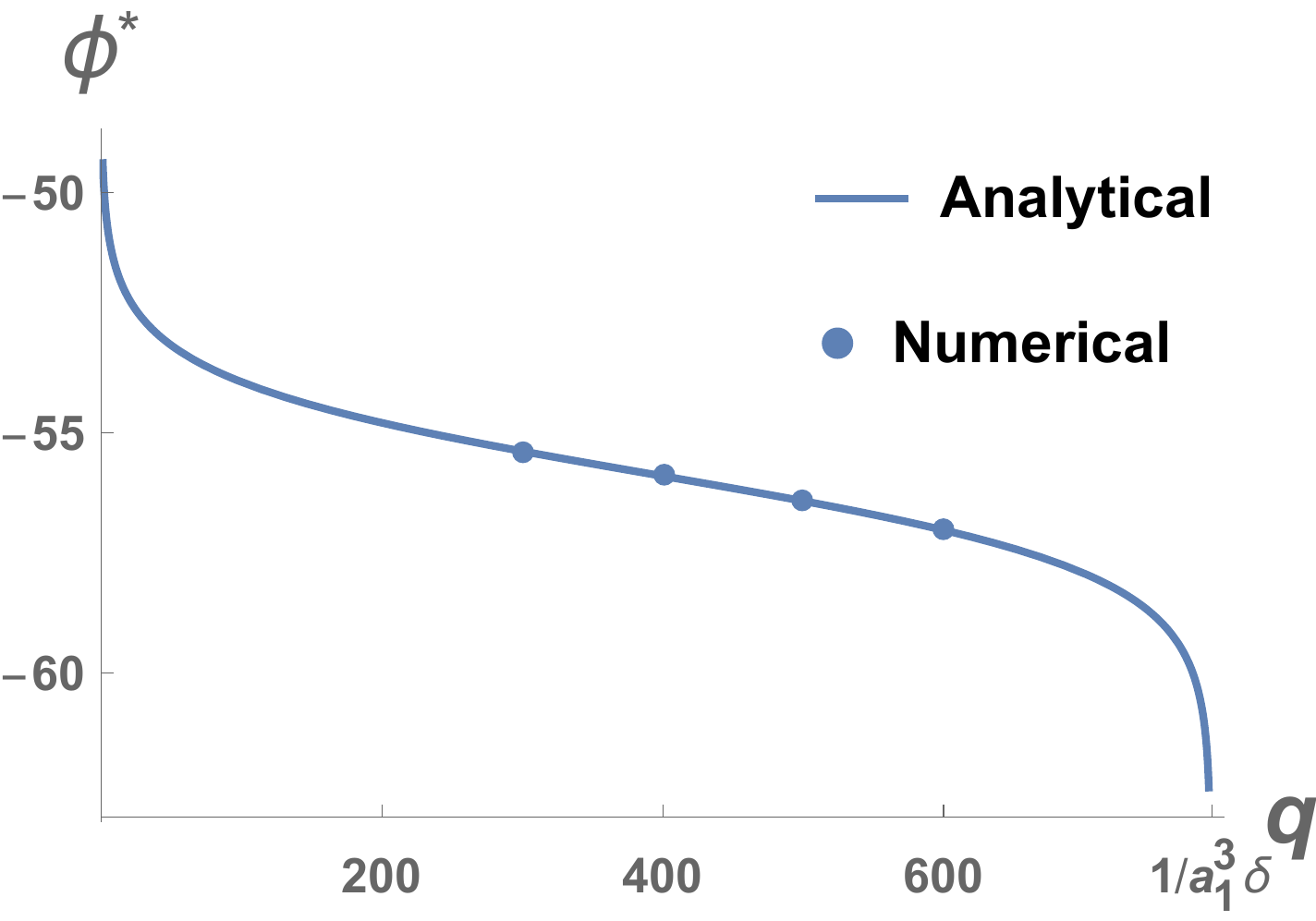} \includegraphics[width=2in]{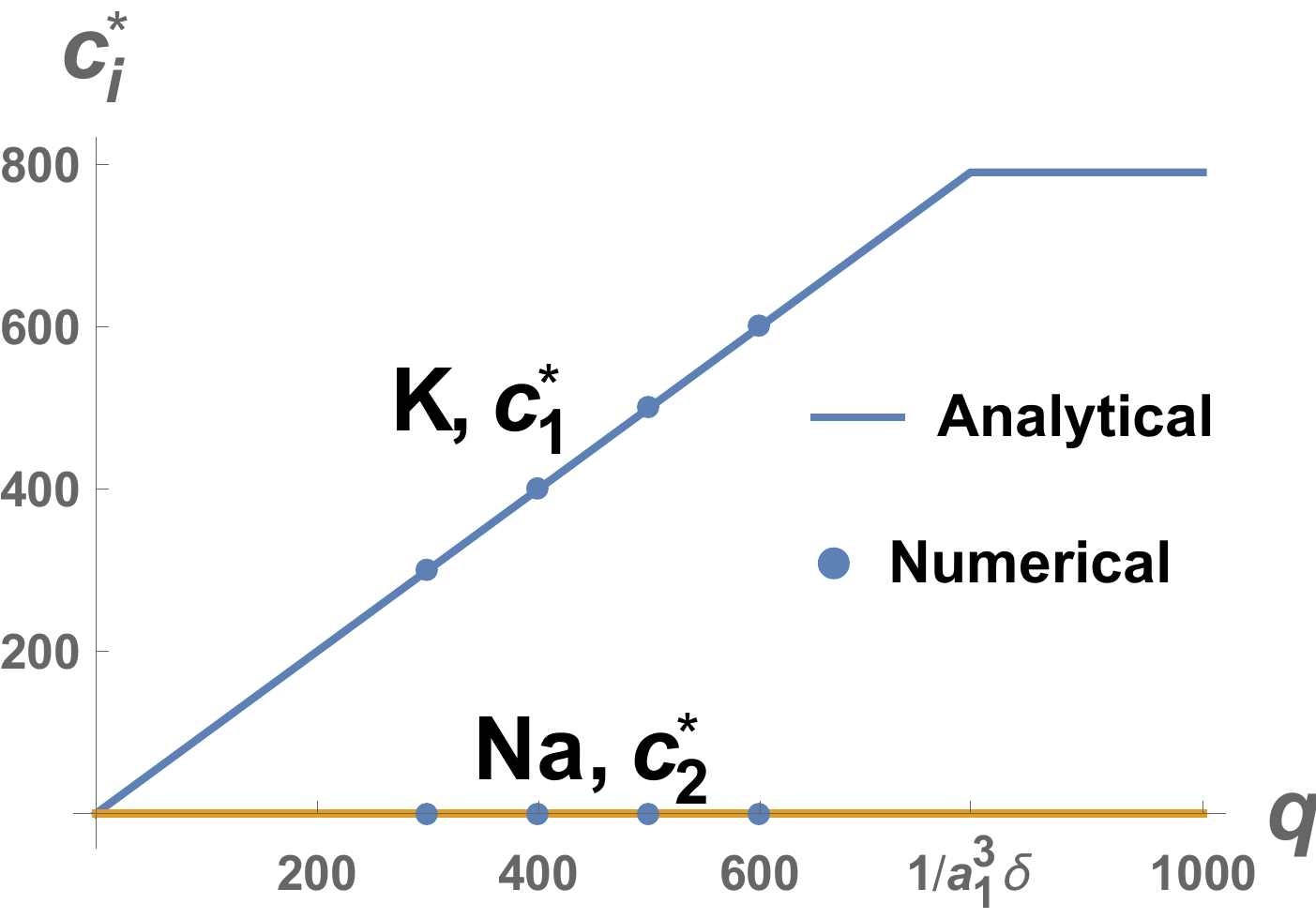} 
\caption{\label{Fig16}Profiles of $\phi$ at steady state for $V=0$ and different $q$, and comparison with analytical results in Figure 2.}
\end{center}
\end{figure}

First, we set $V=0$ and compare the numerical results with analytical results in (\ref{eq20}) (or Figure \ref{fig2}). A series of cases with different $q$ will be simulated. In the simulation, finite-volume method is used with non-uniform mesh points. More mesh points are used in filter, near filter edge, and in regions for large gradient of $\epsilon_r$, and there are totally 273 points. Very small time  step (because of large $q$, small $\epsilon$ and small mesh size) is chosen to ensure stability and accuracy of algorithm. After quite a long time, about 20 h on a computer (processor: 4 GHz, i76700K; memory: 32 GB), the solution tends to some steady state (i.e., all fluxes are almost 0). The profiles of $\phi$ and $c_i$ ($i=1,2,3$) for $q=600$ are shown as red curves in Figure \ref{Fig15}, in comparison with the analytical results in blue curves from Section 3. The numerical and analytical solutions agree very well except a smoothing region near two edges of filter.  For instance, the constant values of $\phi$ in filter show remarkable agreement, i.e., $\phi^\ast=-57.0257,-57.0268$ in numerical and analytical results. One can see that K$^+$ is favored in the filter region, and all the other ions are essentially 0 in filter region. This agrees with results in (\ref{eq20}). To see clearly the dependence on $q$, the profiles of $\phi$ for different $q$ are shown in Figure \ref{Fig16}a, showing that it is constant in filter region. In Figures \ref{Fig16}b and \ref{Fig16}c, the constant values of $\phi,c_1,c_2$ in filter are compared with previous analytical results, where curves are from previous Figure 2 and dots are from numerical results. 

We have also tested different smoothing profiles of $\epsilon_r(x)$ and $A(x)$ and boundary conditions for concentrations, and as long as $\epsilon_r$ is $1/40$ (original value is 2 before scale) in part of filter, the minimum values of $\phi$ will not change. This also verifies the predictions in (\ref{eq20}).


\begin{figure}[h]
\begin{center}
\includegraphics[width=2in]{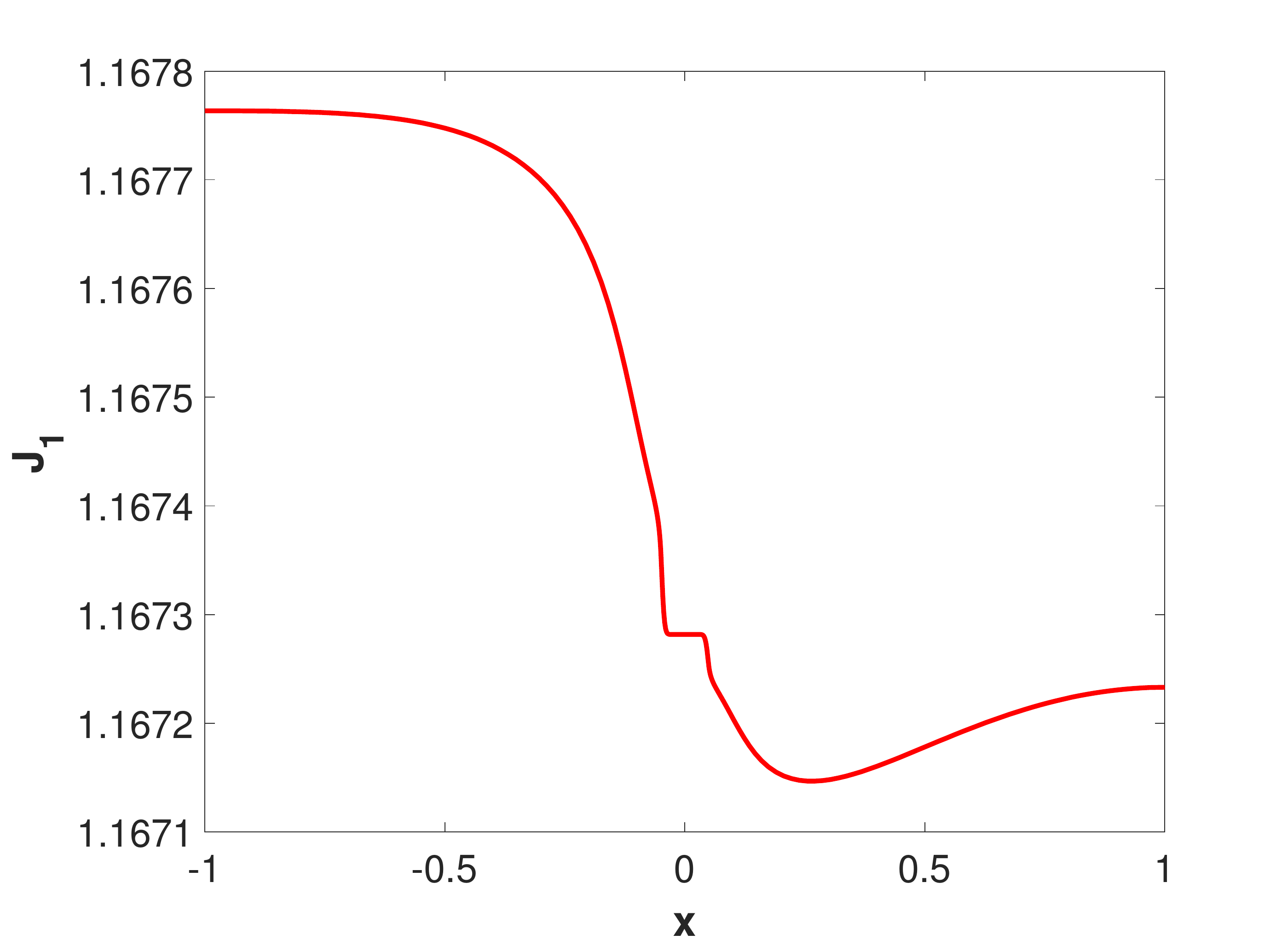} \includegraphics[width=2in]{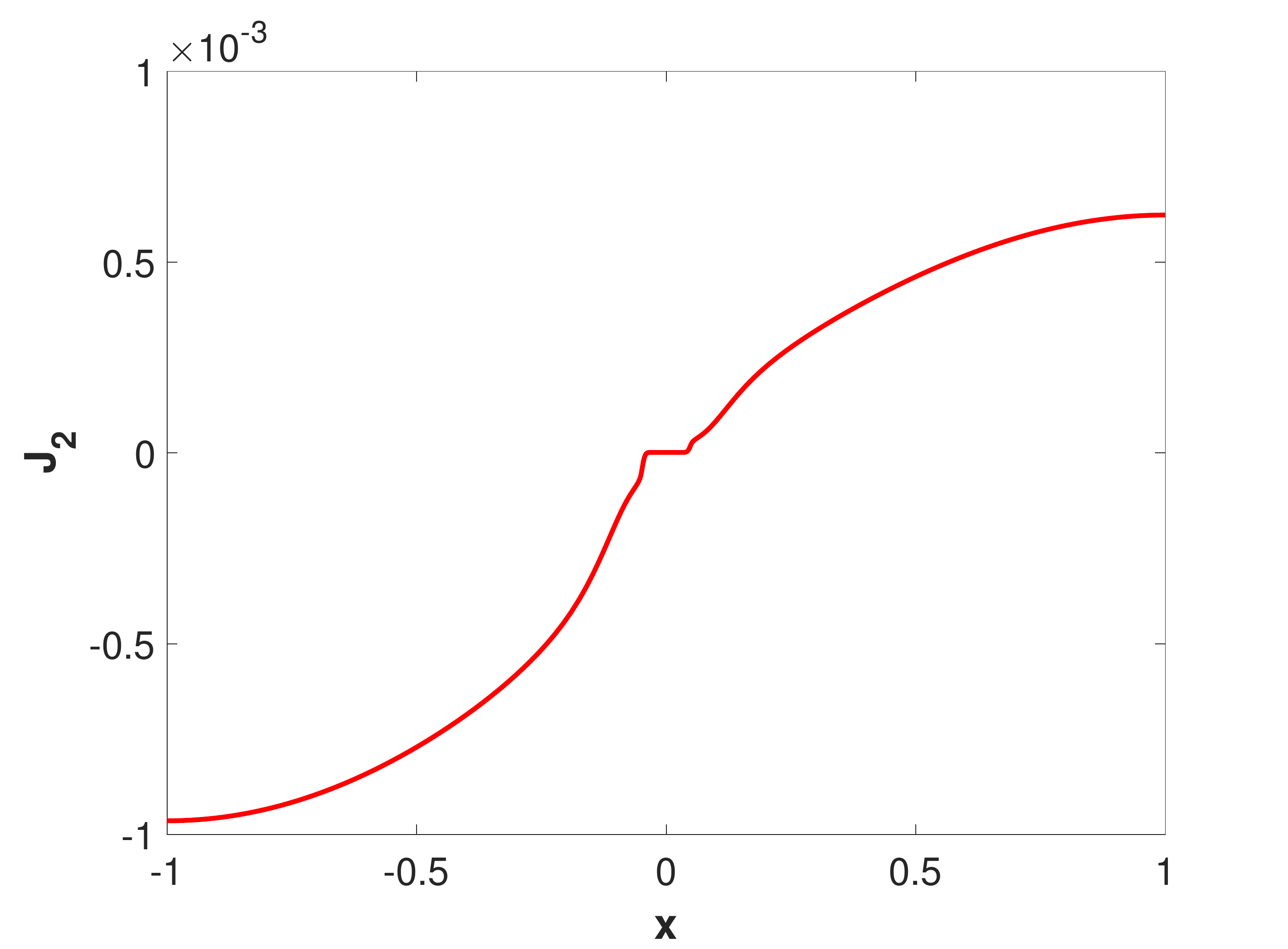}
\includegraphics[width=2in]{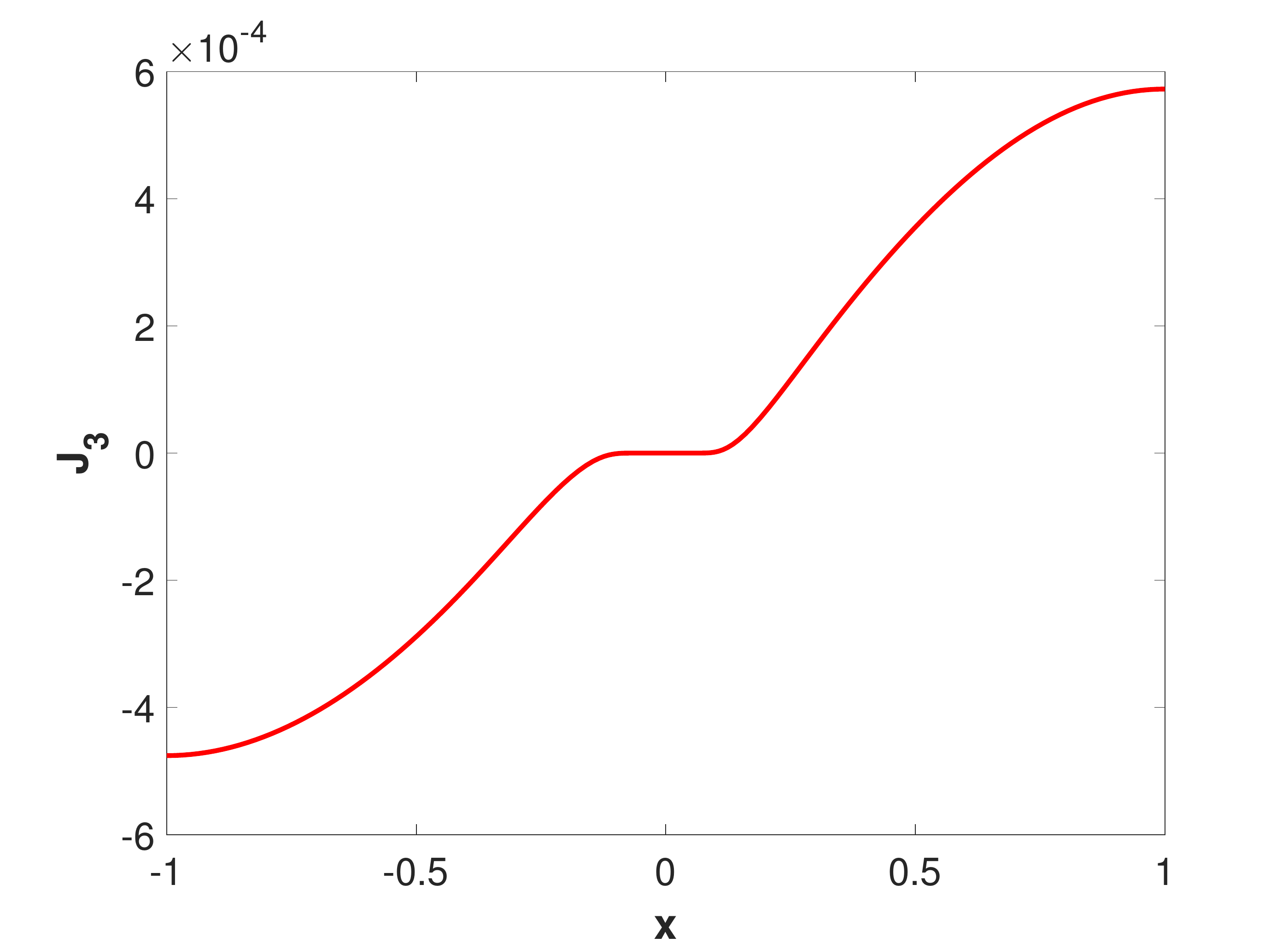}
\caption{\label{Fig17}The fluxes $J_i$ ($i=1,2,3$) near steady state for $V=1,q=600$.}
\end{center}
\end{figure}

\begin{figure}[h]
\begin{center}
\includegraphics[width=2.5in]{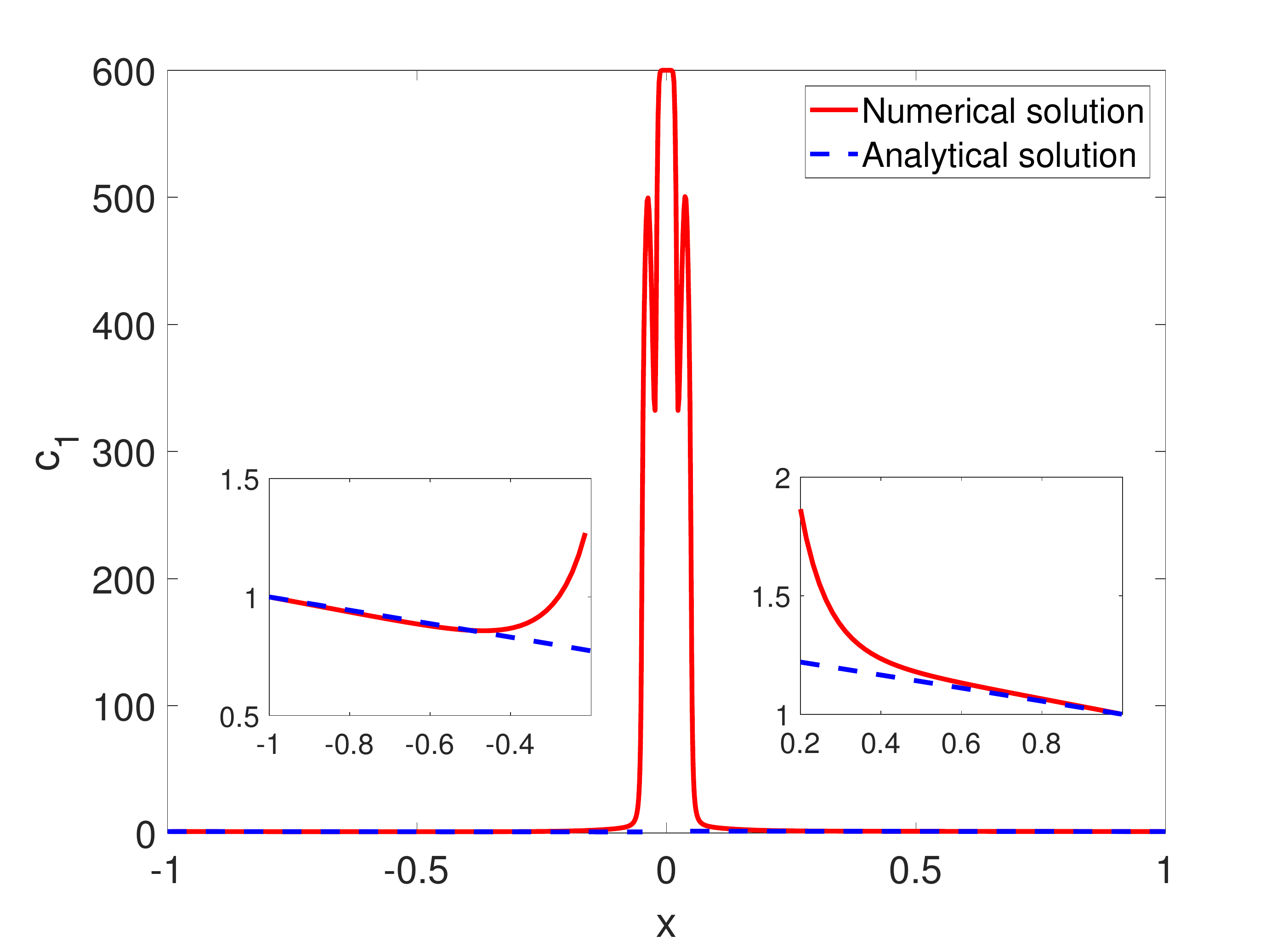} \includegraphics[width=2.5in]{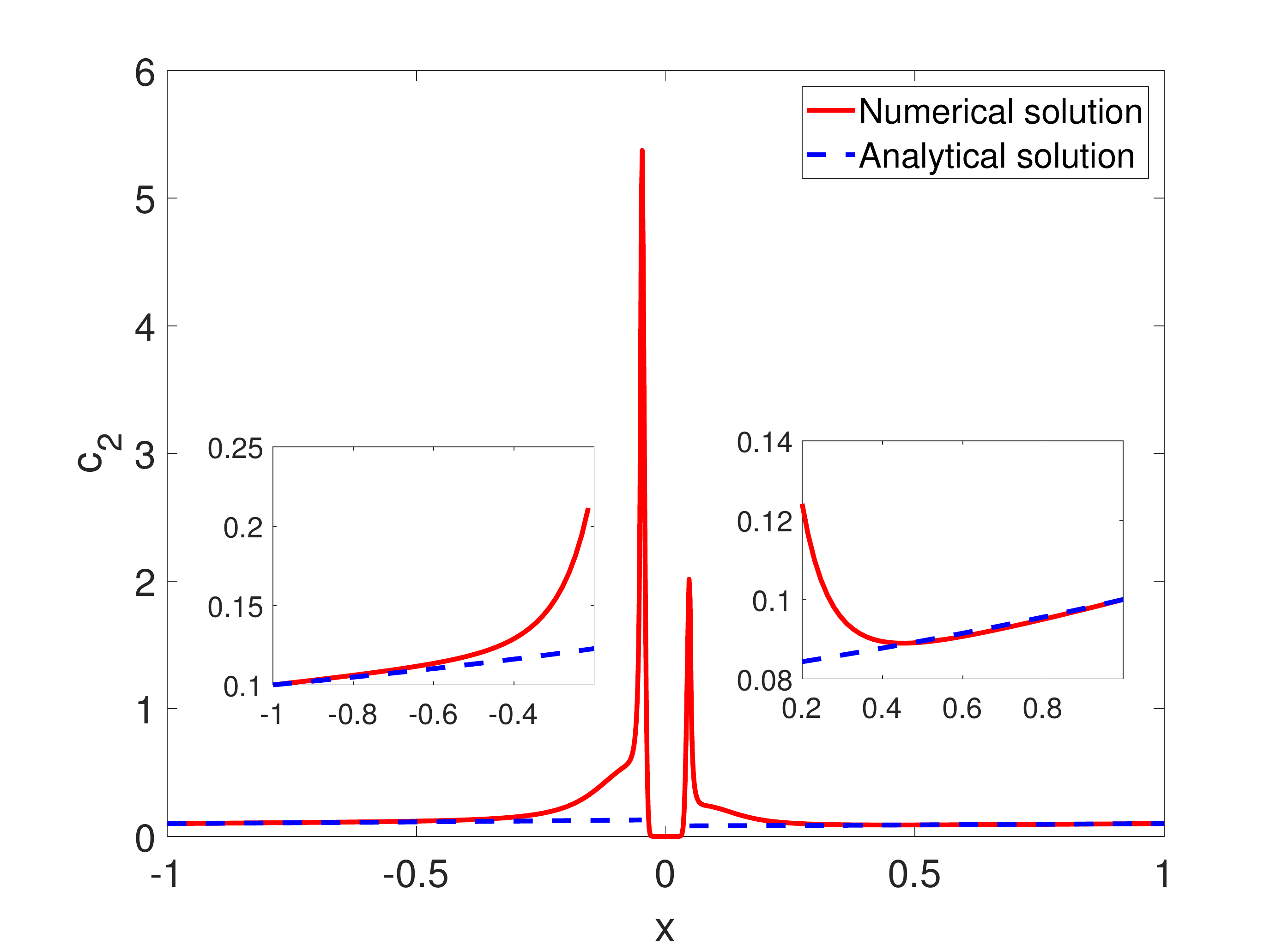}\\
\includegraphics[width=2.5in]{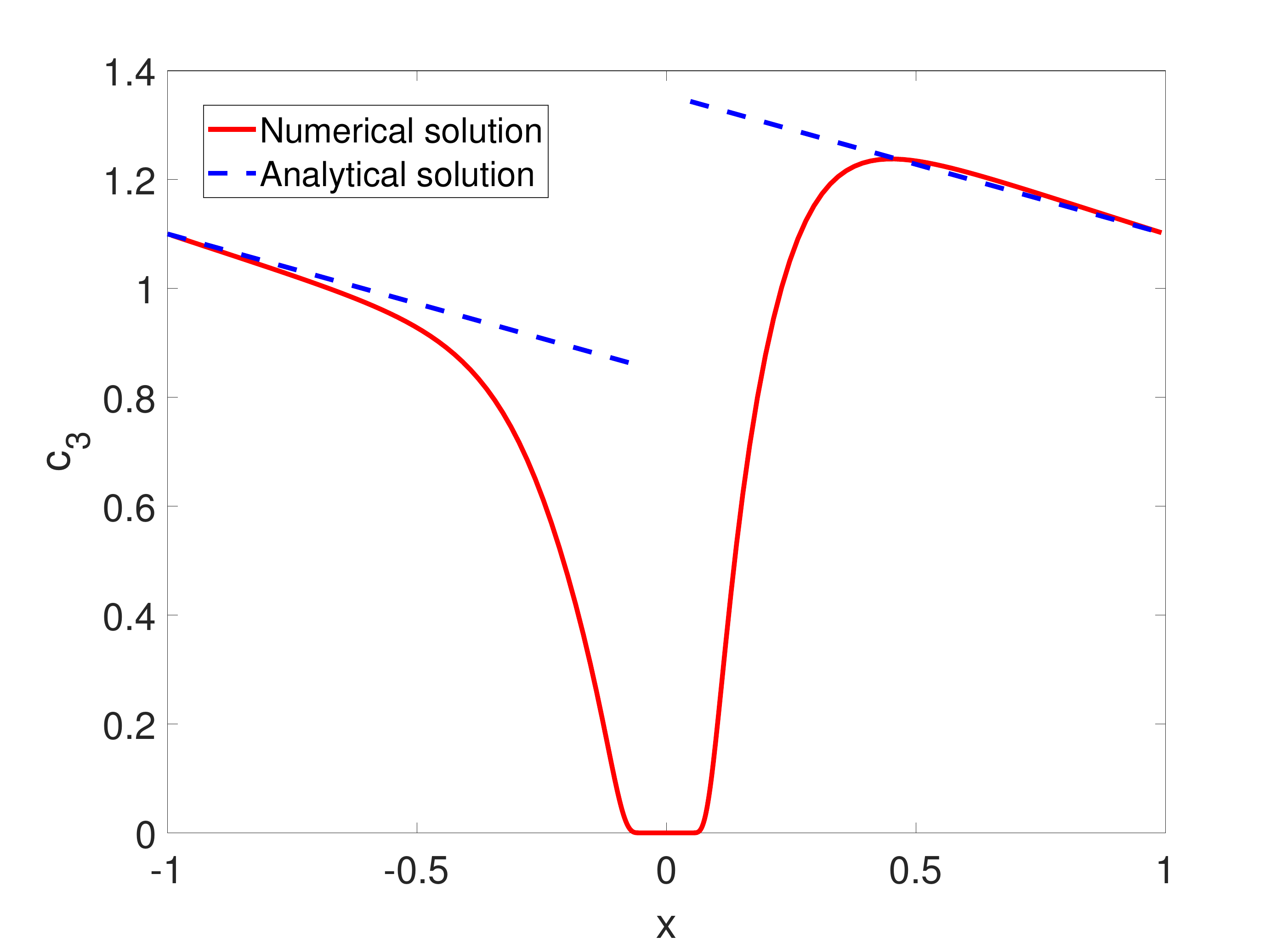} \includegraphics[width=2.5in]{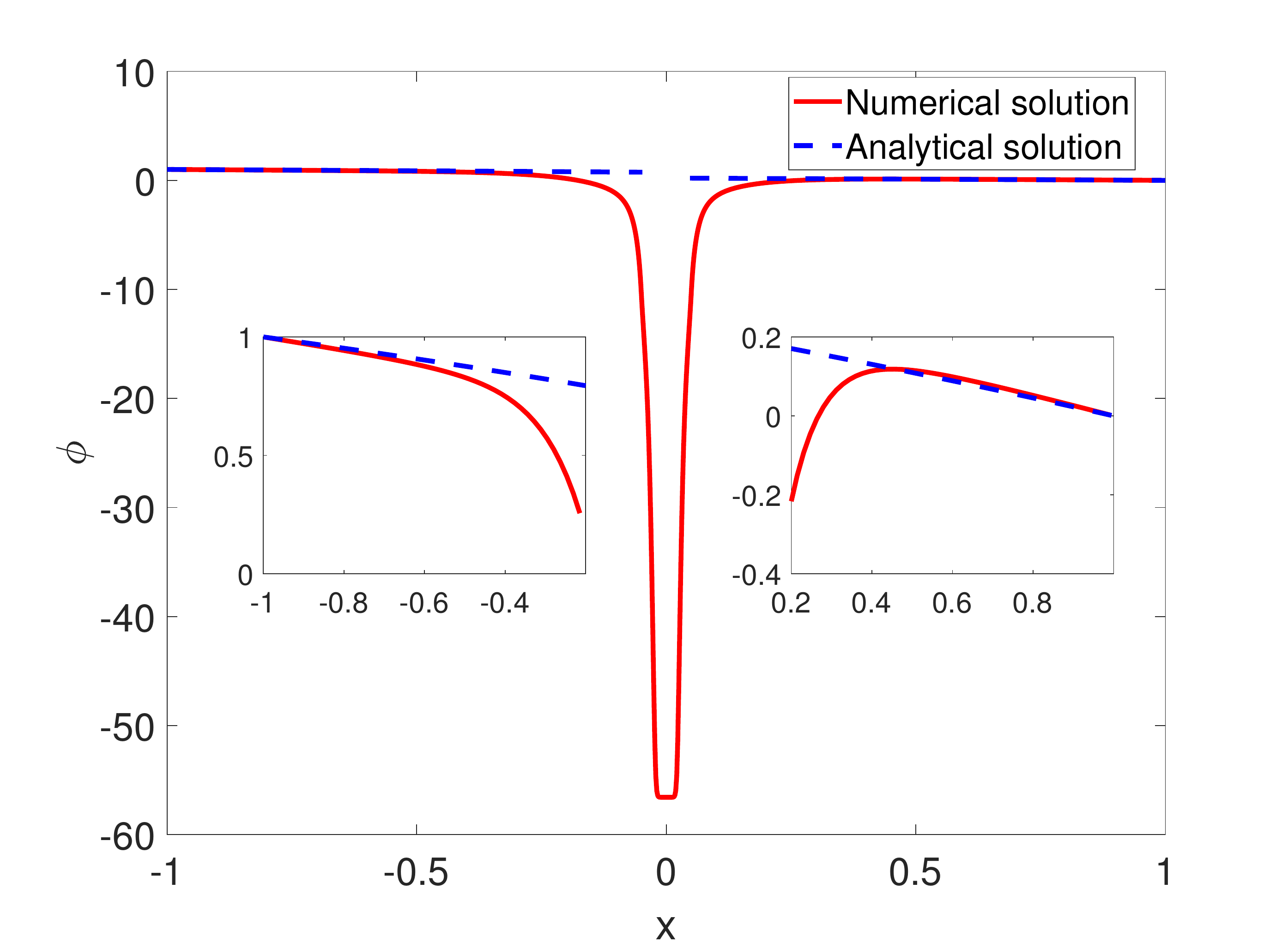}
\caption{\label{Fig18}The profiles of $c_i$ ($i=1,2,3$) and $\phi$ near steady state, for $V=1,q=600$.}
\end{center}
\end{figure}

To test the analysis of non-equilibrium case, we set $V=1,q=600$ and others the same as above. After computation of about 20 h to $t=2$, the system tends to some steady state. The fluxes are shown in Figure \ref{Fig17}, indicating that only flux $J_1$ is nonzero and goes to a constant 1.167 at steady state. This feature agrees with previous analysis. The previous predicted flux by formula (\ref{eq43_1}) with $A(x)=1,c_{2b}=0.1$ is $J\approx 0.51$, and hence $J_1 =D_1 J \approx 1$. They differ by an $O(\epsilon)$ with present $\epsilon \approx 0.13$, as it is natural for previous approximation. In addition the difference is partly due to the the smoothing of $\epsilon_r(x)$ and $A(x)$. The profiles of $c_i,\phi$ ($i=1,2,3$) are shown in Figure \ref{Fig18}.  Some features are similar to the equilibrium case, but the profiles are not symmetric anymore. The profiles of $\mu_i$ ($i=1,2,3$) are shown in Figure 18. The numerical solutions in chamber are also compared with the previous analytical solutions (see Figure \ref{fig7} and Appendix B) in dashed lines of embedded figures.  All the profiles of $c_i,\phi,\mu_i$ ($i=1,2,3$) except $\mu_3$ show agreement with previous analytical results. We have also tested different $V$, and compared with analytical flux-voltage curves in next section.

\begin{figure}[h]
\begin{center}
 \includegraphics[width=2in]{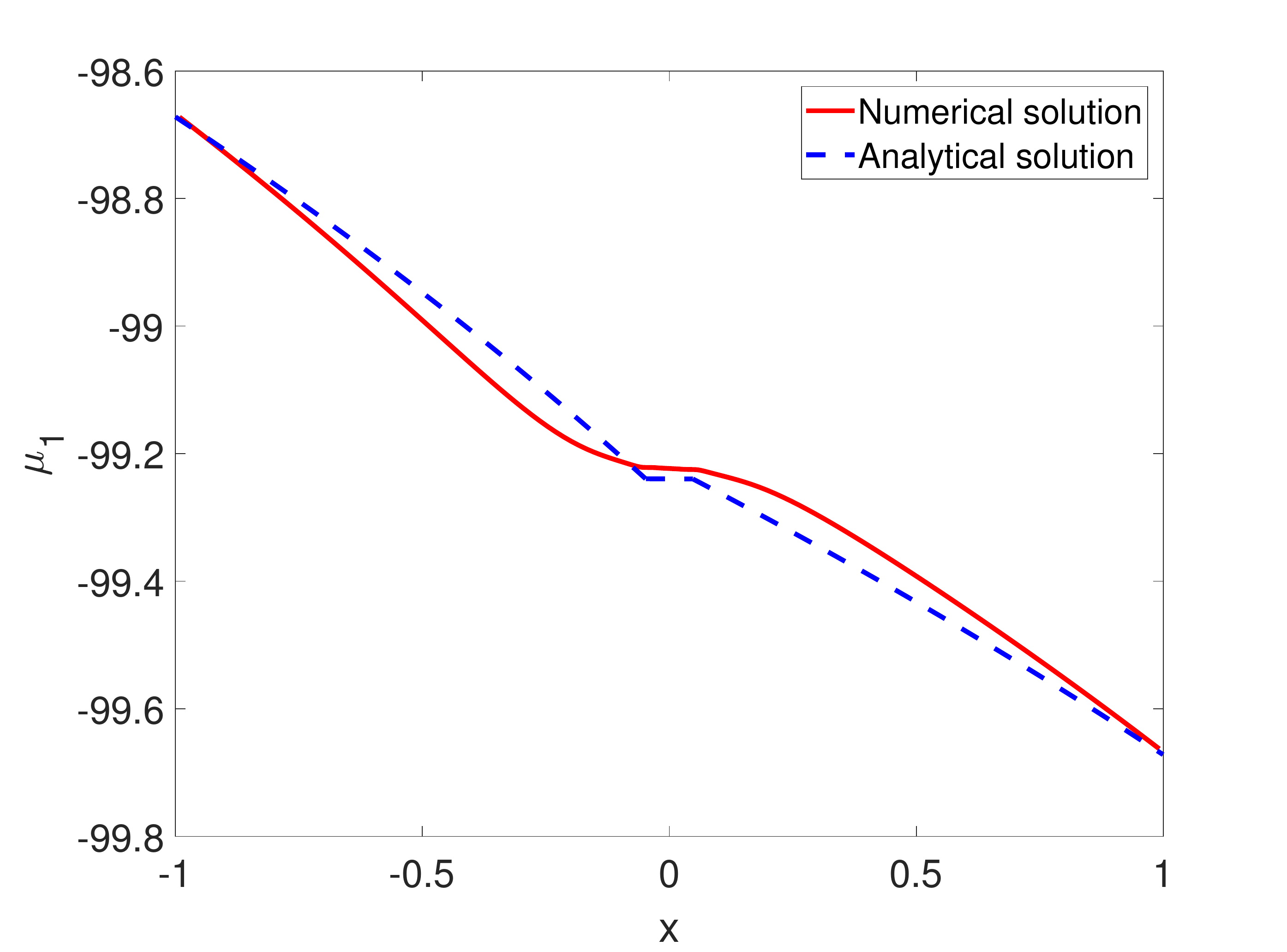}
\includegraphics[width=2in]{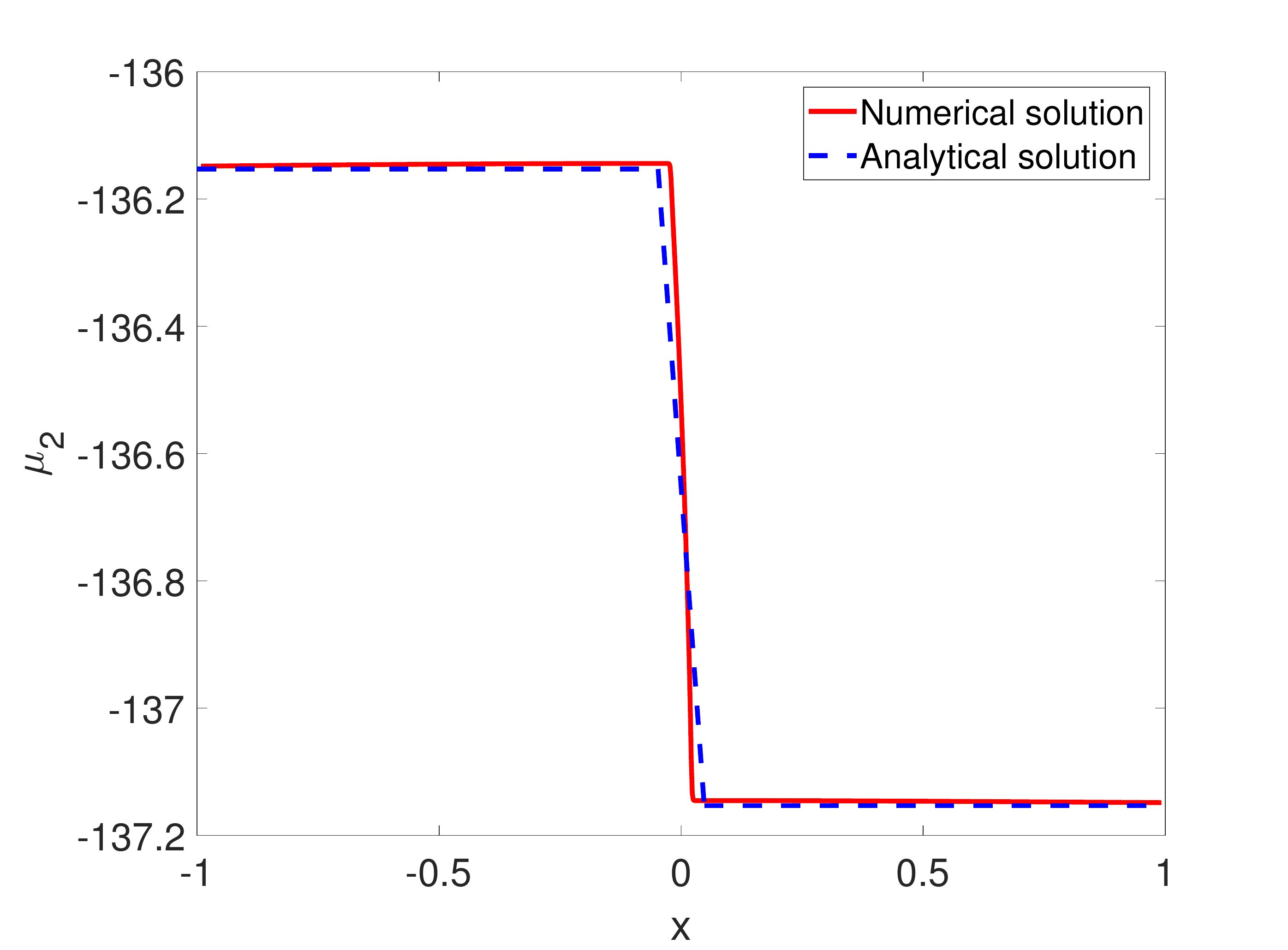}\includegraphics[width=2in]{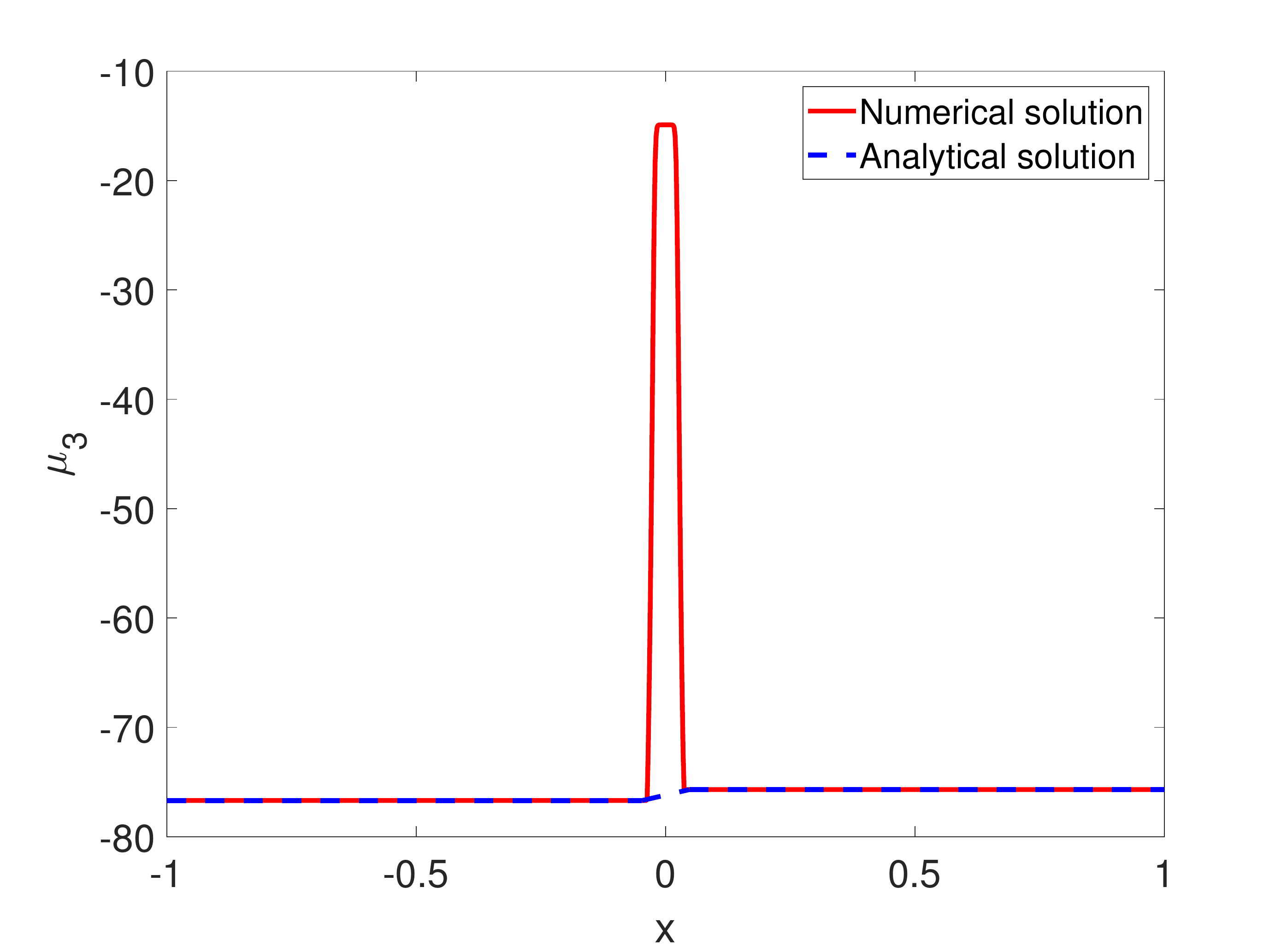}
\caption{\label{Fig19}The $\mu_i$ ($i=1,2,3$) near steady state for $V=1,q=600$.}
\end{center}
\end{figure}

Now we provide some insight and explanation for above wrong $\mu_3$, based on previous analytical results. We can easily prove that $\mu_3$ is monotone in steady state by the positivity of $c_3$.  Thus, Figure \ref{fig8}c  is correct and direct numerical result in Figure \ref{Fig19}c is wrong. By definition of $\mu_3$ and matching with boundary conditions (two  values of $\mu_3$ at boundaries do not differ much), in filter we approximately have
\begin{equation}
\label{Eq53}
\begin{aligned}
&\log c_3 - \log \left(1- \sum_{k=1}^3  c_k(x) a_k^3 \delta\right) \sim -37.5 +\phi \sim -90
\end{aligned}
\end{equation} 
Since the second term is $O(1)$ for present $q=600$ not exceeding the critical value 790, we need $c_3$ to be as accurate as $e^{-90}\sim 10^{-40}$. We know it is almost 0, but to compute correct $\mu_3$ in filter, it has to go to as small as $10^{-40}$. This is partially verified numerically, i.e., when we increase the accuracy of $c_3$ in filter, the values of $\mu_3$ in filter as in Figure \ref{Fig19}c will decrease further (in both cases $V=0$ and $V=1$). In addition, the  accuracy of $c_3$ would also affect other results in filter, to certain degree. For example, if we only keep accurate up to $10^{-10}$, the minimum values of $\phi$ are wrong (differ much from analytical results), and it works for $\phi$ when we keep accurate up to $10^{-15}$ (Figure \ref{Fig16}(b)  is based on this). The inaccuracy of $\mu_3$ is also one reason that the profile of $c_3$ in Figure \ref{Fig15} has a relatively larger discrepancy with analytical resutls.





When Ca$^{2+}$ is present and with above $q=600$, the results and features are very similar to above results (omitted here), and this agrees with previous analysis. We also tried for large $q=1000$ in above 3-ion case and in a case with Ca$^{2+}$, but the computation is very unstable and failed to capture the features in analysis. Now we give some explanation based previous analysis and provide some insight on the numerical difficulty. In such cases, the ions saturate in filter and thus the second term in $\mu_i$ of Eq. (\ref{eq6}) is crucial and requires very high accuracy for $c_i$ in computation. Take the 3-ion case with Ca$^{2+}$ in Section 3.2 for example, one can see that even for the simple case of algebraic equations from (\ref{eq33}) and  (\ref{eq12}, \ref{eq11}), it is not straightforward to determine $\phi$. Originally, the solution depends on identity (\ref{eq11}), and from the solution in (\ref{eq35})  we find that in this case
\begin{equation}
\label{Eq54}
\begin{aligned}
&\log \left(1- \sum_{k=1}^3  c_k(x) a_k^3 \delta\right)  \sim -170.
\end{aligned}
\end{equation} 
This causes the main difficulty of direct numerical simulation, as this term is essential to capture the behaviour in filter. One should be very cautious to calculate $c_i$ directly in simulation, since both Ca$^{2+}$ and K$^{+}$ are in the order $O(q)$ but they need to be accurate to $e^{-170}$ to capture this term. Other difficulty can also come from $\log c_i$ terms, as some ion like Cl$^{-}$ is exponentially small (this is already illustrated in last paragraph for previous case). These difficulties can be avoided if the $c_i$ can be represented by $\phi$, as $\phi$ behaves good in analysis and computation. This can be easily done for equilibrium case with help of formula (\ref{eq12},\ref{eq11}), but not straightforward in dynamic case.

We also briefly mention the 3-ion case of K$^{+}$, Na$^{+}$, Cl$^{-}$ with large large $q$, as in Section 3.1.2.  It is similar for the difficulties from the two log terms in $\mu_i$ in (\ref{eq6}). In addition, analytical solution or Figure 4c shows that there is an internal transition point for $c_1$ in filter, where $c_1$ changes from exponential small to $O(q)$. In some part of filter, on the one hand $c_1$ is exponential small, and on the other $J_1\sim c_1 \partial_x \mu_1$ should be finite. It is not easy to capture the transition or to compute the form $0*\infty$.

\section{Hybrid computational-asymptotic analysis}

When $q$ is large, direct numerical computation becomes challenging and inefficient. In addition, when $\epsilon$ is relatively large (i.e., short chamber length $L$), our analysis for the J-V relation in Section 4 fails since the EN assumption is no longer valid in the chamber. 

In this subsection, we provide an alternative hybrid method by combining asymptotic analysis in the filter with numerical computation in the chamber. We obtain an analytical solution in filter for non-equilibrium case by slightly modifying that from Section 3.1.2, and in the chamber we can simplify the system, which is generally easy to solve numerically (no such difficulties mentioned in last subsection) or relates to some special functions. We could also call the solutions in the subsection as semi-analytical solutions.

We take the three ion case K$^{+}$, Na$^{+}$ and Cl$^{-}$ as illustration, and assume $A=1$ and $\epsilon_r=1$ in the chamber (the general case should not cause essential difficulty). The dimensional length can be either large or small (reflected in parameter $\epsilon$), say $L=10.5$ nm in previous sections or $L=3$ nm in more practical case. The system in the right chamber by neglecting $O(\delta)$ term is 
\begin{equation}
\label{Eq55}
\begin{aligned}
& c_1' (x)+ c_1  \phi'(x)= -J,\\
& c_2' (x)+ c_2  \phi'(x) = 0,\\
& c_3'(x) - c_3  \phi' (x) = 0,\\
& -\epsilon^2 \phi''(x) = c_1 + c_2 -c_3, \quad s<x<1.
\end{aligned}
\end{equation} 
with boundary conditions $c_i= c_{ib},\phi=0$ at $x=1$. Here position $s$ denote the edge of filter. We immediately get $c_2,c_3$ in terms of $\phi$ 
\begin{equation}
\label{Eq56}
\begin{aligned}
c_2= c_{2b} e^{-\phi},\quad  c_3= c_{3b} e^{\phi},
\end{aligned}
\end{equation} 
so that
\begin{equation}
\label{Eq57}
\begin{aligned}
& -\epsilon^2 \phi''(x) = c_1 + c_{2b} e^{-\phi} -c_{3b} e^{\phi}, \quad s<x<1.
\end{aligned}
\end{equation} 
Multiplying $\phi'$ on this equation and using $(\ref{Eq55})_1$, we obtain
\begin{equation}
\label{Eq58}
\begin{aligned}
& c_1(x) = \epsilon^2 \frac{1}{2}[ (\phi'(x))^2 -(\phi'(1))^2] -J(x-1) +c_{1b} - c_{2b} (e^{-\phi}-1) - c_{3b} (e^{\phi} -1).
\end{aligned}
\end{equation}
Substituting into equation (\ref{Eq57}) and with $c_{1b}+c_{2b} = c_{3b}$, we obtain 
\begin{equation}
\label{Eq59}
\begin{aligned}
\epsilon^2 \phi''(x) =  -\frac{1}{2} \epsilon^2 [ (\phi'(x))^2 -(\phi'(1))^2] + J(x-1) + 2c_{3b} (e^{\phi} -1), \quad s<x<1.
\end{aligned}
\end{equation}
Similarly for the left chamber with boundary conditions $c_i =c_{ib}$ and $\phi=V$, we would have
\begin{equation}
\label{Eq60}
\begin{aligned}
\epsilon^2 \phi''(x) =  -\frac{1}{2} \epsilon^2 [ (\phi'(x))^2 -(\phi'(-1))^2] + J(x+1) + 2c_{3b} (e^{\phi-V} -1),\quad -1<x<s.
\end{aligned}
\end{equation}
These two equations are to be solved with help of solution in filter or with some matching connection conditions.

\noindent \textbf{Remark}: The final differential equation for $\phi$ seems complicated, but actually it relates to a special function, defined by Painlev\'e II (PII) equation. Here we would like to bring attention to this connection, as Painlev\'e transcendents have been studied intensively in last decades \cite{Wong}. The reduction of steady state PNP system with $\pm 1$ ions to PII equation was mentioned in \cite{Rubinstein}.  For the present 3-ion case, it is similar and we can adopt the transform 
\begin{equation}
\label{Eq61}
\begin{aligned}
& y = \frac{e^{\phi/2}}{\sqrt{2} (\epsilon J)^{1/3}},\quad z =  \frac{J x+ C }{2 (\epsilon J)^{2/3}},  \quad C= -J + \frac{1}{2} \epsilon^2  (\phi'(1))^2 - 2 c_{3b},
\end{aligned}
\end{equation}
so that equation (\ref{Eq58}) becomes PII equation with parameter 0,
\begin{equation}
\label{Eq62}
\begin{aligned}
& y''(z) = 2 y^3 + z y.
\end{aligned}
\end{equation}
The typical solutions in present setting is that $\phi(x)$ either blows up to $\infty$ or to $-\infty$ at $x=x^\ast$ as $x$ decreases from 1, and this agrees with some features (like poles) of solutions of PII equations.  But the reasonable  solution in current case is connected to the filter solution at $x=s$ before it reaches $x^\ast$.

Next we would like to connect above solutions in chamber with filter solution.  We take $q>{1}/{a_1^3 \delta}$ for example. In general, for non-equilibrium case, one can not express $c_i$ in terms of $\phi$ and then directly construct the solution like Section 3.1.2.  But we make use of the facts that Eq. (\ref{eq12}) still holds in non-equilibrium case. In addition, for selected ions (K$^+$ or K$^{+}$ and Ca$^{2+}$), $\mu_i$ are constants for filter region based on evidence from both analysis and simulation. Thus, the only modification of filter solution in (\ref{eq30}, \ref{eq31}) is that the constant $B_1$ is replaced by $\mu_1(s)$, which relates to chamber solution. We can determine the solutions by using shooting method. Once we fix $J$ and $\phi'(1)$, we can compute the solution of $\phi$ and hence $c_i$ ($i=1,2,3$) upto $x=s$. We treat the solution as a special function of arguments $J,\phi'(1)$. With calculated $B_1=\mu_1(s)$,  the filter solution is known. Then, the connection conditions at $x=s$ are
\begin{equation}
\label{Eq63}
\begin{aligned}
&A_f \sqrt{2 \epsilon_{r0}(G(\phi_s) - G(\phi_0))} = \epsilon \phi'(s),\quad \phi_s=\phi(s),\\
& \sqrt{\frac{\epsilon_{r0}}{2}}  \int_{\phi_0}^{\phi_s} \frac{1}{\sqrt{G(\phi) - G(\phi_0)}} d\phi =(s-s_0)/\epsilon,
\end{aligned}
\end{equation}
where $s_0$ is position of minimum of $\phi$ or $\phi'=0$ in filter. Similarly for the left chamber, with given $V,J,\phi'(-1)$, the get the solutions and then the connection conditions at $x=-s$
\begin{equation}
\label{Eq64}
\begin{aligned}
&A_f \sqrt{2 \epsilon_{r0}(G(\phi_{-s}) - G(\phi_0))} = -\epsilon \phi'(-s),\quad \phi_{-s}=\phi(-s),\\
& \sqrt{\frac{\epsilon_{r0}}{2}}  \int_{\phi_0}^{\phi_{-s}} \frac{1}{\sqrt{G(\phi) - G(\phi_0)}} d\phi =(s+s_0)/\epsilon.
\end{aligned}
\end{equation}
Note that we have $s_0=0$ for the equilibrium case $V=0$, but in general the solution is not exactly symmetric. The final condition is
\begin{equation}
\label{Eq65}
\begin{aligned}
&\mu_1(s) = \mu_1(-s).
\end{aligned}
\end{equation}
In brief, with given boundary value $V$, we have 7 nonlinear equations for 7 unknowns $\phi_0$, $\phi(\pm s)$, $\phi'(\pm1)$,$J$ and $s_0$. The case $q<{1}/{a_1^3 \delta}$ is simpler, and we do not need the two integral conditions $(64)_3$ and $(65)_3$ anymore, which are replaced by
\begin{equation}
\label{Eq66}
\begin{aligned}
& \phi_0 =  \mu_1(s)- W_1(0)+ \log (1- a_1^3 q\delta) - \log q.
\end{aligned}
\end{equation} 
Then, we have 6 nonlinear equations for 6 unknowns $\phi_0$, $\phi(\pm s)$, $\phi'(\pm1)$,$J$.

\begin{figure}[h]
\begin{center}
\includegraphics[width=2.5in]{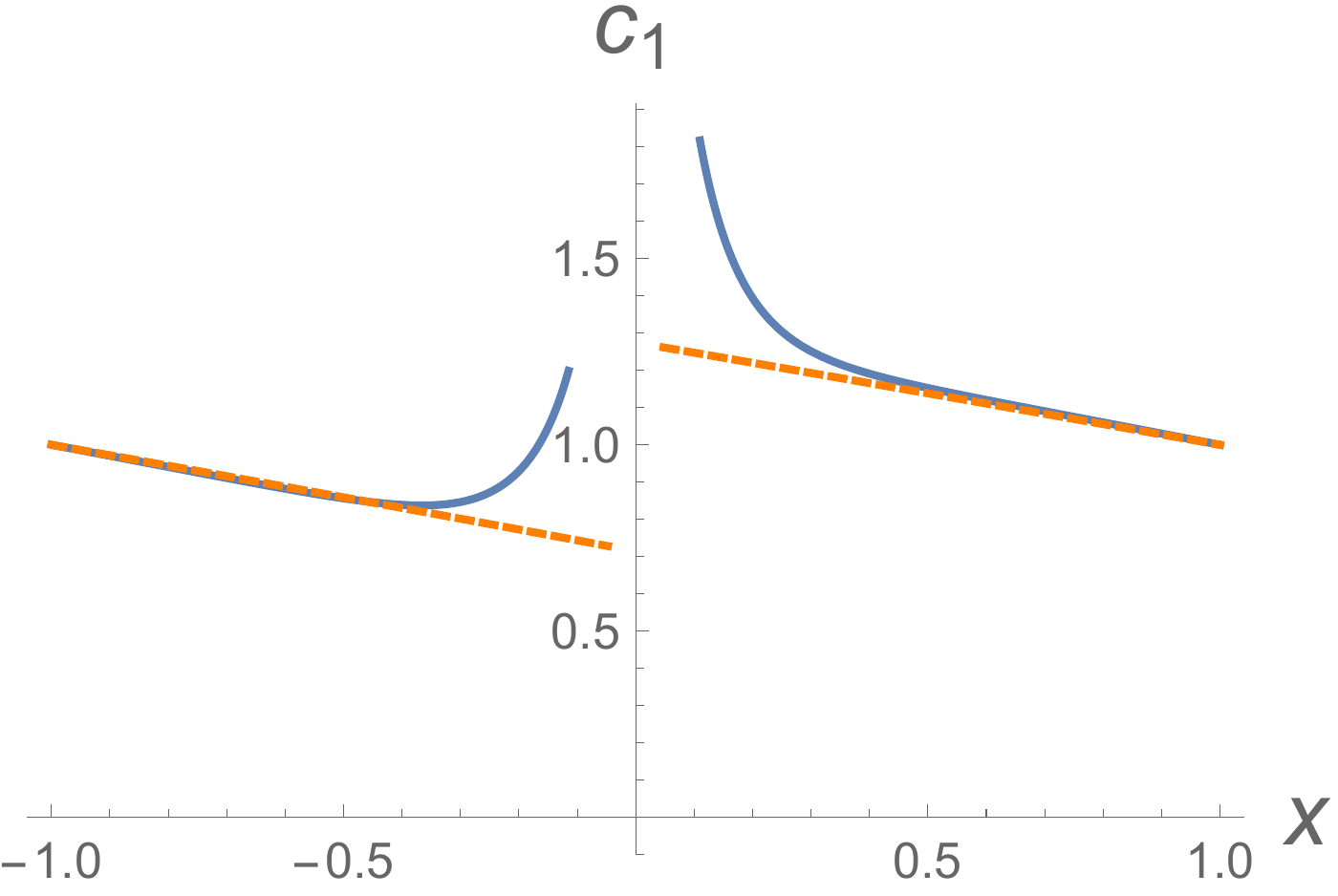} \includegraphics[width=2.5in]{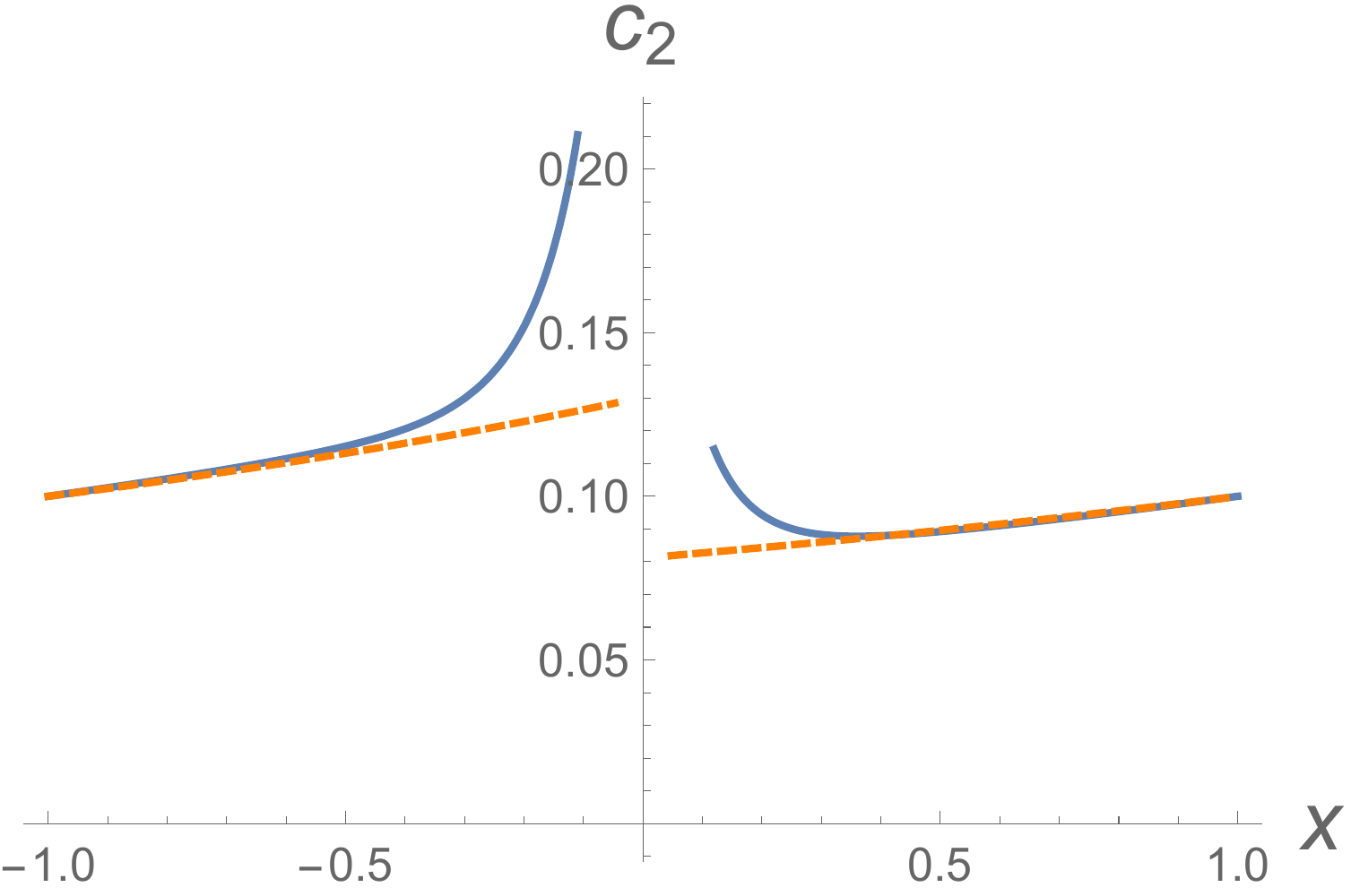}\\
\includegraphics[width=2.5in]{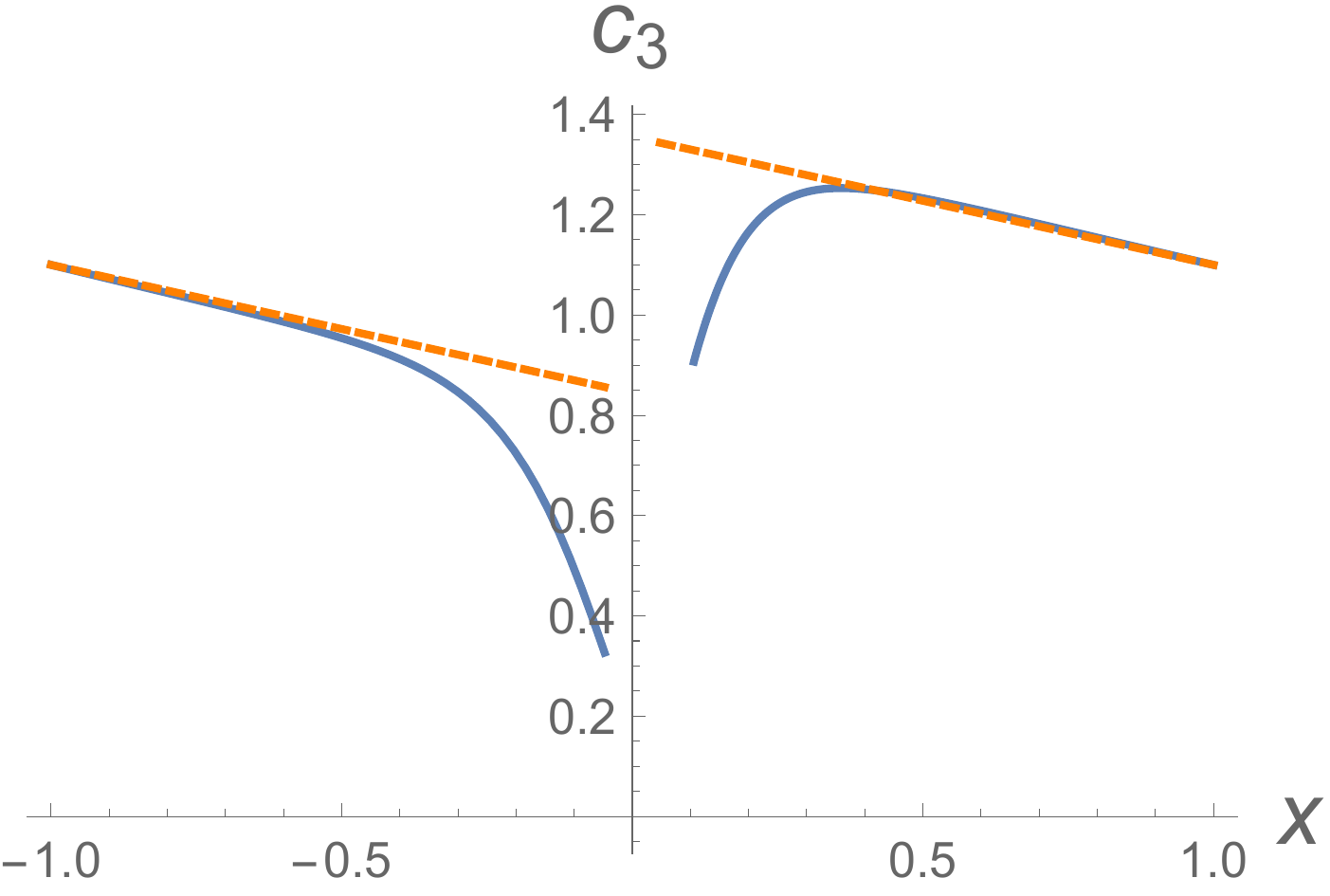} \includegraphics[width=2.5in]{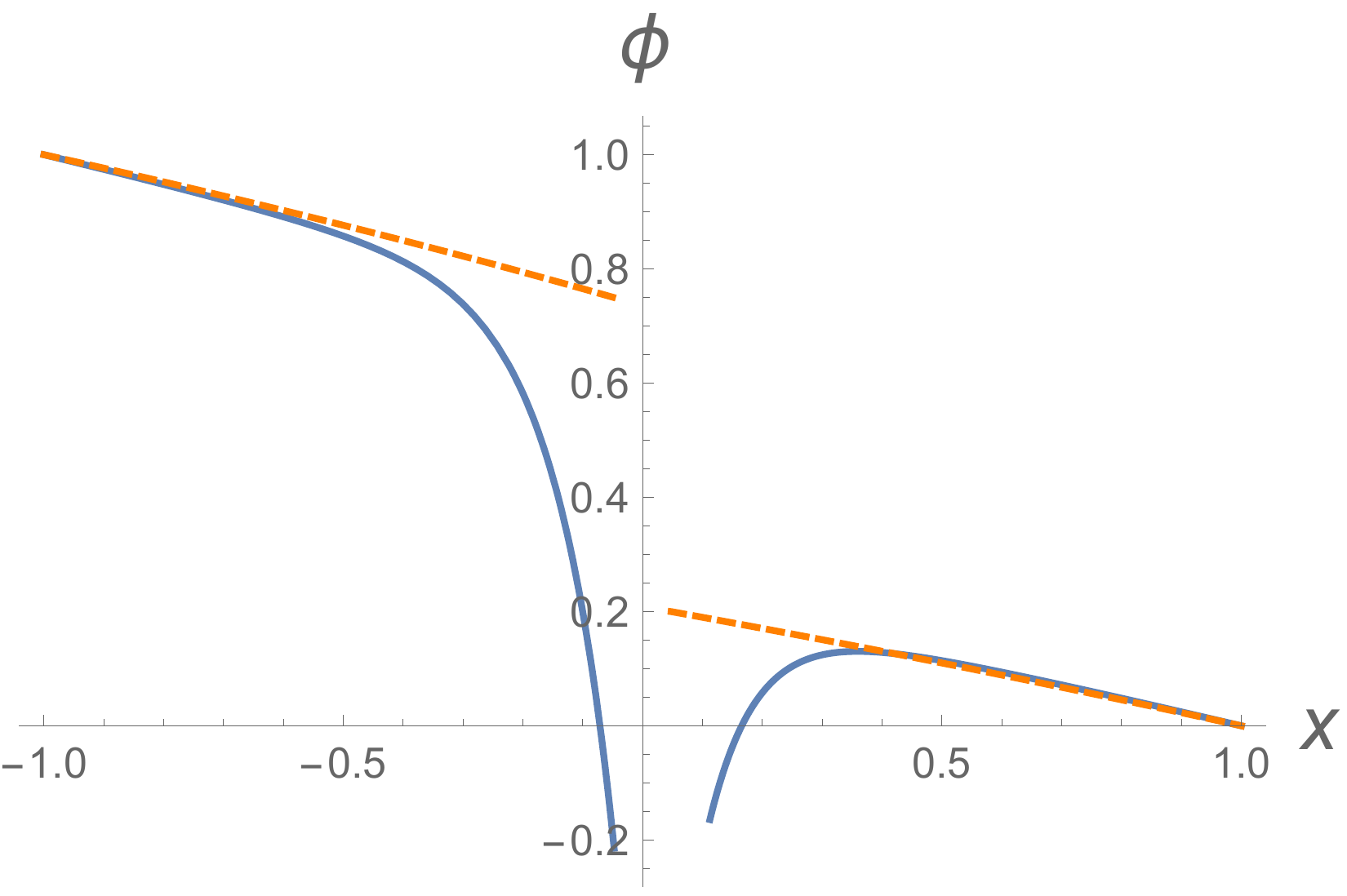}
\caption{\label{Fig20}The profiles of $c_i$ and $\phi$ with $V=1$ and $q=600$.}
\end{center}
\end{figure}

\begin{figure}[h]
\begin{center}
 \includegraphics[width=3in]{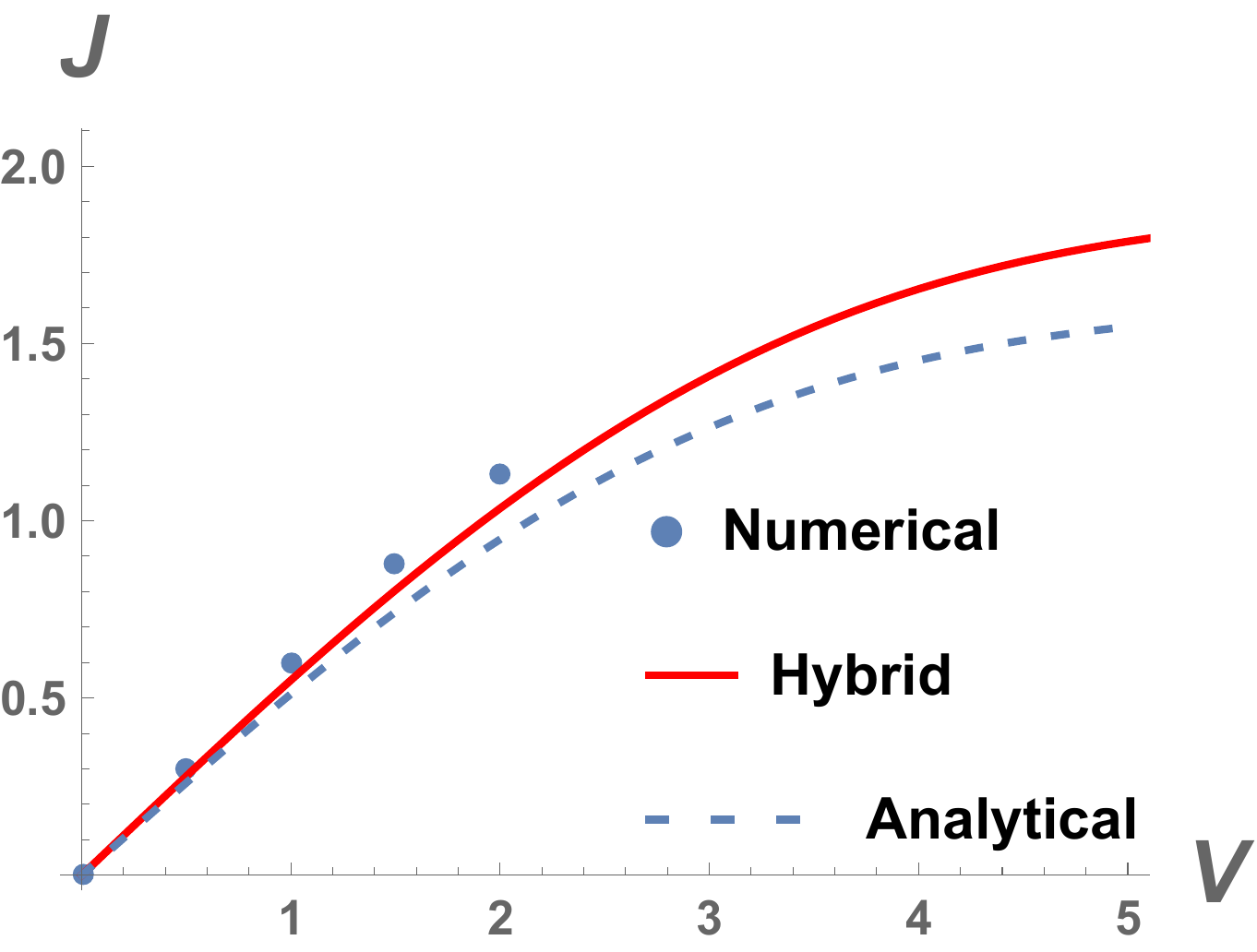} \includegraphics[width=3in]{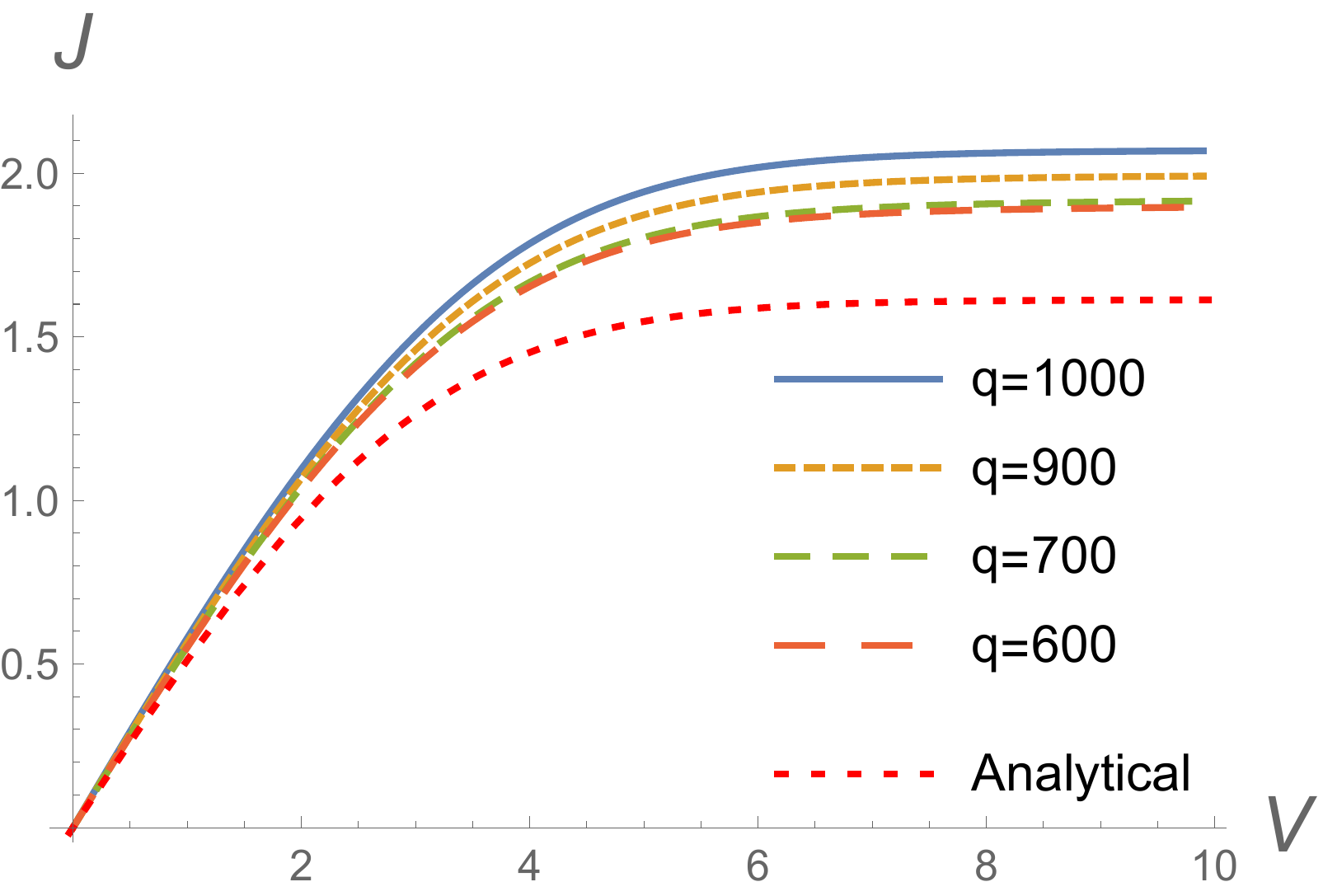}
\caption{\label{Fig21} The $J$-$V$ relations with small $\epsilon$ (dimensional length $L=10.5$ nm): (a) comparison of different methods (b) different q.}
\end{center}
\end{figure}

The above algorithm can be easily achieved in Mathematica (or Matlab) with only a few lines of code, the solutions for given $V$ can be computed by finding roots of the 6 or 7 nonlinear equations. The computation is super quick, and the solution is found within seconds on a laptop. This is verified with $V=0,q=1000$ and data in (\ref{eqA3}), and it coincides with previous results in Section 3.1.2. For previous case  $q=600$, $V=1$ in Section 4.1, the solutions are computed for comparison. The profiles of $\phi$ and $c_i$ ($i=1,2,3$) are shown in Figure \ref{Fig20} with dashed lines from previous analytical solution, showing good agreement away from filter. The flux computed here is $J\approx 0.55$ (or $J_1\approx1.08$), also indicating that the previous approximation $J\approx 0.51$ in Section 4.1 slightly underestimates the flux. 

\noindent \textbf{Remark}: There is a fictitious singularity in the integrals in (\ref{Eq63},\ref{Eq64}), i.e., the integrand is singular at $\phi=\phi_0$, but the integral is like $\int_0^a \frac{1}{\sqrt{x}} dx$. We have used a little trick in pratical computation to ensure stability and accuracy, i.e., replace $\phi_0$ by $\phi_0 + \delta_0$, say $\delta_0 = 10^{-10}$ (this would be helpful if one wants to repeat above computation). For quite small $\epsilon$ (long dimensional $L$) and large $c_{2b}$, the solution of $\phi$ is sensitive to boundary conditions $\phi'(\pm1)$. It can easily blow up to $\pm \infty$, and only a narrow interval of $\phi'(\pm1)$ with given $J$ leads to solution of $\phi$ in whole interval $[s,1]$.

For different $V$, the flux-voltage (IV) relations by three different methods are compared in Figure \ref{Fig21}a, where red curve is from current section, and dots and dashed lines are from previous numerical and analytical solutions. Although different approximations regarding boundary layer near filter or parameters $A(x),\epsilon_r(x)$ are made, the three methods provide similar results and trend for IV curve. The analytical solution underestimates the flux, due to the neglect of boundary layer near edge, while the slight difference between numerical and hybrid methods are due to the smoothing of $\epsilon_r(x)$ and $A(x)$ used in numerical solutions. For different $q$, the flux-voltage $J$-$V$ relations are computed by varying $V$, shown in Figure \ref{Fig21}b, with reference curve from analytical result in Section 4.1. The flux in each curve saturates for large $V$, and as $q$ increases the flux will increase.

\begin{figure}[h]
\begin{center}
\includegraphics[width=3in]{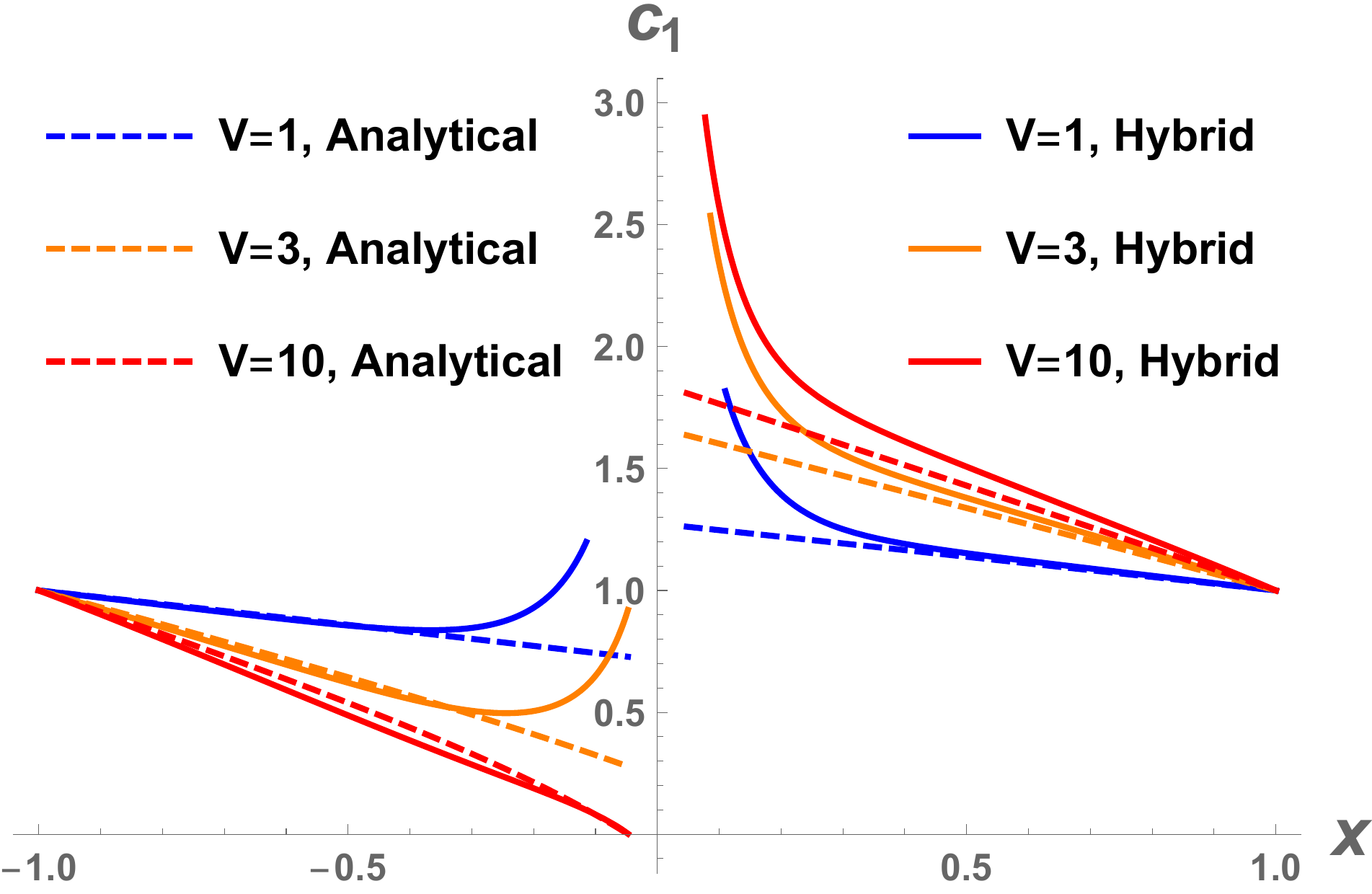}
\caption{\label{fig12} The profiles of $c_1$ in chamber for  different $V$.}
\end{center}
\end{figure}

The saturation of flux is certainly a consequence of selectivity of filter, which is originally due to parameters $\epsilon_r$ and $q$. Without filter, the flux-voltage relations will be totally different, as indicated at the end of Section 4.1. With filter, the most important condition is continuity of $\mu_i$ for selected ions. To see the direct reason of saturation of flux for the K$^{+}$/Na$^+$ case, we analyze the profiles of $c_1$ in chamber for different $V$, obtained by both analytical and hybrid methods. Figure \ref{fig12} shows the profiles of $c_1$ with parameters $c_{2b}=0.1,q=600$ and three different $V$.  The dashed lines from analytical results provide reasonable approximation in the region away from filter, but not as accurate as the solid lines near filter, which also capture the BL.  Both indicate that $c_1$ approaches 0 near the left edge of filter as $V$ increases, and one can easily see this trend from the analytical expressions in Appendix \ref{appendixB}.  The left edge of filter is important here since the flux is from left to right with positive $V$ (otherwise we should analyze the right edge). As $c_1$ can not be negative, this is the main restriction for the saturation of scaled flux $J$.  Also note that the original flux $J_1$ is controlled by diffusion constant $D_1$, one may think the saturation is related to the diffusion limit \cite{Miller2002,Allen2011}. When $c_1$ is near 0 at left edge, there are not enough ions available to go through the filter even with large $V$. As $c_{2b}$ increases, $c_1$ will be more likely to reach this critical value, resulting in smaller saturation flux $J$. The reason of saturation of both fluxes for the case with Ca$^{2+}$ in Section 4.2 is similar, except that the two fluxes are restricted by values of both $c_1$ and $c_2$ at the edge of filter (both approach 0 as $V$ increases).

\begin{figure}[h]
\begin{center}
 \includegraphics[width=2.7 in]{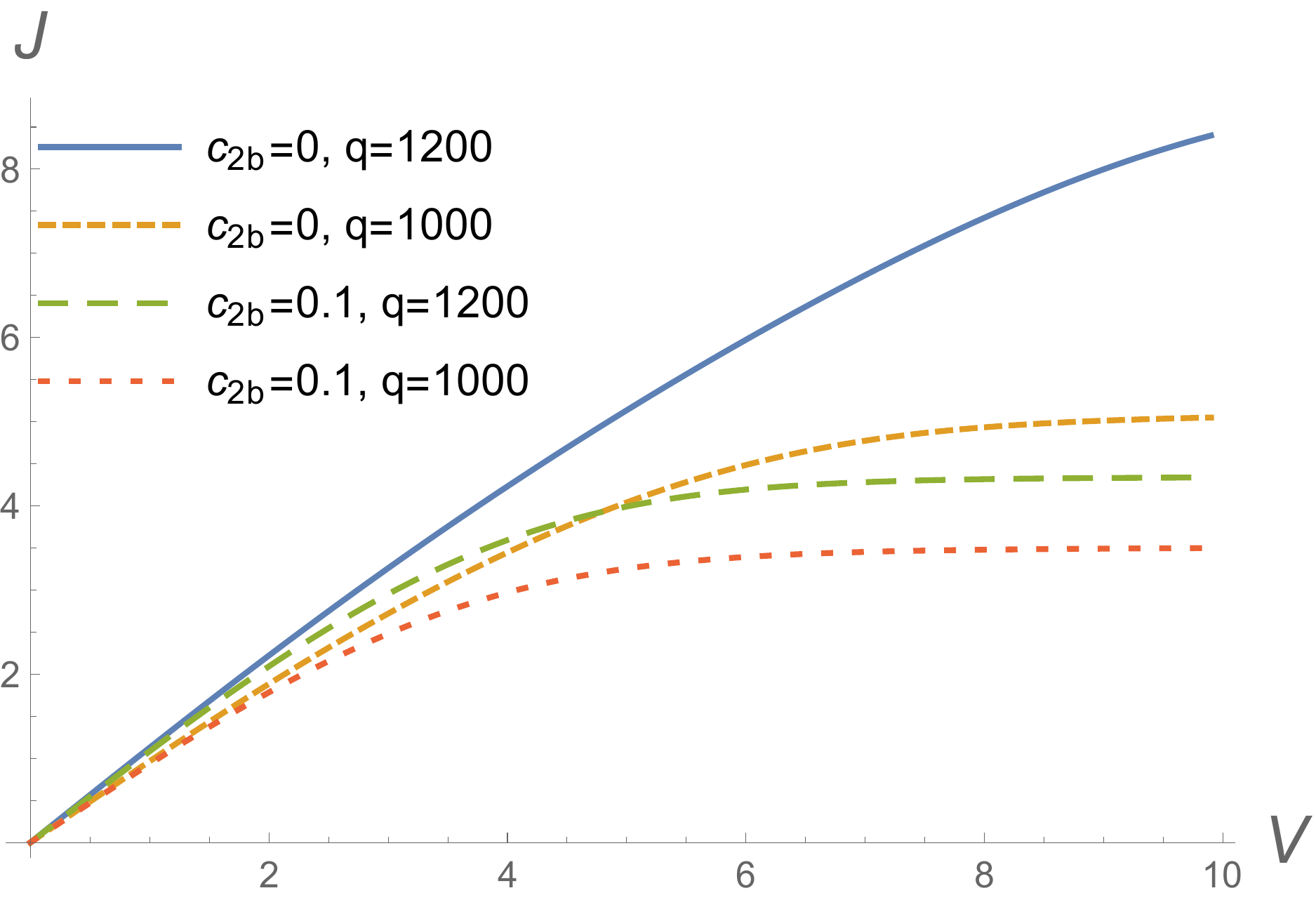}  \includegraphics[width=2.7 in]{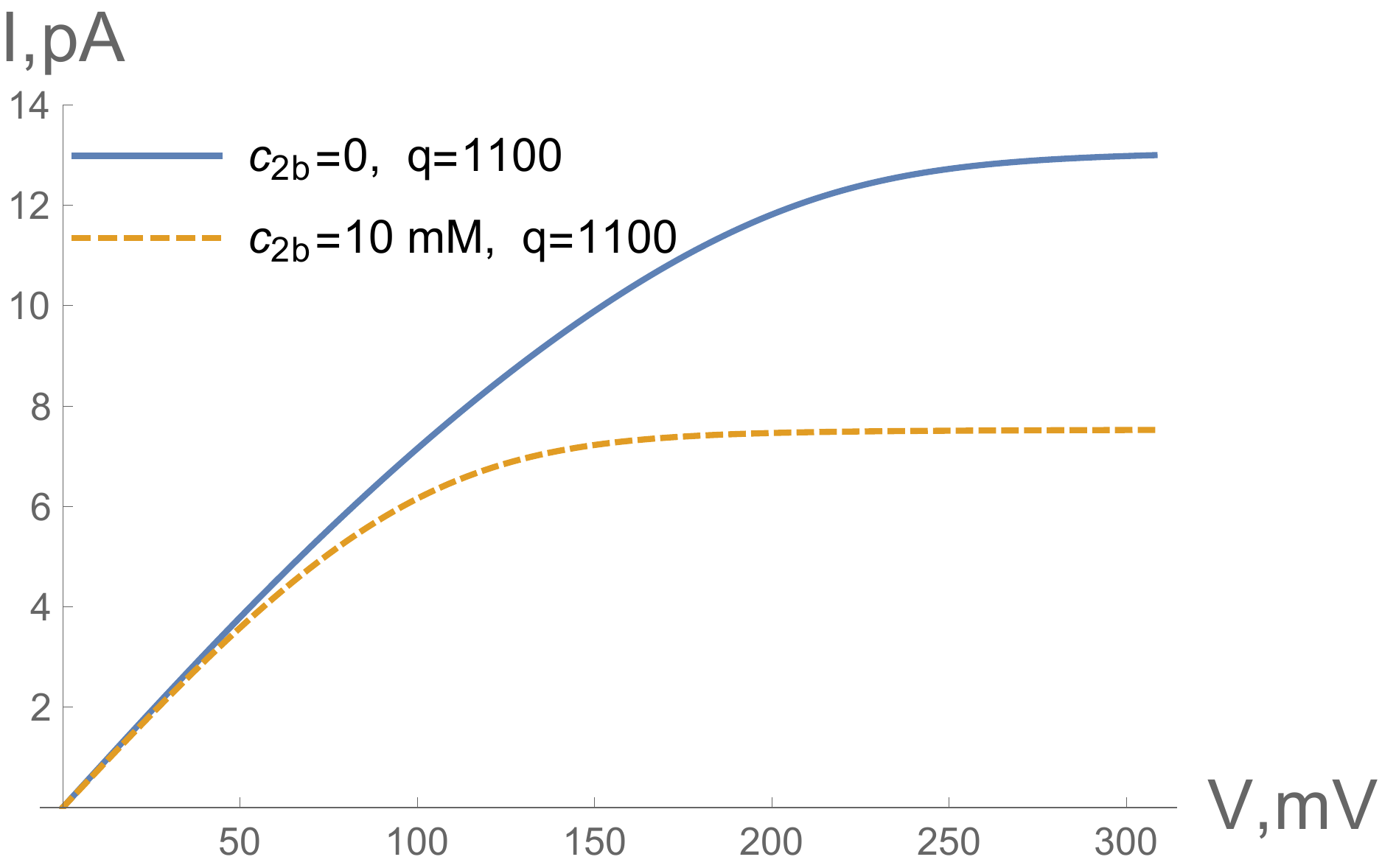}
\caption{\label{Fig22}The $J$-$V$ relations with different $q,c_{2b}$ and relatively large $\epsilon$ (small dimensional length $L=3$ nm).}
\end{center}
\end{figure}

The hybrid method in this section has the advantages of both efficiency and accuracy for the IV relation. The direct numerical computation is extremely time-consuming, even for one point in the IV  curve of Figure \ref{Fig21}(a), thus it can hardly be used to compare IV relations with experiments. The hybrid method can produce IV curves efficiently, say 20 min for one smooth curve in Figure \ref{Fig21}(b). It also includes the boundary layer effect near filter of edge, and does not have the restriction for parameters (like $\epsilon$ or length $L$), in contrast to analytical approximations.  Thus it can be readily used to compare with experiments or estimate parameters in the model. 

The data in (\ref{eqA3}) of Appendix \ref{appendixA} corresponds to  relatively long dimensional length $L=10.5$ nm. But in more realistic case, $L$ is much shorter based on molecular structure of KcsA channel. In order to compare with experiments, we adopt the dimensional length $L=3$ nm (i.e., $L_{b}=2.5$ nm), which leads to $\epsilon\approx 0.46$. We also compute the $J$-$V$ relations for  $c_{1b}=1$ and different $c_{2b}$ and $q$, shown in Figure \ref{Fig22}(a). As $c_{2b}$ increases, the flux will decrease, while the flux will increase as $q$ increases. From the present formulation, the dimensional flux and current are scaled by
\begin{equation}
\label{Eq67}
\begin{aligned}
& \frac{A_b D_0 c_0}{L} =  6.02 * 10^{6} /s,\quad \frac{e_0 A_b D_0 c_0}{L} =  0.96 \, \textrm{pA}
\end{aligned}
\end{equation} 
where $L=3$ nm is used. Note also $J_1=D_1 J$ where $D_1 = 1.96$. Figure  \ref{Fig22}b shows the I-V relations with physical units for $q=1100$, which are in similar order to figure 2B of experiment paper \cite{Miller2002}. One could also make it more comparable by adjusting other parameters, say the cross section area $A(x)$. 

The idea in this subsection can be applied to more general cases, say general $A(x)$, slowly varying $\epsilon_r(x)$ in chamber or with ion Ca$^{2+}$. The formulation and solving process are quite similar, except that we might solve more than one equation in chamber region. We will not repeat this here.

\section{Concluding remarks}

We have studied the selectivity of KcsA potassium channel and the current-voltage (IV) relation. With a 1D modified PNP system by keeping essential elements, many features of the channel have been demonstrated by both analytical formulas and numerical simulations. The selectivity among K$^+$ and other ions are clearly illustrated with analytical formulas. Saturation of IV curve is captured by various methods, and explanation is provided. We hope these methods in current work can be applied to other types of ion channels, and provide insights into the selectivity and IV relations. More work is needed to make comparison with experiments or calibrate some parameters in the model for different channels.  Some feature in detailed 3D simulations such as pile-up of ions near filter may be missed in current 1D framework. This could be due to the boundary charge distribution (instead of local source charge) in filter  and complex geometry of the channel. More work under 3D framework is ongoing as an extension of current work.

\section*{Acknowledgment}

This work was initiated when Dr. Tzyy-Leng Horng was a Fields Research Fellow at Fields Institute. 

\appendix

\section{Parameter values}
\label{appendixA}

The data in this Appendix are mainly from \cite{jinnliangliu2014, Hille2001, song2018, Malmivuo1995}. For dimensional system, the vacuum permittivity $\epsilon_0$, elementary charge $e_0$, Boltzmann constant $k_B$ and absolute temperature $T$ are 
\begin{equation}
\label{eqA1}
\begin{aligned}
&\epsilon_0 = 8.854\times 10^{-12} \, \textrm{C}/(\textrm{V}\cdot \textrm{m}),\quad e_0 = 1.602 \times 10^{-19} \, \textrm{C}, \\
&  k_B = 1.38 \times 10^{-23}\, \textrm{J}/\textrm{K},\quad T = 300\, \textrm{K}.
\end{aligned}
\end{equation}
Some typical values are adopted as 
\begin{equation}
\label{eqA2}
\begin{aligned}
& \phi_0 = \frac{k_B T}{e_0} \approx 24\, \textrm{mV},\quad c_0 = 100 \, \textrm{mM} =  6.022 \times 10^{25}\, \textrm{m}^{-3},\quad D_0 = 10^{-9} \, \textrm{m}^2/\textrm{s},\\
& a_0=3 \, \textrm{\AA},\quad L_b=10 \, \textrm{nm},\quad L_f = 1 \, \textrm{nm}, \quad L= 10.5 \, \textrm{nm},\\
& \epsilon_{rb}=80,\quad \epsilon_{rf} = 2,\quad A_b = 30\,  \textrm{\AA}^2,\\
& a_{\textrm{K}} = 2.76 \, \textrm{\AA}, \quad a_{\textrm{Na}} = 2.04 \, \textrm{\AA},\quad a_{\textrm{Ca}} = 1.98 \, \textrm{\AA},\quad a_{\textrm{Cl}} = 3.62 \, \textrm{\AA}, \quad a_{Ba} = 2.70 \, \textrm{\AA}.
\end{aligned}
\end{equation}
If we think of exact sphere instead of cube, the factor $(\pi/6)^{1/3}\approx 0.8$ should be multiplied to above effective diameters of ions $a_i$.

For dimensionless system, we have the estimates of dimensionless parameters
\begin{equation}
\label{eqA3}
\begin{aligned}
&\epsilon  \approx 0.13,\quad \delta = a_0^3 c_0 \approx 1.6\times 10^{-3}, \quad W_0 =  \frac{e^2}{8 \pi \epsilon_0 a_0 k_B T}\approx 93,\\
& \frac{1}{40} \le \epsilon_r \le 1, \quad A_f \le A \le 1,\quad L_f = 0.095,\\
& D_{\textrm{K}} = 1.96,\quad D_{\textrm{Na}} = 1.33,\quad D_{\textrm{Ca}} = 0.79,\quad D_{\textrm{Cl}} = 2.03,\\
& a_{\textrm{K}} = 0.92, \quad a_{\textrm{Na}} = 0.68,\quad a_{\textrm{Ca}} = 0.66 ,\quad a_{\textrm{Cl}} = 1.21,\quad a_{\textrm{Ba}} = 0.9.\\
\end{aligned}
\end{equation}
The permanent charge and cross section area are estimated from a 3D Poisson-Boltzmann computation based on realistic molecular structure of KcsA. The corresponding dimensionless quantities for $q$ and $A_f$ are
\begin{equation}
\label{eqA5}
\begin{aligned}
&  q \sim 10^3 , \quad \textrm{e.g.,} \quad [1000, 2000]\\
& A_f\sim \frac{1\, \textrm{\AA}^2}{30 \, \textrm{\AA}^2 }= \frac{1}{30}.
\end{aligned}
\end{equation}

\section{Some solutions and expressions}
\label{appendixB}

From definition (\ref{eq6}), we get
\begin{equation}
\label{Eq72}
\begin{aligned}
\frac{ c_i}{1- \sum_{k=1}^n \delta c_k a_k^3} = e^{ \mu_i-W_i- z_i \phi }, \quad i=1,..,n,
\end{aligned}
\end{equation}
then by multiplication of $a_i$ and summation, we obtain
\begin{equation}
\label{Eq73}
\begin{aligned}
&\frac{ C}{1-  C\delta } = \sum_{i=1}^n a_i^3 e^{ \mu_i - W_i- z_i \phi} \equiv F, \quad C= \sum_{i=1}^n c_i a_i^3,
\end{aligned}
\end{equation}
which implies
\begin{equation}
\label{Eq74}
\begin{aligned}
& C= \frac{F}{1+ F\delta }, \quad c_i = \frac{e^{ \mu_i-W_i- z_i \phi }}{(1+ F \delta)}.
\end{aligned}
\end{equation}

The solution of (\ref{eq27}) in chamber region is given by
\begin{equation}
\label{eqB1}
\begin{aligned}
& \phi(X) = 2 \log \left( \frac{e^{\sqrt{2}X} + m}{e^{\sqrt{2}X} - m}  \right),\quad m =\frac{e^{\sqrt{2} S} (e^{\phi_s/2} -1)}{e^{\phi_s/2} +1}, \quad S<X<\infty.
\end{aligned}
\end{equation}

For the system (\ref{eq41}), we get for the left-half chamber $-1<x<0$
\begin{equation}
\label{eqB2}
\begin{aligned}
& c_3(x) = 1+c_{2b}- \frac{J}{2} (x+1),\quad \phi(x) = \log  \frac{c_3(x)}{1+ c_{2b}}  + V,\\
& c_2(x) =\frac{c_{2b} (1+c_{2b})}{c_3(x)},\quad c_1(x) = c_3(x) - c_2 (x), 
\end{aligned}
\end{equation} 
and for the right-half chamber  $0<x<1$
\begin{equation}
\label{eqB3}
\begin{aligned}
& c_3(x) = 1+c_{2b}- \frac{J}{2} (x-1),\quad \phi(x) = \log  \frac{c_3(x)}{1+ c_{2b}},\\
& c_2(x) =\frac{c_{2b} (1+c_{2b})}{c_3(x)},\quad c_1(x) = c_3(x) - c_2 (x).
\end{aligned}
\end{equation} 
Based on the solutions, we get the $\mu_1(x)$ for left chamber
\begin{equation}
\label{eqB3_1}
\begin{aligned}
\mu_1(x) =& \log c_1 + \phi + W_1\\
= & \log \left(c_3(x) - \frac{c_{2b} (1+c_{2b})}{c_3(x)}\right) + \log  \frac{c_3(x)}{1+ c_{2b}} +V+ W_1\\
=& \log \left(\frac{c_3^2(x)}{1+ c_{2b}} - c_{2b} \right) +V+ W_1\\
=& \log \left(\frac{[1+c_{2b}- \frac{J}{2} (x+1)]^2}{1+ c_{2b}} - c_{2b} \right) +V+ W_1,
\end{aligned}
\end{equation} 
substituting $x=0$ give the left-hand side of (\ref{eq42}) except the $W_1$ term. 

For general $A(x)$, the linear terms $x+1,x-1$ in $c_3(x)$ in (\ref{eqB2},\ref{eqB3}) should be replaced by 
\begin{equation}
\label{eqB4}
\begin{aligned}
& \int_{-1}^x \frac{1}{A(s)} ds,\quad \int_{1}^x \frac{1}{A(s)} ds,
\end{aligned}
\end{equation} 
and all the other expressions are the same. The final result for $J$-$V$ relation is almost the same except that $J$ is multiplied by a factor $\int_{L_f/2}^1 \frac{1}{A(s)} ds$.

The system (\ref{eq46}) is equivalent to a system for functions of $\phi$
\begin{equation}
\label{eqB5}
\begin{aligned}
& \dot{c}_1 + {c}_1 = -\tilde{J}_1 \dot{x},\quad \dot{c}_2 +2 c_2 = -\tilde{J}_2 \dot{x},\\
& \dot{c}_3 - c_3 = 0,\quad c_1 + 2c_2 - c_3 = 0,
\end{aligned}
\end{equation} 
where dot represents derivative with respect to $\phi$. Then the solutions $x_R(\phi)$ and $c_{iR}(\phi)$ ($i=1,2,3$) for right-half interval $0<x<1$ (i.e., $\phi_{0R}<\phi<0$ or $0<\phi<\phi_{0R}$) are
\begin{equation}
\label{eqB6}
\begin{aligned}
&   c_{3R}(\phi) =   (2 c_{2b}+1) e^{\phi},\\
  &  c_{2R}(\phi) = \frac{(3 c_{2b} \tilde{J}_1-2 \tilde{J}_2)
   e^{\lambda \phi}}{3
   \tilde{J}_1+4 \tilde{J}_2}+\frac{2 (2 c_{2b}+1)
   \tilde{J}_2 e^{\phi}}{3 \tilde{J}_1+4 \tilde{J}_2},\quad \lambda = -\frac{2
   (\tilde{J}_1+\tilde{J}_2)}{\tilde{J}_1+2 \tilde{J}_2},\\
    &  c_{1R}(\phi)   =    c_{3R}(\phi) - 2    c_{2R}(\phi), \\
& x_R(\phi)= 1+ \frac{3
   c_{2b}+2}{\tilde{J}_1+\tilde{J}_2}  -\frac{6 (2 c_{2b}+1) e^{\phi}}{3
   \tilde{J}_1+4 \tilde{J}_2}+\frac{(3 c_{2b}
   \tilde{J}_1-2 \tilde{J}_2) e^{\lambda \phi}}{(\tilde{J}_1+\tilde{J}_2) (3 \tilde{J}_1+4
   \tilde{J}_2)}.
\end{aligned}
\end{equation} 
The solutions $c_{iL}(\phi)$ and $x_L(\phi)$ for left-half interval $-1<x<0$ (i.e., $V<\phi<\phi_{0L}$ or $\phi_{0L}<\phi<V$) are
\begin{equation}
\label{eqB7}
\begin{aligned}
&   c_{iL}(\phi) =  c_{iR}(\phi-V), \quad i=1,2,3,\\
& x_L(\phi)= x_R(\phi-V)-2.
\end{aligned}
\end{equation} 
For the general case of $A(x)$, one only needs to make a transformation $y= \int_{\pm1}^x \frac{1}{A(s)}ds$ for right and left chamber equations. The only modifications of above solutions are 
\begin{equation}
\label{eqB8}
\begin{aligned}
& y_R(\phi)= \frac{3 c_{2b}+2}{\tilde{J}_1+\tilde{J}_2}  -\frac{6 (2 c_{2b}+1) e^{\phi}}{3
   \tilde{J}_1+4 \tilde{J}_2}+\frac{(3 c_{2b}
   \tilde{J}_1-2 \tilde{J}_2) e^{\lambda \phi}}{(\tilde{J}_1+\tilde{J}_2) (3 \tilde{J}_1+4
   \tilde{J}_2)}, \quad y_L(\phi) = y_R(\phi-V).
\end{aligned}
\end{equation} 
For flux voltage relations, the equations in (\ref{eq47}) will not change and the equations in (\ref{eq48}) change to  
\begin{equation}
\label{eqB9}
\begin{aligned}
& y_R(\phi_{0L})=\int_{1}^{L_f/2} \frac{1}{A(s)}ds, \quad y_L(\phi_{0L}) = \int_{-1}^{L_f/2} \frac{1}{A(s)}ds.
\end{aligned}
\end{equation}

\bibliographystyle{plain}
\bibliography{reference_size}


\end{document}